\documentclass[letterpaper,11pt]{article}
\usepackage[letterpaper,top=2cm,bottom=2cm,left=3cm,right=3cm,marginparwidth=1.75cm]{geometry}
\usepackage{array}
\usepackage{float}
\usepackage{amsmath}
\usepackage{authblk}
\usepackage{amssymb}
\usepackage{graphicx}
\usepackage{subcaption}
\usepackage{xcolor}
\usepackage[numbers,sort&compress]{natbib}
\usepackage{hyperref}
\usepackage{comment}
\definecolor{darkgreen}{RGB}{0, 150, 0}

\newcommand{\pt}{p_T}
\newcommand{\Dzero}{D^0}
\newcommand{\Qsquared}{Q^2}
\newcommand {\jpsi}{J/\psi}
\title{Inclusive open charm photoproduction in ultraperipheral collisions at the LHC with G$\gamma$A-FONLL}

\author[1,2]{Matteo Cacciari}
\author[3]{Gian Michele Innocenti}
\author[4]{Anna M. Sta\'sto}

\affil[1]{\small \it Sorbonne Université, CNRS, Laboratoire de Physique Théorique et Hautes Énergies, LPTHE, F-75005 Paris, France}
\affil[2]{\small \it
Université Paris Cité, LPTHE, F-75006 Paris, France}
\affil[3]{\small \it Massachusetts Institute of Technology, Cambridge, MA 02139,  USA}
\affil[4]{\small \it Department of Physics, Penn State University, University Park, PA 16802, USA}

\begin{document}
\maketitle
\begin{abstract}
We compute the inclusive $D^{0}$ production cross section in ultraperipheral Pb–Pb collisions at the LHC as a function of the $D^{0}$ transverse momentum and rapidity. These calculations are carried out within the new G$\gamma$A–FONLL (Generalized Photon–Nucleus FONLL) framework, which can predict photonuclear cross sections for charm and beauty hadrons in electron–proton, electron–nucleus, and ultraperipheral heavy-ion collisions. The framework relies on FONLL (Fixed-Order Next-to-Leading Logarithm) to model heavy-quark production in photonuclear collisions and employs a photon-flux reweighting procedure to describe the production cross sections in ultraperipheral heavy-ion collisions. The G$\gamma$A calculations are first validated against the photoproduction cross sections of $D^{*}$ in electron–proton collisions at HERA. The predictions for the $D^{0}$ production cross section in ultraperipheral Pb–Pb collisions at the LHC are then presented and compared to the first experimental results obtained by CMS at $\sqrt{\rm s_{\scriptscriptstyle NN}}=5.36~\text{TeV}$. The predictions are benchmarked against different choices of nuclear parton distribution functions, fragmentation functions, and renormalization and factorization scales.
\end{abstract}
\section{Introduction}
\label{sec:intro}
\begin{figure}
    \centering
    \raisebox{4pt}{
    \begin{subfigure}{0.49\textwidth}
    \includegraphics[width=0.9\textwidth]{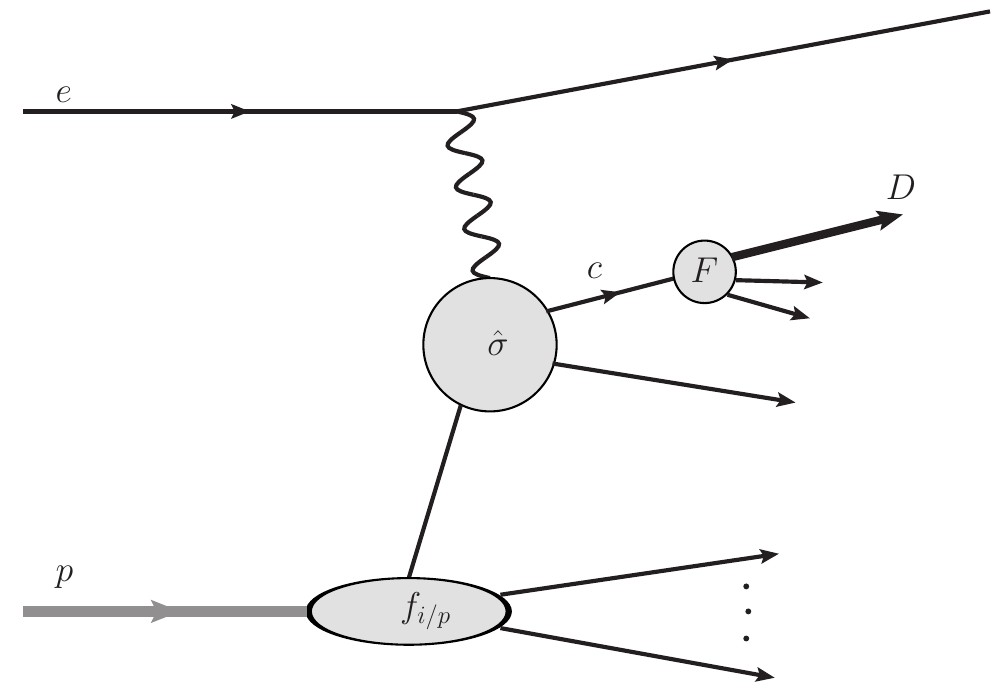}
    \end{subfigure}}
    \hfill
    \begin{subfigure}{0.49\textwidth}
    \includegraphics[width=0.9\textwidth]{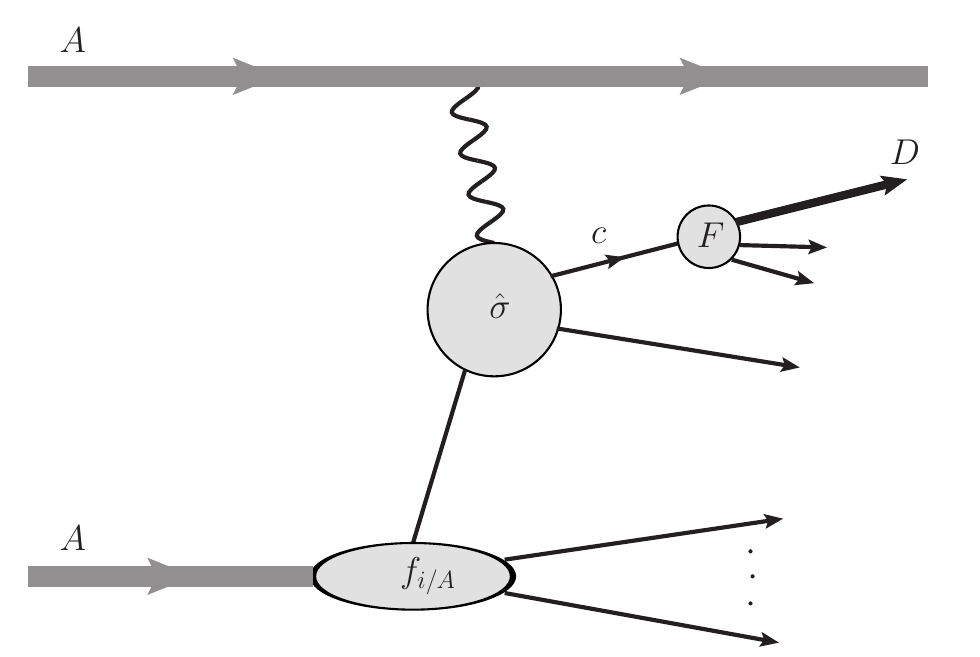}
    \end{subfigure}
    \caption{Schematic description of the inclusive $D$  meson production  in electron-proton collisions (left) and in ultraperipheral heavy-ion collisions (right). Here: $f_{i/p}$ and $f_{i/A}$ denotes PDF in the proton or nucleus, $F$ fragmentation function and $\hat{\sigma}$ hard scattering process. } 
    \label{fig:sketches}
\end{figure}

Ultraperipheral collisions (UPCs) of nuclei at very high energies provide unique opportunity to characterize the properties of nuclear structure, and perform tests of perturbative quantum-chromo dynamics (QCD) in a clean experimental environment~\cite{Bertulani:2005ru,Strikman:2005yv,Baltz:2007kq}. In UPCs, two Lorentz-contracted nuclei scatter at impact parameters that exceed the sum of the radii of the two nuclei. As a result, UPCs are dominated by electromagnetic processes, while hadronic scatterings are suppressed. At high energies, the strong electromagnetic fields surrounding the ultrarelativistic ions can be viewed as a flux of quasi-real photons, which is calculable thanks to the use of the equivalent photon approximation (EPA) method~\cite{Fermi:1924tc,Fermi:1925fq,vonWeizsacker:1934nji,Williams:1934ad}.
UPCs lead to large cross sections for photonuclear interactions. For processes involving heavy-quark production, the cross sections can be calculated using standard collinear factorization theorems in terms of perturbatively calculable partonic cross sections and nuclear parton distribution functions (nPDF) (see for example ~\cite{Ball:2001pq}).  \\[4pt]
Extensive experimental \cite{ALICE:2021gpt,ALICE:2023jgu,CMS:2016itn,CMS:2023snh,LHCb:2022ahs} and theoretical~\cite{Goncalves:2011vf,Goncalves:2017wgg,Guzey:2016piu,Guzey:2016qwo,Eskola:2022vpi,Cepila:2017nef,Kopeliovich:2020has,Mantysaari:2017dwh,Schenke:2024gnj,Adeluyi:2012ds,Adeluyi:2012ph} efforts have been devoted to the study of charmonium production in UPCs. In particular, measurements of the exclusive production of $J/\psi$ mesons in coherent photon-Pb scatterings~\cite{ALICE:2021gpt,ALICE:2023jgu,CMS:2016itn,CMS:2023snh,LHCb:2022ahs} have provided experimental indications of a strong suppression of the lead PDF for gluons at very small values of $x$. Despite the unprecedented low-$x$ coverage, the constraining power of these observables remains limited by the complex theoretical description of coherent $\jpsi$ photoproduction and by the fact that these measurements probe gluon density at scales restricted to the mass of the vector meson. \\[4pt]
The study of ``open’’ heavy-flavor (charm and beauty) production in UPCs offers an outstanding opportunity to extend the experimental constraints into unexplored kinematic regimes. To date, theoretical calculations of heavy quark production in ultra-peripheral collisions (UPCs) have been performed only at leading order, either within collinear factorization~\cite{Klein:2002wm,Goncalves:2001vs,Goncalves:2003is,Adeluyi:2012ds} or using the Color Glass Condensate framework~\cite{Goncalves:2003is,Gimeno-Estivill:2025rbw}. 
Charmed hadrons such as $D^0$ and $D^*$, produced in photonuclear reactions, can be exploited as powerful probes of nuclear structure. Since both the hard scale and the parton momentum fraction (Bjorken-$x$) can be inferred from the transverse momenta and rapidity of the final-state charm quarks, measurements of charmed-meson cross sections as a function of $\pt$ and rapidity ($y$) provide unique constraints on the gluon properties in the nucleus across various $x$ and $\Qsquared$ regimes. While the photoproduction of $D^*$ mesons was extensively measured in electron–proton collisions at HERA \cite{H1:1998csb,ZEUS:1998wxs,H1:2011myz}, it is only recently that the feasibility of similar measurements in UPCs has been demonstrated. The CMS collaboration presented the first analysis \cite{CMS:2025jjx} of inclusive $D^0$ production in ultraperipheral PbPb collisions at $\rm \sqrt{s_{\scriptscriptstyle NN}}=5.36$ TeV. The measurement, carried out as a function of $p_T$ and $y$ in the ranges $2<p_T<12$ GeV and $-2<y<2$, is sensitive the gluon density in the nucleus for scales in the region $Q^2 \simeq 30-200\;\text{GeV}^2$ and for values of $x$ down to about $10^{-3}$ (see ~\ref{sec:nPDF} for more detail). These new experimental opportunities underscore the need for accurate theoretical calculations to describe open charm and beauty production in UPCs and in high-energy electron-ion collisions at the Electron-Ion Collider~\cite{Accardi:2012qut,AbdulKhalek:2021gbh} as well as in  proposed facilities, such as Large Hadron electron Collider (LHeC)~\cite{LHeCStudyGroup:2012zhm,LHeC:2020van}. \\[4pt]
\noindent
In the present work, we compute the inclusive $D^0$ production cross section in ultraperipheral PbPb collisions at the LHC using the  newly developed G$\gamma$A-FONLL (Generalized-Photon-Nucleus Fixed-Order Next-to-Leading Logarithm) framework. It generalizes the calculation of photonuclear heavy-flavor production implemented in FONLL for electron-proton and electron-nucleus to UPCs. 
The FONLL (Fixed-Order Next-to-Leading Logarithm) framework ~\cite{Cacciari:1998it,Cacciari:2001td} is used to compute the photoproduction of charm in the fixed next-to-leading order (FO) and fixed order plus next-to-leading logarithmic enhancements of $\ln \frac{p_T}{m}$ at high $p_T$. Since the photoproduction of charm was extensively measured at the HERA collider, we have validated our calculations against data from the H1 \cite{H1:1998csb,H1:2011myz} and ZEUS \cite{ZEUS:1998wxs} experiments. As part of this validation, the earlier FONLL calculations from \cite{Frixione:2002zv} were updated with the latest parton distribution functions for protons (PDFs). The framework developed to reproduce charmed-meson cross sections in electron–proton collisions was subsequently adapted to describe the corresponding cross sections measured in UPCs. Although photoproduction processes in UPCs are in many ways analogous to those in electron–proton and electron–ion collisions, important differences remain (see Fig.~\ref{fig:sketches}).
\noindent
The most evident difference is the photon flux in electron–nucleus versus electron–proton collisions. Although the flux from the nucleus is enhanced by square  of the atomic number Z, its functional dependence on the photon energy differs significantly, producing a generally softer photon spectrum than in the electron case. To account for this distinction, our calculations are modified to reflect the photon flux expected in nuclear collisions. Additional differences arise from the experimental strategy used to select clean photonuclear events in UPCs. The event selection applied in UPCs includes the condition that one of the two colliding nuclei breaks up and the other remains intact (Xn0n) by using the information provided by the zero-degree calorimeters (ZDCs). The `0n' requirement means that the event is not fully inclusive unlike theoretical calculation of photoproduction. In addition, the presence of soft electromagnetic interactions between the two colliding nuclei can lead to breakup of the photon-emitting nucleus, leading to a reduction the rate of selected Xn0n events. To allow for a direct comparison with the data, a dedicated correction was computed assuming the factorization of the soft-electromagnetic interactions and the hard scattering. \\[4pt]
The Xn0n selection further requires at least one neutron in the opposite ZDC, which partially suppresses diffractive production in the data. Since the nuclear diffractive component is included in the theoretical calculation of inclusive charm production, it is important to note that this component is partially removed in the measurement. A rough estimate based on nuclear diffractive parton distributions from the Frankfurt–Guzey–Strikman model \cite{Frankfurt:2011cs} suggests that this effect amounts to about a 10$\%$ contribution in the relevant kinematic range. However, a more detailed study of the diffractive component goes beyond the scope of the present analysis. \\[4pt]
\noindent
The paper is organized as follows: Section~\ref{sec:theory} describes the main ingredients of the theoretical calculation of the photoproduction cross section in electron–proton collisions and its extension to ultraperipheral heavy-ion collisions (UPCs). 
\noindent
In Section~\ref{sec:hera}, the resulting predictions for photoproduction in electron–proton collisions at HERA are presented and compared to H1 and ZEUS measurements. Section~\ref{sec:upcfinebin} provides the G$\gamma A$-FONLL predictions for $D^0$ production in UPCs, and these predictions are then confronted with recent  CMS measurements in Section~\ref{sec:cms}. Finally, our conclusions are presented in Section~\ref{sec:conclu}, where prospects for future theoretical and experimental efforts are also discussed. In the Appendix we include additional plots with calculations for UPCs.

 \section{Generalized $\gamma$A FONLL framework for photonuclear charm production}
 \label{sec:theory}
The results presented in this paper are obtained within the new G$\gamma$A-FONLL framework for photonuclear charm photoproduction. This framework has been developed to provide a comprehensive description of heavy-flavor hadron production in photon–proton, photon–nucleus, and ultraperipheral heavy-ion collisions. In particular, it calculates photoproduction cross sections using standard collinear factorization theorems in terms of perturbatively calculable partonic cross sections, PDFs, and fragmentation functions. To ensure consistency with UPC measurements, the calculations are modified to incorporate the softer photon flux expected in nuclear collisions and to account for the experimental strategies used to select clean photonuclear events at colliders. In this section, a detailed description of each ingredient is provided, including a separate subsection that covers the strategy adopted to account for the bias introduced in UPCs by soft-photon exchanges between the ultrarelativistic ions.

\subsection{Photoproduction cross section for heavy quarks}
\label{sec:cross}
For the calculation of the differential photoproduction cross section for heavy quarks we  use the code for the FONLL calculation~\cite{Cacciari:2001td}.
In that work a resummation approach was developed which allows for the description of the transverse momentum distribution of heavy quarks produced in photoproduction on a proton or nuclear target. The FONLL approach connects the region of small transverse momenta where the fixed order approach is valid to the region with large transverse momenta where the logarithmically enhanced terms are resummed. To be precise the formalism includes  terms up to NLO order, that is terms $\alpha_{\rm em} \alpha_s$ and $\alpha_{\rm em} \alpha_s^2$. These terms are included exactly, with mass effects. In addition, all logarithmically enhanced terms $\alpha_{\rm em} \alpha_s (\alpha_s \ln(p_T/m))^k$ and $\alpha_{\rm em} \alpha_s^2 (\alpha_s \ln(p_T/m))^k$ are included, 
with the possible exception of terms that are suppressed by powers of $m/p_T$.
The matching resummed FONLL cross section  can be expressed as \cite{Cacciari:2001td,Frixione:2002zv}
\begin{equation}
   {\rm  FONLL=FO+(RS-FOM0)} \times G(m,p_T) \; .
    \label{eq:fonllmatching}
\end{equation}
In this formula the  FO  is fixed order, that is exact NLO calculation \cite{Ellis:1988sb,Smith:1991pw}. RS is the resummed result with all the logarithmically enhanced contributions, up to  terms suppressed by powers of $m/p_T$. FOM0 is the massless limit of fixed order FO without  terms suppressed by powers of $m$, while logarithms of the mass are retained. Finally, $G(m,p_T)$ is  an arbitrary damping function, that must be regular in $p_T$, and must approach unity up to terms suppressed by powers of $m/p_T$ at large transverse momenta.
\\[4pt] \noindent
The photoproduction cross section contains both {\em direct} and {\em resolved } contributions. The direct (or pointlike) contribution has the form
\begin{equation}
\frac{d\sigma}{dydp_T}\bigg|_{\rm dir} \; = \; \sum_j \int dx_p f_{j/H}(x_p) \, \frac{d\hat{\sigma}_{j\gamma}}{dydp_T}(P_{\gamma},x_p P_H) \; , 
    \label{eq:direct}
\end{equation}
where $f_{j/H}(x_p)$ is the distribution of the parton $j$ in the hadron H and $\frac{d\hat{\sigma}_{j\gamma}}{dydp_T}(P_{\gamma},x_p P_H)$ is the partonic cross section for scattering of parton $j$ with photon $\gamma$ producing a heavy quark in the final state. The resolved (or hadronic) component has the form
\begin{equation}
\frac{d\sigma}{dydp_T}\bigg|_{\rm res} \; = \; \sum_{kj} \int dx_{\gamma} \, dx_p f_{k/\gamma}(x_\gamma) \, f_{j/H}(x_p) \, \frac{d\hat{\sigma}_{kj}}{dydp_T}(x_\gamma P_{\gamma},x_p P_H) \; , 
    \label{eq:resolved}
\end{equation}
where $f_{k/\gamma}$ is the parton density in the photon and $\frac{d\hat{\sigma}_{kj}}{dydp_T}(x_\gamma P_{\gamma},x_p P_H)$ is the partonic cross section for  scattering of two partons $kj$ resulting with a heavy quark in the final state. In the above we have suppressed the dependence on the factorization and renormalization scales.  Both terms  given by Eqs.~\eqref{eq:direct} and Eqs.~\eqref{eq:resolved} have expansion   as series of $\alpha_{\rm em}\alpha_s^k$  if one takes into account that the parton density in the photon carries a factor $\alpha_{\rm em}/\alpha_s$. It is worth noting that only the sum of direct and resolved contributions is an observable quantity, as both these contributions are related beyond the leading order, see \cite{Cacciari:2001td}.

\subsection{Parametrization of the Parton Distribution Functions for protons and nuclei}
\label{sec:nPDF}
\noindent
For the calculation of the photoproduction cross sections discussed in previous section, which are implemented in the FONLL code, one needs to select the parton distribution function (PDF) for nucleus or a proton. 
\noindent
For the nuclear PDF we take the EPPS21 \cite{Eskola:2021nhw} set as well as the  nNNPDF3.0 \cite{AbdulKhalek:2022fyi} set. In addition we also perform and compare the calculations with the EPPS21 proton baseline PDF CT18ANLO \cite{Hou:2019efy}, to estimate the role of the nuclear effects in the kinematics for $D^0$ production at CMS. \\[4pt] \noindent
In the EPPS21  \cite{Eskola:2021nhw} set    the nuclear modification factor is parametrized  and  for the baseline proton PDF the CT18ANLO \cite{Hou:2019efy}  is used. The value of charm mass for this PDF set is $m_c=1.3 $ GeV. The EPPS21 parametrization uses the same data sets as EPPS16 \cite{Eskola:2016oht} and in addition 5 TeV double-differential CMS dijet and LHCb D-meson data, as well as 8 TeV CMS W boson data from pPb collisions. These new data sets lead to significantly better-constrained gluon distributions at small and intermediate values of the momentum fraction $x$. In addition new electron-nucleus data from JLab were used in the fit, which impact the large $x$ and small virtualities in the nuclear PDFs. Another improvement in the EPPS21 analysis is the inclusion into the Hessian framework the errors of the baseline proton PDFs and their propagation into the uncertainties of the  nuclear PDFs. \\[4pt] 
\noindent
The nNNPDF3.0 \cite{AbdulKhalek:2022fyi} set  extends the  NNPDF methodology  which uses machine learning methods in the global analysis and extraction of PDFs. In particular the  artificial neural networks  are used as universal  interpolants to parameterize the $x$ and $A$ dependence of the nPDFs. This methodology allows for the reduction of model parametrization bias. The value of charm mass in nNNPDF3.0 is $m_c=1.5 $ GeV. The nNNPDF3.0 set used all the data in the nNNPDF2.0 \cite{AbdulKhalek:2020yuc} as well as the additional ones, from  the LHC pPb collisions. New data sets included: forward and backward rapidity fiducial cross-sections for the production of W bosons at $\sqrt{s}$ = 5.02 TeV measured by ALICE experiment; forward and backward rapidity fiducial cross-sections for the production of Z bosons measured by ALICE and LHCb at 5.02 TeV and 8.16 TeV; the differential cross-section for the production of Z bosons measured by CMS at 8.16 TeV; CMS dijet at 5 TeV and ATLAS prompt photon data at 8.16 TeV and also the LHCb data on $D^0$ production at 5 TeV.  We note that, at present the data on the exclusive $J/\psi$ production in UPC are not used in the global analysis of the nuclear PDFs. 
Both nPDF sets exhibit very distinctive nuclear modifications of the parton distribution functions. In particular they confirm the presence of shadowing (suppression) of gluons at small $x$, and anti-shadowing (enhancement) at large $x$. For the purpose of benchmarking and comparison of charm production in $ep$ scattering at HERA, presented in Sec.~\ref{sec:hera}, we also need to select proton PDFs. For calculation in Sec.~\ref{sec:hera} we used CT18ANLO \cite{Hou:2019efy} and nNNPDF3.0p \cite{AbdulKhalek:2022fyi} 
 which are proton baselines for the EPPS21 and nNNPDF3.0  nuclear PDF sets respectively.
 In addition, for the $ep$ case we also performed comparisons using HERAPDF2.0 \cite{H1:2015ubc} set.
\noindent
\subsubsection{Estimate of the $x$ and $Q^2$ Coverage for $D^0$ Photoproduction Measurements}
A qualitative estimate of the $(x, Q^2)$ coverage accessible through open-charm observables in ultra-peripheral collisions (UPCs) can be obtained by approximating the transverse momentum scale $Q^2$ with the squared transverse mass of the charm quark $m_T^2 = p_T^2 + m_c^2$ and the target’s longitudinal momentum carried by the parton as $x \approx m_T / {\rm\sqrt{s_{\scriptscriptstyle NN}} }\, e^{-y_c}$, where $m_c$ is the charm quark mass, ${\rm \sqrt{s_{\scriptscriptstyle NN}}}$ is the center-of-mass energy of the nucleon-nucleon system, and $p_T$ is the transverse momentum of the charm quark.
\begin{figure}
    \centering
    \includegraphics[width=0.6\textwidth]{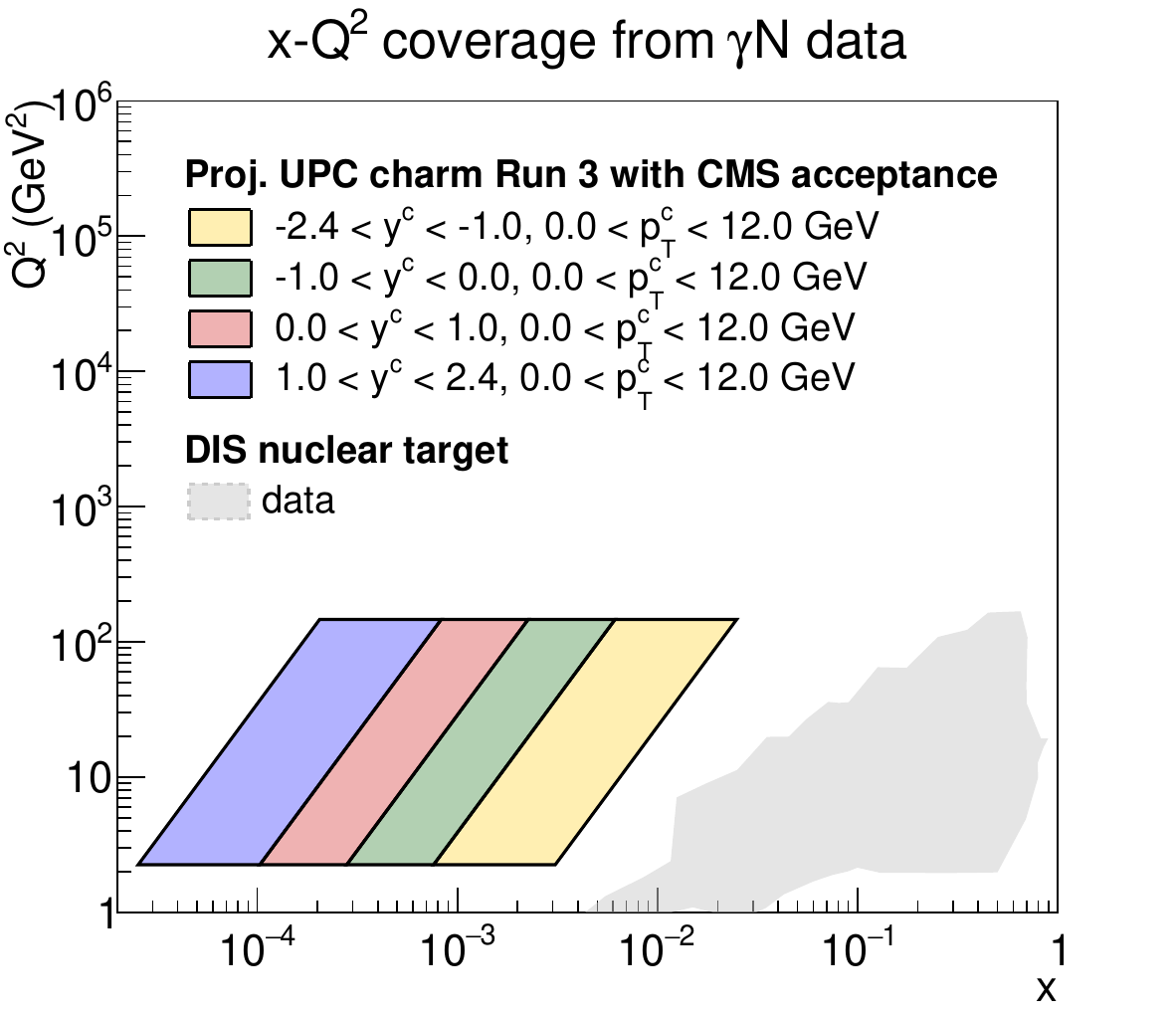}
    \caption{$x$ and $Q^2$ coverage charm production in UPCs collisions. In grey, the existing coverage from fixed-target deep-inelastic photonuclear measurements is presented~\cite{AbdulKhalek:2019mzd}.}
    \label{fig:xQ2coverage}
\end{figure}
In Fig.~\ref{fig:xQ2coverage}, the predicted $(x, Q^2)$ region probed by open-charm measurements in UPCs at ${\rm \sqrt{s_{\scriptscriptstyle NN}}} = 5.36$~TeV is shown, corresponding to charm-quark transverse momentum in the range $0 < p_{T,c} < 12$~GeV and rapidity $-2.4 < y_{c} < 2.4$. This kinematic region, which is accessible with existing LHC detectors, enables the exploration of $x$ values from approximately $0.01$ down to below $10^{-4}$, and $Q^2$ values ranging from $m_c^2$ up to around  ~$\sim 100\,$GeV$^2$. 

\subsection{Photon flux parametrization in electron-proton, electron-nucleus collisions and PbPb UPCs}
\label{sec:flux}

In order to compute the cross section in electron–proton collisions at HERA or in UPCs at the LHC, the photoproduction cross section must be convoluted with the photon flux from the electron or the nucleus, respectively. For UPCs, an effective parametrization was developed to account for the geometrical properties of the collisions. To enable direct comparison with experimental measurements, the impact of electromagnetic dissociation on the photon-emitting nucleus is also taken into account. In this section, a detailed description of the adopted parametrizations in electron-proton, electron-ion and UPCs lead-lead collisions is provided. 
\noindent
\subsection{Photon flux in electron-proton and electron-nucleus collisions} 
 For the description of electron-proton and electron-ion collisions, the original parametrization implemented in the FONLL code was used. The photon flux from electrons implemented in the FONLL code for the electroproduction  is given by the following formula
 \begin{equation}
f_{\gamma/e}(z) = \frac{\alpha_{em}}{2\pi} \bigg[ \frac{1+(1-z)^2}{z} \log\bigg(\frac{Q_{\rm max}^2(1-z)}{(m_e z)^2}\bigg) + 2 m_e^2 z\bigg(\frac{1}{Q_{\rm max}^2} -\frac{1-z}{(zm_e)^2}\bigg) \bigg] \; .
\label{eq:eleflux}
\end{equation}
where $z$ is the fraction of the energy of the electron carried by the photon, $m_e$ is the electron mass, $Q^2_{\rm max}$ is the limit on the (negative) photon virtuality, and $\alpha_{em}$ is the electromagnetic coupling.
The second term in the above expression is a subleading correction, see   Ref.~\cite{Frixione:1993yw}. In Fig.~\ref{fig:fluxes}, the resulting parametrizations for the photon flux from electrons at HERA are shown. The solid black curve corresponds to \( Q^2_{\rm max} = 0.01 \, \mathrm{GeV}^2 \), chosen to match the experimental conditions of the {\it ETAG33} H1 data sample~\cite{H1:1998csb}. The dashed blue line uses \( Q^2_{\rm max} = 2 \, \mathrm{GeV}^2 \), to enable comparison with the 2012 H1 data~\cite{H1:2011myz}. The electron flux is very flat over entire range of $z$, and has non-negligible dependence on $Q^2_{\rm max}$. The difference in flux between $Q_{\max}^2=0.01 \, \rm GeV^2$ and  $Q_{\max}^2=2 \, \rm GeV^2$ is of the order of 30 to 60\% depending on the fraction of the energy $z$. We have checked that the second term in Eq.~\ref{eq:eleflux} is a small correction of  about 3-7\% to the leading logarithmic term.

\subsubsection{Photon flux in ultra-peripheral heavy-ion collisions} 
For the case of the UPCs, the photon flux from electrons was replaced with the expected photon flux from the nucleus. The parametrization considered is the analytic expression presented in~ \cite{Bertulani:1987tz}, obtained in the pointlike approximation:
\begin{equation}
f_{\gamma/A}(z) = \int d^2 \mathbf{b} \, \tilde{f}_{\gamma/A}(z,\mathbf{b}) \,\theta(|\mathbf{b}|-b_{\min}) =\frac{2Z^2 \alpha_{em}}{\pi z} \left[\eta K_0(\eta) K_1(\eta) - \frac{\eta^2}{2} (K_1(\eta)^2-K_0(\eta)^2)\right]  ,
\label{eq:nucflux}
\end{equation}
with the dimensionless variable $\eta$ defined as 
\begin{equation}
    \eta = \frac{z \, m_p \, b_{\min} }{\hbar c} \; .
\end{equation}
In Eq.~\eqref{eq:nucflux} $ \tilde{f}_{\gamma/A}(z,\mathbf{b})$
is the photon flux of the nucleus evaluated at the transverse distance $\mathbf{r}=\mathbf{{b}}$ where $\mathbf{b}$ is vector connecting the centers of the two nuclei in the transverse plane of the collision. This is appropriate in the pointlike approximation (see for example \cite{Eskola:2024fhf}).  
Here $z$ is the fraction of the energy of nucleus carried by the photon, $m_p$ is the proton mass, $Z$ atomic number ($Z=82$ for lead), $b_{\min}$ distance between the nuclei, and $K_0,K_1$ are modified Bessel functions. Following~\cite{Nystrand:2004vn,Guzey:2018dlm,Eskola:2024fhf} we take the impact parameter $b_{\min} = 2 R = 14.2$ fm, with $R = 7.1$ fm  being the nuclear hard-sphere radius, to account for the finite size of the colliding nuclei. 
\begin{figure}
    \centering
    \includegraphics[width=0.6\textwidth]{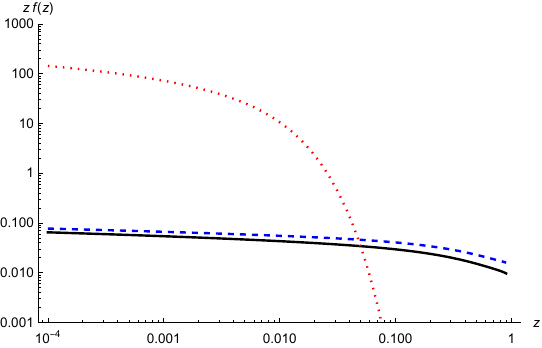}
\caption{Photon flux scaled by $z$ as a function of the fractional photon energy $z$. The electron flux of Eq.\eqref{eq:eleflux} is shown for $Q_{\text{max}}^{2}=0.01~\text{GeV}^{2}$ (solid black) and $Q_{\text{max}}^{2}=2~\text{GeV}^{2}$ (dashed blue), while the lead–nucleus flux of Eq.\eqref{eq:nucflux} appears as a red dotted line.}
\label{fig:fluxes}
\end{figure}
The parametrization for the photon flux from nuclei, presented in Eq.~\ref{eq:nucflux}, is shown as a red dotted line in Fig.~\ref{fig:fluxes}, and compared to the fluxes from electrons at HERA discussed previously. At small \( z \), the photon flux from nuclei is much larger than that from electrons. At intermediate \( z \), it exhibits a sharp cutoff due to the presence of the Bessel functions \( K_0 \) and \( K_1 \). As a result, the photon flux from nuclei becomes negligible for \( z \gtrsim 0.1 \). These differences between electron and nucleus fluxes are important when comparing the kinematics of HERA and UPCs at the LHC, as well as when comparing various HERA data sets collected with different kinematic cuts on photon virtuality and \( z \) ranges (see Table~\ref{table:names_numbers} in Sec.~\ref{sec:hera}).
\subsubsection{Modeling of the no-break up probability in UPCs}
\label{sec:emd}

\begin{figure}
    \centering
    \includegraphics[width=0.95\textwidth]{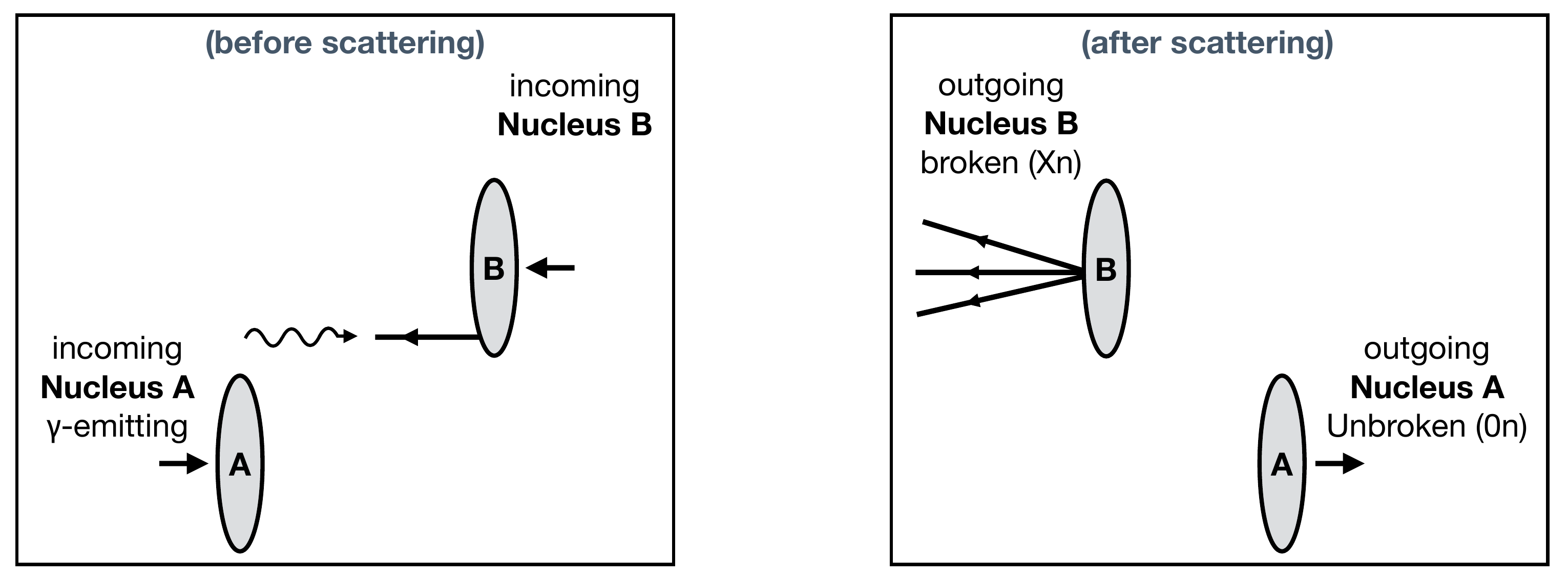}
    \caption{Sketch of an ultraperipheral heavy-ion collision before (left) and after (right) the hard scattering.}
    \label{fig:sketch}
\end{figure}
Photonuclear events, in which one of the two colliding nuclei breaks up and the other remains intact, are selected in UPC measurements by requiring that there are no forward neutrons in the zero-degree calorimeter (ZDC) in one direction (0n) and at least one neutron (Xn) in the opposite direction (see Fig.~\ref{fig:sketch}). This selection, coupled with the requirement of a large rapidity gap in the photon-going direction, allows for the suppression of the contamination of hadronic events (XnXn) and of two-photon (0n0n) processes. \\[4pt]
The presence of a 0nXn requirement, thus, poses some additional difficulties for theoretical calculations. In UPCs, indeed, the presence of soft electromagnetic interactions between the two colliding nuclei (also known as electromagnetic dissociation or EMD) can lead to breakup of the photon-emitting nucleus~\cite{CMS:2025jjx}. To facilitate the comparison with UPCs data, therefore, one has to account for the probability that a genuine photonuclear event, in which the photon-emitting nucleus is initially left unbroken (0nXn), could be rejected as a consequence of the presence of soft-electromagnetic interactions that lead to the dissociation of the photon-emitting nucleus. 
\noindent
Under the hypothesis that the soft-excitation probability factorizes from the hard interaction~\cite{Baltz:2002pp}, the no-breakup (survival) probability can be folded in the calculation of the effective photon flux. For this study, we considered the calculation of the photon flux in the presence of EM dissociation calculated in~\cite{Eskola:2024fhf,Paakkinen:2024qunb}. 
The validity of this theoretical approach, and its ability to accurately capture the impact of EMD in ultraperipheral collisions, has been demonstrated in a recent measurement of jet photoproduction in Pb–Pb collisions at the LHC. In that analysis, the no-breakup probability as a function of the momentum fraction carried by the partons in the emitted photon, as measured by ATLAS, is well described by a calculation based on the same theoretical framework adopted here. Within this approach, the photon flux in the presence of EMD is modeled as:
\begin{equation}
  f_{\gamma/A}^{\text{eff}}(z)
  \;=\;
  \int
  d^2 \textbf{b} \,P_{\text{no-EM}}(\textbf{b})\;
  \tilde{f}_{\gamma/A}(z, \textbf{b})\;
  \theta(|\mathbf{b}|-b_{\min}),
\label{eq:nucfluxEMD}
\end{equation}
where the factor $P_{\text{no-EM}}$ is the probability that the electromagnetic breakup occurs in terms of the vector $\textbf{b}$. 
\begin{figure}
    \centering
    \includegraphics[width=0.45\textwidth]{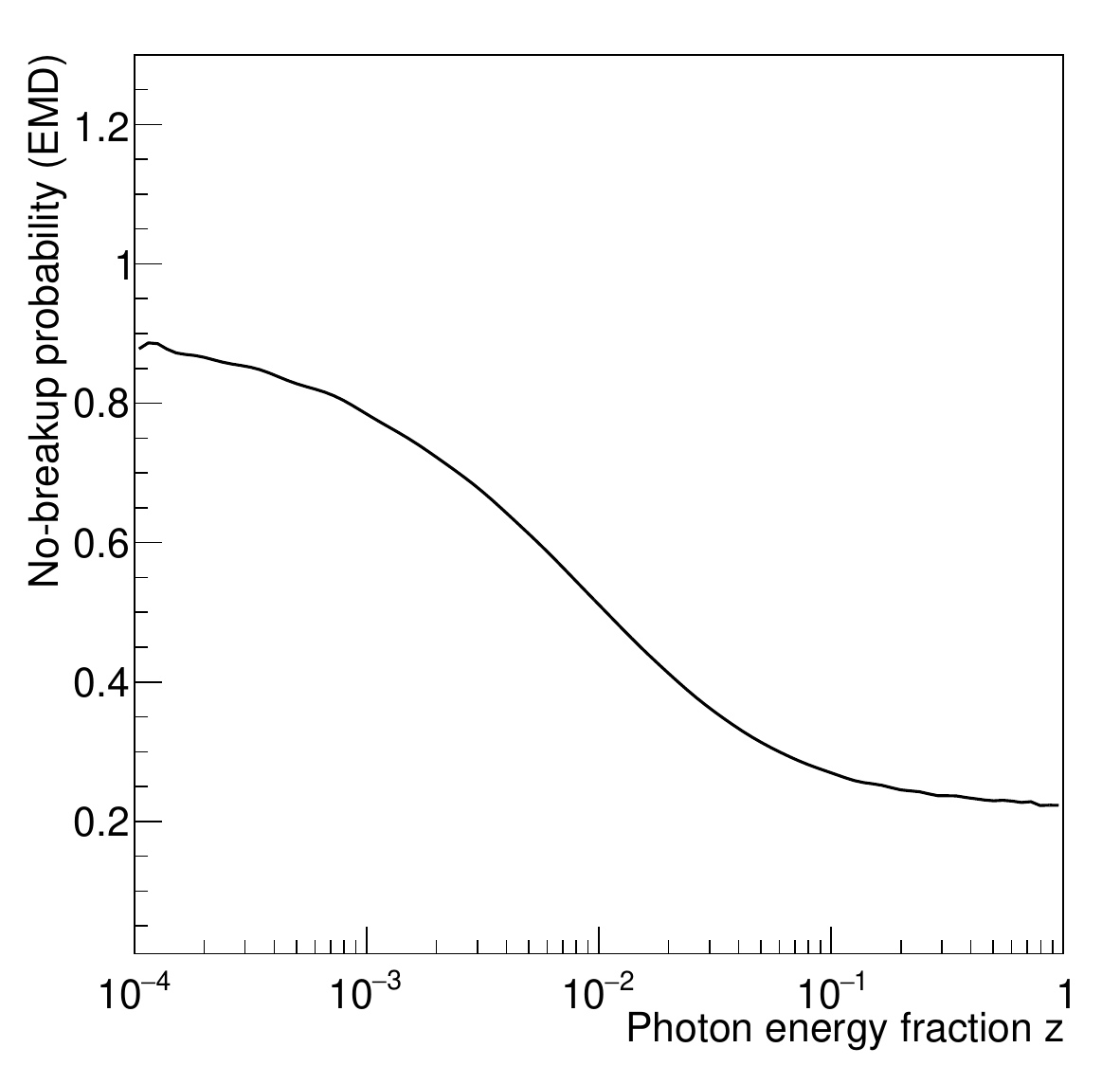}
    \includegraphics[width=0.45\textwidth]{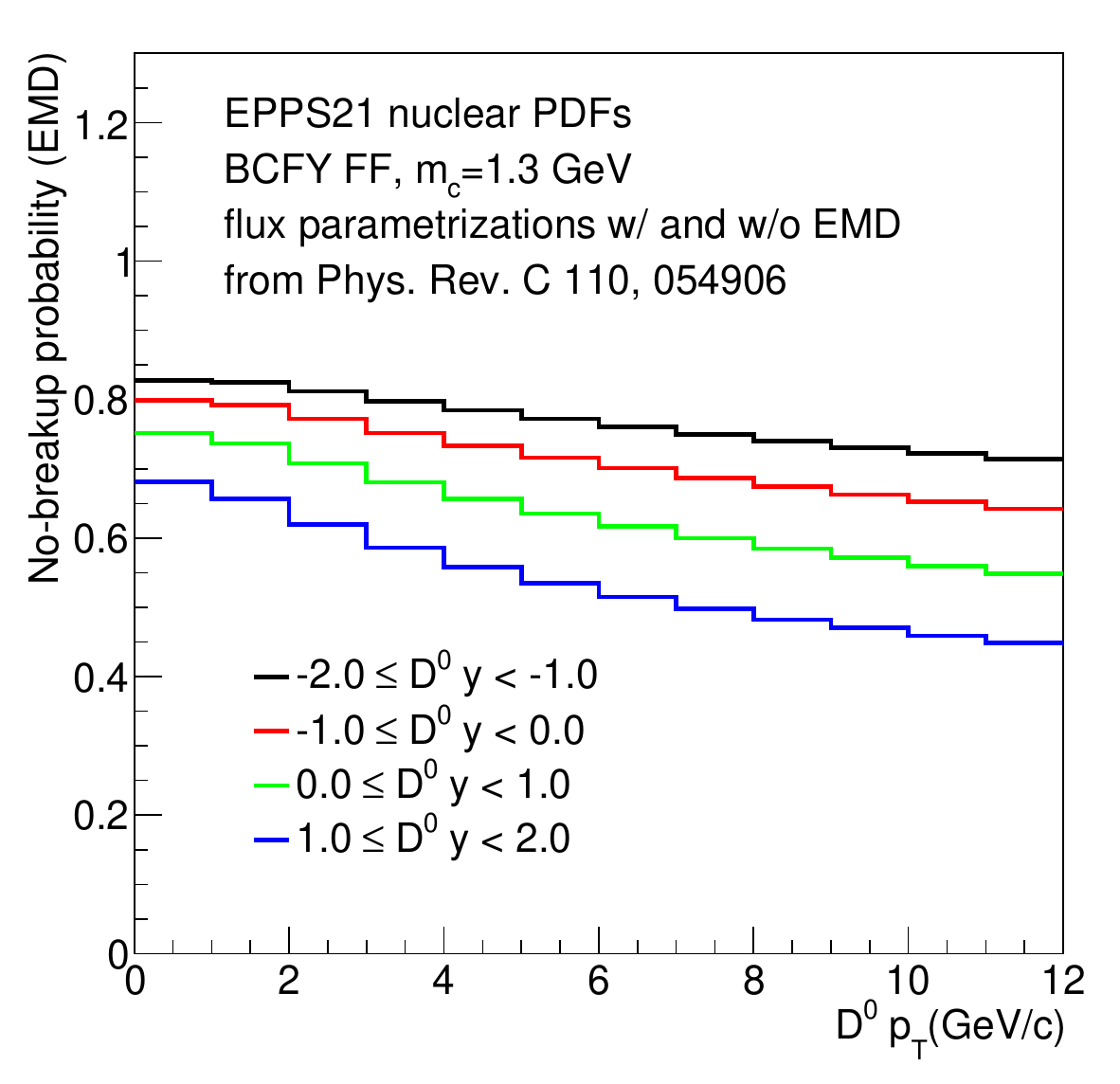}
    \caption{(Left) No-breakup probability (EMD) as a function of the photon energy fraction \( z \). (Right) No-breakup probability (EMD) as a function of the $\Dzero~\pt$, in different intervals of the $\Dzero$ y. The interpolated Chebyshev parametrizations of the photon fluxes used to compute these ratios are taken from~\cite{Eskola:2024fhf}.}
    \label{fig:ratiofluxEMD}
\end{figure}
\noindent
In the left panel of Fig.~\ref{fig:ratiofluxEMD}, the no-breakup probability computed as the ratio of the photon fluxes described in Eq.~\ref{eq:nucflux} and Eq.~\ref{eq:nucfluxEMD}, is shown as a function of the photon energy fraction \( z \). For this study, we relied on the interpolated Chebyshev parametrizations described in~\cite{Eskola:2024fhf}. As expected, the ratio between the two fluxes decreases significantly with increasing photon energy, reflecting the enhanced probability of EM dissociation in small impact-parameter UPCs. \\[4pt] 
\noindent
In the right panel of the same figure, we present the ratio of the predicted double-differential cross section with EMD effects to the one obtained without this correction. The calculation uses EPPS21 nuclear PDFs, the BCFY fragmentation function (see next section), and a charm-quark mass of $m_c = 1.3~\text{GeV}$. The resulting no-break-up probability, which we will employ throughout this manuscript to rescale the $\Dzero$ double-differential cross sections in UPCs, ranges from 0.83 to 0.68 for $\Dzero$ mesons with $0 < p_T < 1~\text{GeV}$ and from 0.71 to 0.45 for those with $11 < p_T < 12~\text{GeV}$. \\ [4pt]
\noindent
\subsection{Fragmentation functions for heavy quarks}
\label{sec:frag}

The charm-meson production cross section is computed within the QCD factorization framework by combining the partonic charm-quark production cross section with the quark-to-hadron fragmentation function. The latter one is a non-perturbative function which needs to be parametrized. Fragmentation of the heavy quark differs from the light quarks in that the heavy flavored meson should retain large fraction of the momentum of the original heavy quark. Therefore the fragmentation function for heavy quark should be peaked at large longitudinal momentum fraction $z$. One approach is to constrain such parametrization using the electron-positron data from LEP and then assume that they are universal and can be used for the other  collision systems, including hadrons. Whether this procedure is applicable for hadronic collisions is still an open question, since the hadronization can be affected by the interactions with hadronic beam remnants, especially at low $p_T$ and high rapidity. In the present analysis, we assess the impact of fragmentation by evaluating the cross section with two different parametrizations: the Peterson-Schlatter-Schmitt-Zerwas (PSSZ) \cite{Peterson:1982ak} and Braaten–Cheung–Fleming–Yuan (BCFY) \cite{Braaten:1994bz} fragmentation functions. To describe the HERA $D^{*}$ photoproduction data, we employ the PSSZ fragmentation function \cite{Peterson:1982ak}
\begin{equation}
D(z) = {\cal N} \, \frac{1}{z} \left( 1-\frac{1}{z} -\frac{\varepsilon}{1-z}\right)^{-2} \; ,
    \label{eq:peterson}
\end{equation}
where ${\cal N}$ is a normalization constant. 
The parameter $\varepsilon$ originally has the interpretation of being proportional to the ratio of squares of light to heavy quark masses, see \cite{Peterson:1982ak}. In practice, however, it is commonly treated as a free parameter to obtain the best description of the data. 
We perform calculations with two choices of $\varepsilon$ parameter: $0.02$ and $0.035$. These two values have been used in Ref.~\cite{Frixione:2002zv} for FONLL case and in Ref.~\cite{H1:2011myz} for fixed order NLO (FMNR calculation) respectively. The PSSZ function is normalized to unity, and then the fragmentation fraction is taken to include the fragmentation into the specific type of meson. \\[4pt] 
The Braaten–Cheung–Fleming–Yuan (BCFY) fragmentation function \cite{Braaten:1994bz} was employed in Ref.~\cite{Cacciari:2003zu} to model charm-meson production in hadronic collisions. In the latter study, the $D^{0}$ fragmentation function was constructed as a linear combination of two contributions: the component arising from $D^{*}$ to $D^{0}$ decays and the so-called ``primary'' $D^{0}$ mesons produced directly in the hadronization of the charm quark. Schematically, one can write
\noindent
\begin{equation}
    F(c\rightarrow D^0) = F_p(c\rightarrow D^0) + F(c\rightarrow D^{*+})F(D^{*+}\rightarrow D^0)+F(c\rightarrow D^{*0})F(D^{*0}\rightarrow D^0)  \; .
    \label{eq:FFcd}
\end{equation}
The relation above  contains contributions from the direct $c\rightarrow D^0$ hadronization and indirect $c\rightarrow D^*\rightarrow D^0$ channel. 
$F_p$ is the fragmentation function for the primary $D^0$ production, $F(c\rightarrow D^{*+})$ and $F(c\rightarrow D^{*0})$ are fragmentation functions for production of $D^*$ states, and functions $F(D^{*+}\rightarrow D^0)$ and $F(D^{*0}\rightarrow D^0)$ describe a decay of $D^*$ states into $D^0$.
The branching fractions for charm quarks fragmenting into  $D^0$ and $D^*$ are taken  respectively \cite{Cacciari:2003zu} 
\begin{equation}
BR(c\rightarrow D_p^0) = 0.168\,   ,  \; BR(c\rightarrow D^*) = 0.235\, .
\label{eq:branching}
\end{equation}
Taking into account that the branching fractions of $D^{*0}$ and $D^{*+}$ into $D^0$ meson
are
\begin{equation}
BR(D^{*0}\rightarrow D^0) = 1.0\, , \; \;  BR(D^{*+}\rightarrow D^0) = 0.677 \, ,
\end{equation}
one can write Eq.~\eqref{eq:FFcd} as (see \cite{Cacciari:2003zu})
\begin{equation}
D^{c\rightarrow D^0}(z,r)=0.168 D^{P}(z,r)+0.39 \tilde{D}^{V}(z,r) \; ,
    \label{eq:bcfyadd}
\end{equation}
where
\begin{equation}
   \tilde{D}^{V}(z,r) \; = \; \Theta\bigg(\frac{m_D}{m_{D^*}}-z\bigg) D^{(V)}\big(\frac{m_{D^*}}{m_D}z,r\big) \frac{m_{D^*}}{m_D} \;.
\end{equation}
Functions $D^{P}(z,r)$ (for pseudoscalar state) and $\tilde{D}^{V}(z,r)$ (for vector state) are given explicitly in \cite{Braaten:1994bz}. They depend on a single non-perturbative parameter $r$, which is the same for both functions. In the original model \cite{Braaten:1994bz} this parameter can be interpreted as the ratio of the constituent mass of the light quark to the mass of the meson.
Following \cite{Cacciari:2003zu,Cacciari:2012ny}, we take the value of    parameter $r= 0.1$.
For the case of $D^*$ production the fragmentation function  is 
\begin{equation}
    D^{c\rightarrow D^*}(z,r)= BR(c\rightarrow D^*) \, {D}^{V}(z,r) \; .
    \label{eq:bcfydstsr}
\end{equation}
We also use this fragmentation function for the $D^*$ production at HERA and compare the results with the ones obtained using PSSZ fragmentation function.

\section{Benchmark with HERA data on inclusive $D^*$ production}
\label{sec:hera}

Comprehensive study aimed on comparison of FONLL calculation with the  HERA data on charm photoproduction \cite{H1:1998csb,ZEUS:1998wxs} was performed in \cite{Frixione:2002zv}. 
Later data from H1 \cite{H1:2011myz}  were also compared with several theoretical calculations. Among them was the fixed order  NLO calculation in the massive scheme  by Frixione-Mangano-Nason-Ridolfi (FMNR)  \cite{Frixione:1994dv,Frixione:1995qc} and  the Generalized-Mass-Variable-Flavour-Number-Scheme (GMVFNS) calculation \cite{Kniehl:2009mh,Kramer:2003jw}. In the latter approach, which unifies  fixed flavor massive scheme with   zero-mass variable flavor scheme, large logarithmic corrections are resummed in universal parton distribution and fragmentation functions and finite mass terms were taken into account. It was found that both NLO calculations reproduced  well the $p_T$ and rapidity dependence of the experimental data from H1 experiment \cite{H1:2011myz}, within their uncertainties.
\\[4pt] 
\noindent
To ensure the consistency of calculation between HERA and UPC at LHC, we shall first revisit FONLL  predictions \cite{Frixione:2002zv} for HERA kinematics  with more up-to-date proton PDFs which serve as proton benchmarks for the nuclear PDFs used for calculations at UPC LHC, and two different fragmentation functions. We use FONLL code \cite{Cacciari:1993mq,Cacciari:2001td} for  calculations
which provides both resummed (FONLL) and fixed order (FO) options. We perform the comparison  with four sets of HERA data on $D^*$ photoproduction \cite{H1:1998csb,H1:2011myz,ZEUS:1998wxs}. In table \ref{table:names_numbers} we list the kinematic cuts on photon virtuality $Q_{\rm max}^2$, range for photon energy  fraction of the electron energy $(z_{\rm min},z_{\rm max})$, transverse momentum $p_T$ and rapidity $y$ or pseudorapidity $\eta$ for these different data sets. Earlier data from H1 \cite{H1:1998csb} and ZEUS \cite{ZEUS:1998wxs} were taken at $E_e \times E_p = 27.5 \rm \, GeV \times 820 \rm \, GeV$, while 2012 data from H1 \cite{H1:2011myz} were taken at $E_e \times E_p = 27.6 \rm \, GeV \times 920\, \rm GeV$ energy combination. 
The two data samples from \cite{H1:1998csb} labeled {\it ETAG33} and {\it ETAG44} were taken for two case when the electron is in the electron tagger at 33 m and 44 m respectively. The average $\gamma p$ centre-of-mass energy for the selected data was \(\langle W \rangle = 194 \) GeV for {\it ETAG33} and \(\langle W \rangle = 83 \) GeV for {\it ETAG44}.
\begin{table}[ht]
\centering
\begin{tabular}{|c|c|c|c|c|c|}
\hline
Data set & $Q^2_{\rm max}$  & $z_{\rm min}$  & $z_{\rm max}$  & $p_T$  & (pseudo)rapidity \\ \hline
H1 \textit{ETAG44} & 0.009 & 0.02 &  0.32 &  $p_T>2 \, \rm GeV$ & $|y|<1.5$ \\ \hline
H1 \textit{ETAG33}   & 0.01 &  0.29 & 0.62 & $p_T>2 \,\rm GeV$ & $|y|<1.5$ \\ \hline
ZEUS & 1 & 0.187 & 0.869 & $p_T>2 \, \rm GeV$ &  $|\eta|<1.5$ \\ \hline
H1 2012  & 2 & 0.09 & 0.8 & $p_T>1.8 \, \rm GeV$ & $|\eta|<1.5$ \\ \hline
\end{tabular}
\caption{Kinematic cuts in data sets for $D^*$ photoproduction. Data are from : H1 (\textit{ETAG33} and \textit{ETAG44} scenarios) \cite{H1:1998csb}, ZEUS \cite{ZEUS:1998wxs} and H1 (2012 data) \cite{H1:2011myz}. }
\label{table:names_numbers}
\end{table}
\\[4pt] 
\noindent
For the photon PDF we tested two distributions AFG  \cite{Aurenche:1994in}
and GRV \cite{GRVPhotonPDF}, originally implemented in FONLL code. For the proton PDFs we used CT18ANLO \cite{Hou:2019efy}, which is a proton baseline for nuclear PDF set EPPS21 \cite{Eskola:2021nhw}. We also compared it with calculations based on   nNNPDF3.0p proton baseline set \cite{AbdulKhalek:2022fyi} and HERAPDF2.0 \cite{H1:2015ubc} set.
Following \cite{Frixione:2002zv} we used $\mu_F=\mu_R=\sqrt{p_T^2+m_c^2}=\mu_0$ as a  central choice of factorization and renormalization scale. We performed   variation of factorization scale $\mu_F/\mu_0=0.5,1.0,2$ as well as   variation of renormalization scale  $\mu_R/\mu_0=0.5,1.0,2.0$, to obtain the uncertainty bands.   We used charm mass \(m_c=1.3 \, \rm GeV \), as this is the value used in CT18ANLO PDF set, but we also performed calculations with  \(m_c=1.5 \, \rm GeV\), the value for nNNPDF3.0 set.  As discussed in previous section, we used PSSZ \eqref{eq:peterson}  and BCFY \eqref{eq:bcfydstsr} fragmentation functions. In PSSZ function  parameter was set to $\varepsilon=0.02$ (used in \cite{Frixione:2002zv} for FONLL calculation) and $\varepsilon=0.035$ (used in FMNR calculation  in \cite{H1:2011myz}). 
For BCFY fragmentation function  we used $r=0.1$.
The calculation is then multiplied by $0.47$ which stems from the fragmentation fraction of $c\rightarrow D^*$  to be $23.5\%$ and the fact that experiment measures $D^{*+}+D^{*-}$ contributions, see \cite{Frixione:2002zv}. The presented cross sections are for electroproduction ($\sigma_{ep\rightarrow D^*X}$), therefore we multiplied the data by the factor coming from the  photon flux from electron, integrated over given range of energies,   (see discussion in \cite{H1:1998csb} and \cite{Frixione:2002zv}).

\begin{figure}
\centering
\begin{subfigure}{0.49\textwidth}
    \centering
    \includegraphics[width=\textwidth]{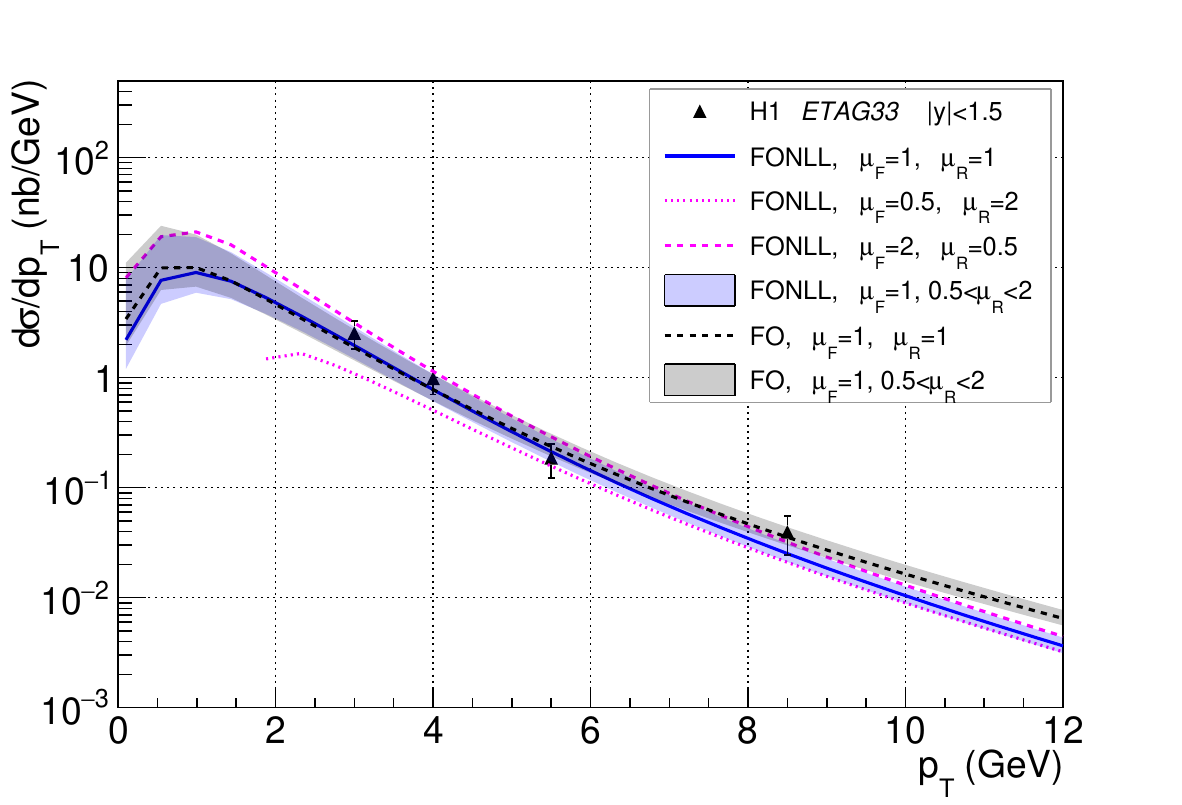}
    \caption{H1 \textit{ETAG33}}
    \label{fig:h1eta33pt}
\end{subfigure}
\hfill
\begin{subfigure}{0.49\textwidth}
    \centering
    \includegraphics[width=\textwidth]{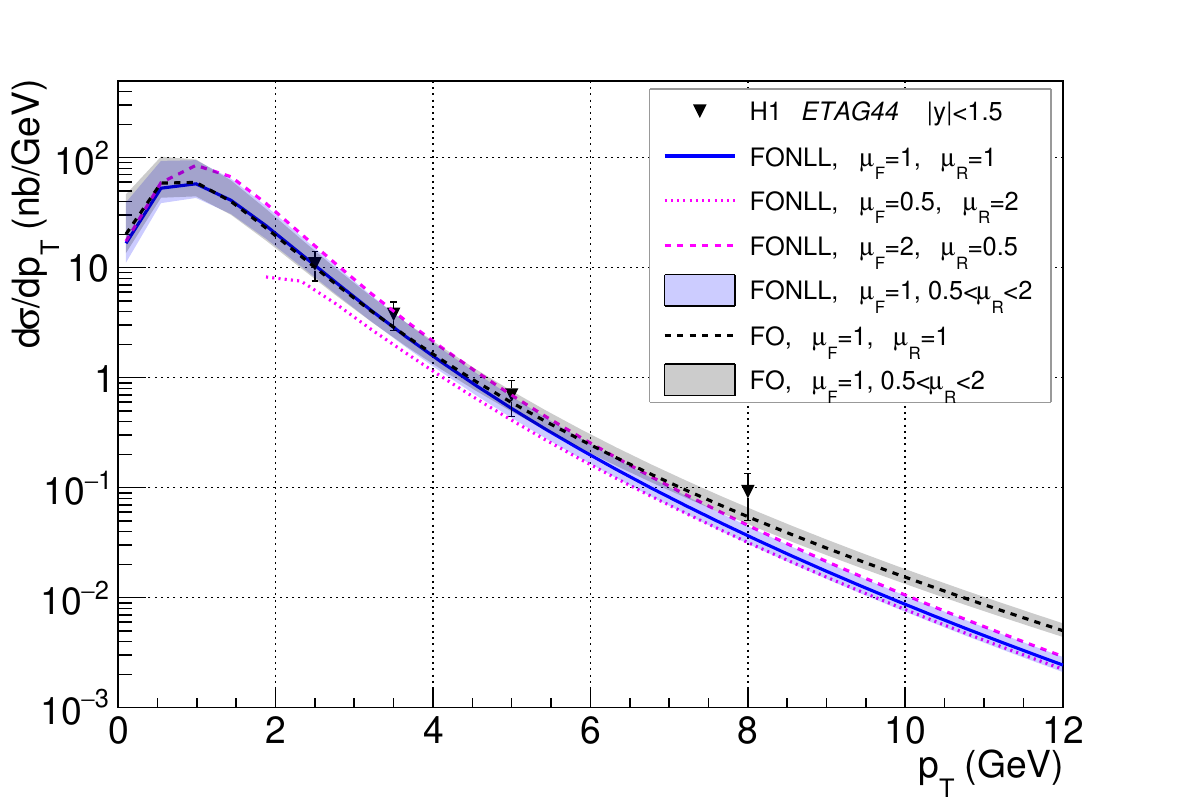}
  \caption{H1 \textit{ETAG44}}
    \label{fig:h1eta44pt}
\end{subfigure} 
\begin{subfigure}{0.49\textwidth}
    \centering
    \includegraphics[width=\textwidth]{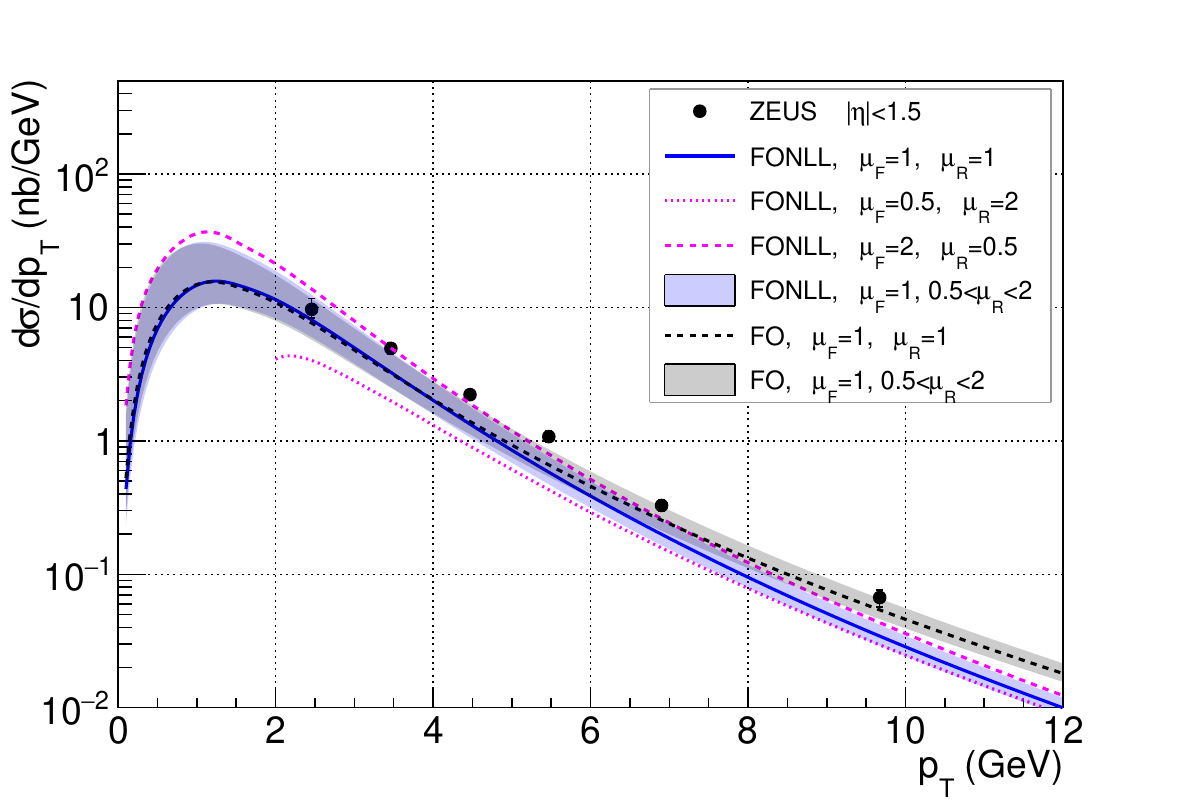}
    \caption{ZEUS}
    \label{fig:zeus}
\end{subfigure}
\hfill
\begin{subfigure}{0.49\textwidth}
    \centering
    \includegraphics[width=\textwidth]{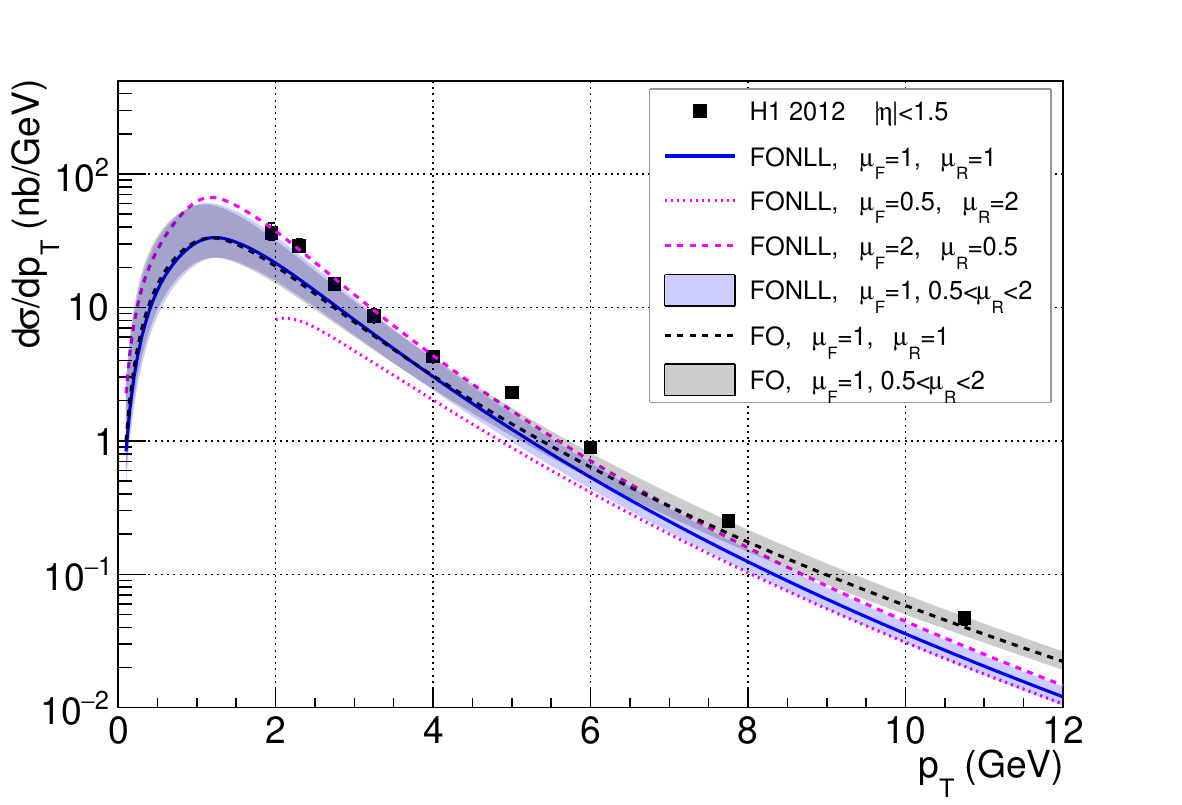}
    \caption{H1 2012}
    \label{fig:h12011}
\end{subfigure}
 \caption{Transverse momentum distribution $D^*$ mesons in photoproduction in electron-proton collisions.  FONLL calculation \cite{Frixione:2002zv} $\mu_F=\mu_R=1$ (blue solid). Shaded blue indicates variation of $0.5<\mu_R<2$ while $\mu_F=1$ is fixed; magenta-dashed $\mu_F=2.0, \mu_R=0.5$, magenta-dotted $\mu_F=0.5,\mu_R=2.0$. 
 FO calculation \cite{Frixione:2002zv} $\mu_F=\mu_R=1$ (black solid), and grey band corresponds to the variation of the renormalization scale. PSSZ fragmentation function is used with $\varepsilon=0.02$ for FONLL and $\varepsilon=0.035$ for FO.
 Compared with data from H1 \cite{H1:1998csb} \textit{ETAG33} (a) and \textit{ETAG44} (b), ZEUS  \cite{ZEUS:1998wxs} (c) and H1 2012 \cite{H1:2011myz} (d).}
 \label{fig:heradata}
\end{figure}

\begin{figure}
\centering
\begin{subfigure}{0.49\textwidth}
    \centering
    \includegraphics[width=\textwidth]{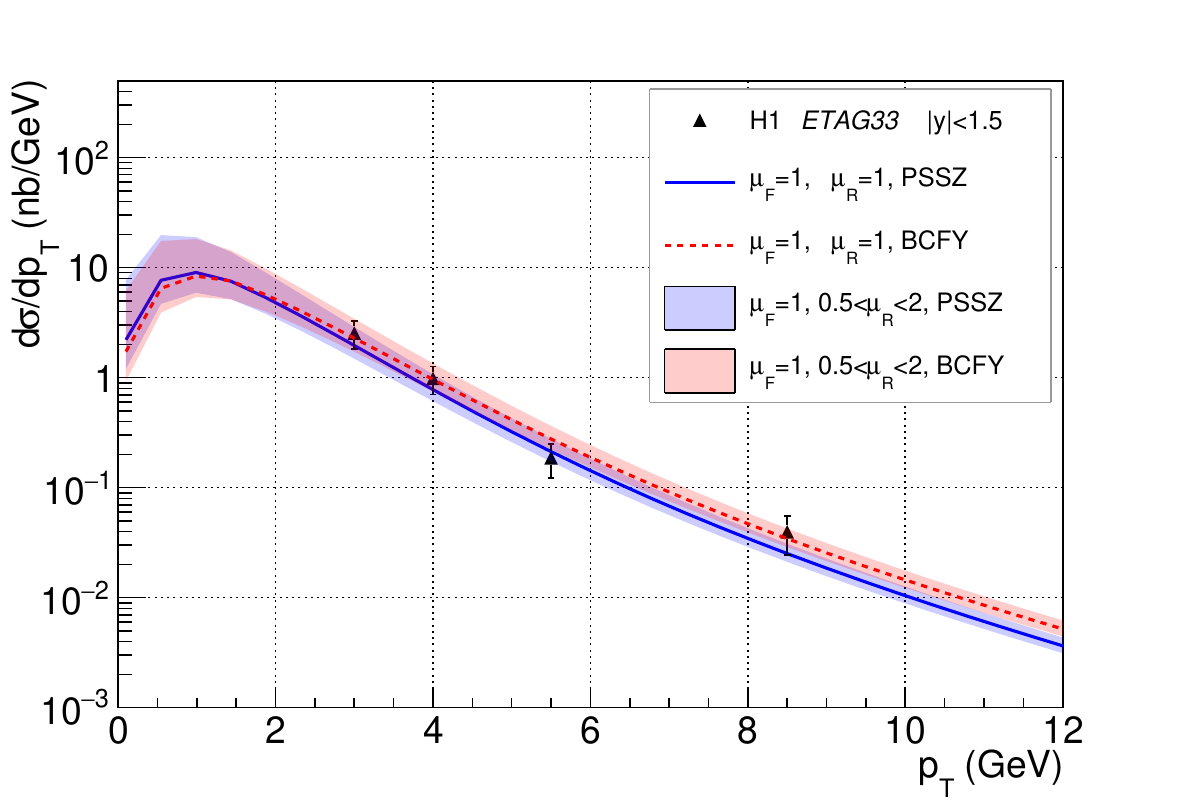}
    \caption{H1 \textit{ETAG33}}
    \label{fig:h1eta33pt_bcfy}
\end{subfigure}
\hfill
\begin{subfigure}{0.49\textwidth}
    \centering
    \includegraphics[width=\textwidth]{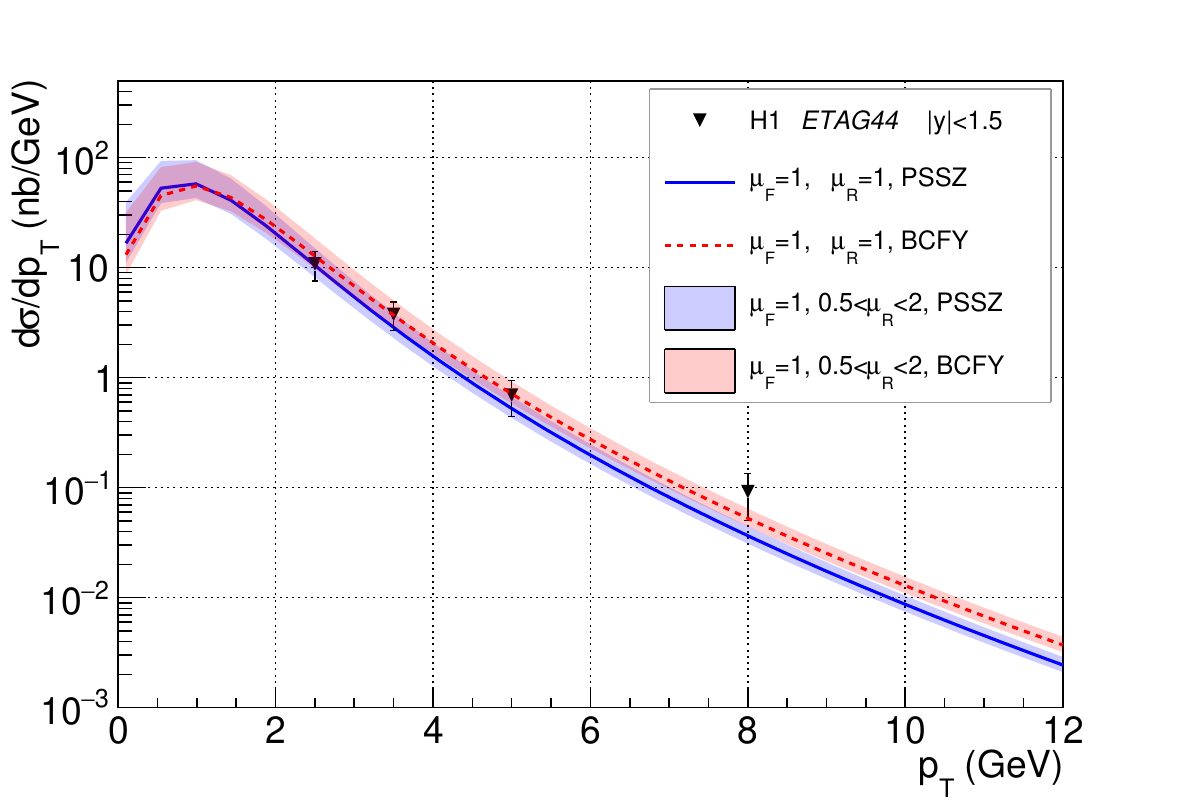}
  \caption{H1 \textit{ETAG44}}
    \label{fig:h1eta44pt_bcfy}
\end{subfigure} 
\begin{subfigure}{0.49\textwidth}
    \centering
    \includegraphics[width=\textwidth]{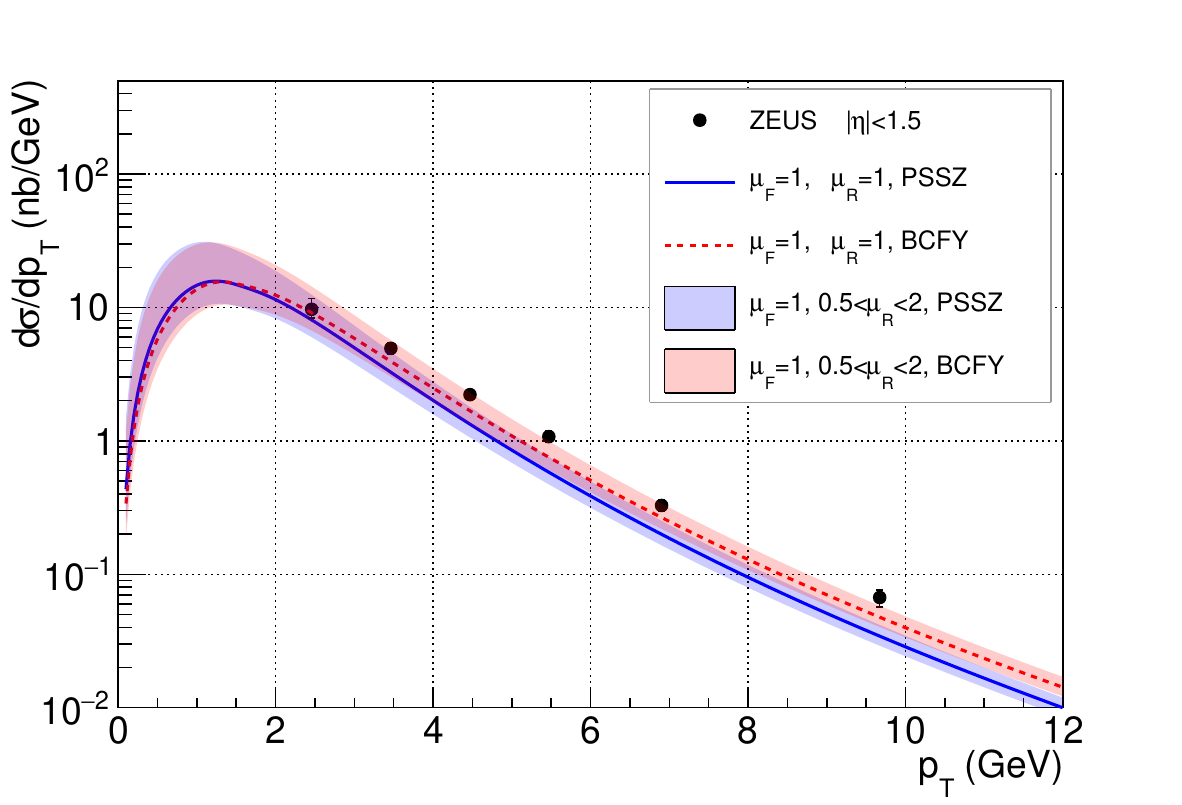}
    \caption{ZEUS}
    \label{fig:zeus_bcfy}
\end{subfigure}
\hfill
\begin{subfigure}{0.49\textwidth}
    \centering
    \includegraphics[width=\textwidth]{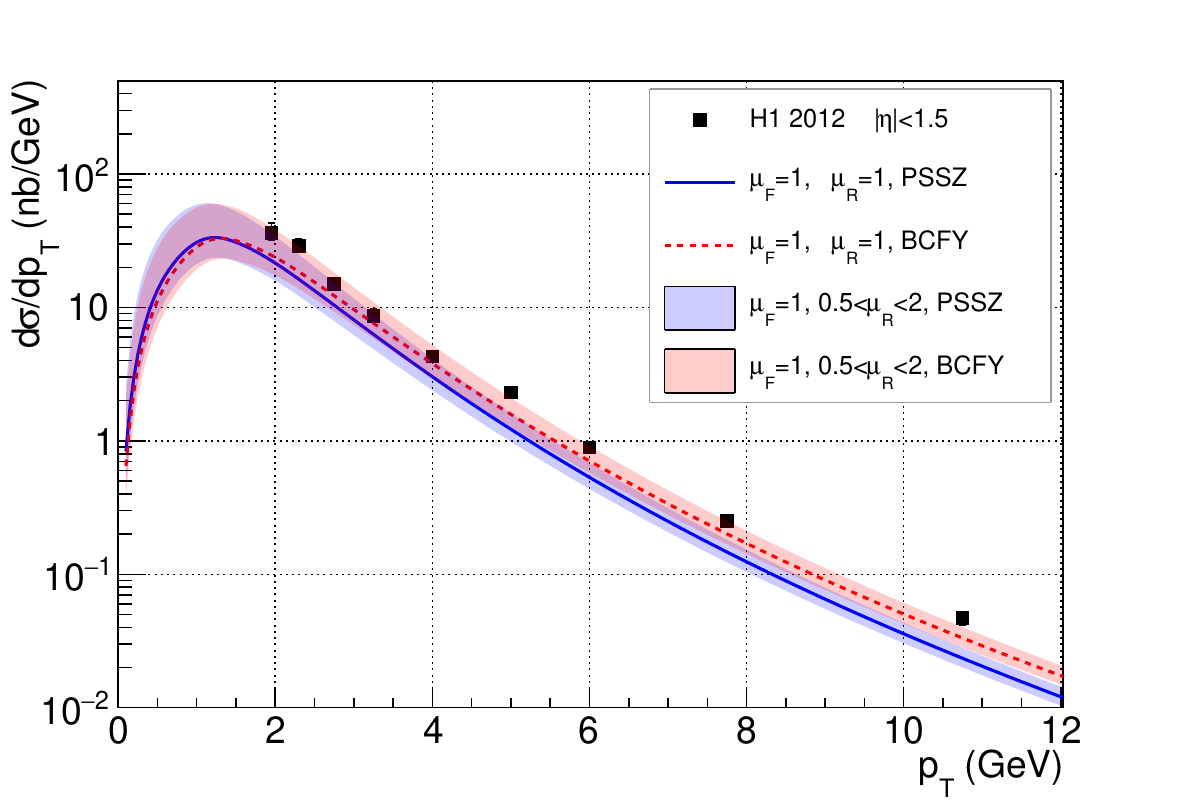}
    \caption{H1 2012}
    \label{fig:h12011_bcfy}
\end{subfigure}
 \caption{Transverse momentum distribution $D^*$ mesons in photoproduction in electron-proton collisions obtained from  FONLL  \cite{Frixione:2002zv}. Bands indicate the   variation of $0.5<\mu_R<2$ while $\mu_F=1$ is fixed. Blue band: PSSZ fragmentation function with $\varepsilon=0.02$, red band: BCFY fragmentation function with $r=0.1$. 
 Compared with data from H1 \cite{H1:1998csb} \textit{ETAG33} (a) and \textit{ETAG44} (b), ZEUS  \cite{ZEUS:1998wxs} (c) and H1 2012 \cite{H1:2011myz} (d).}
 \label{fig:heradata_bcfy}
\end{figure}

\begin{figure}
\centering
\begin{subfigure}{0.49\textwidth}
    \centering
    \includegraphics[width=\textwidth]{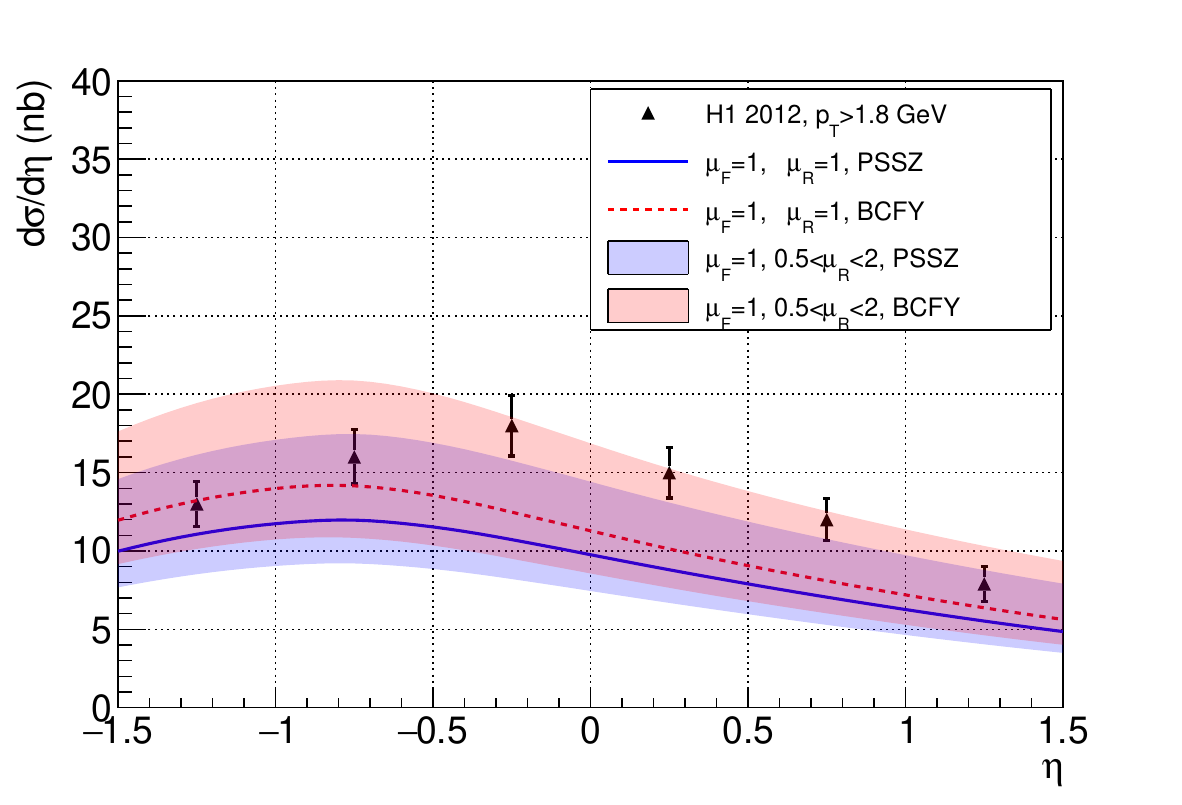}
    \label{fig:h1eta_bcfy}
\end{subfigure}
\hfill
\begin{subfigure}{0.49\textwidth}
    \centering
    \includegraphics[width=\textwidth]{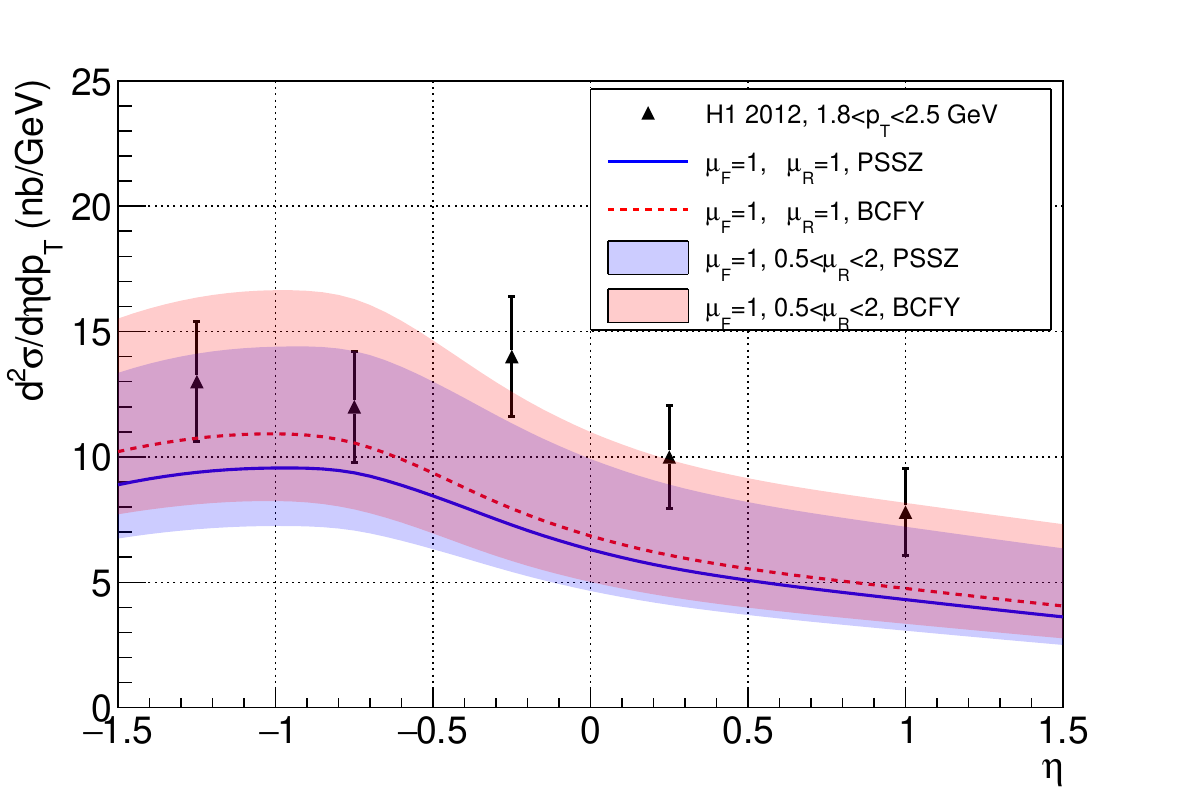}
    \label{fig:h1eta1_bcfy}
\end{subfigure} 
\begin{subfigure}{0.49\textwidth}
    \centering
    \includegraphics[width=\textwidth]{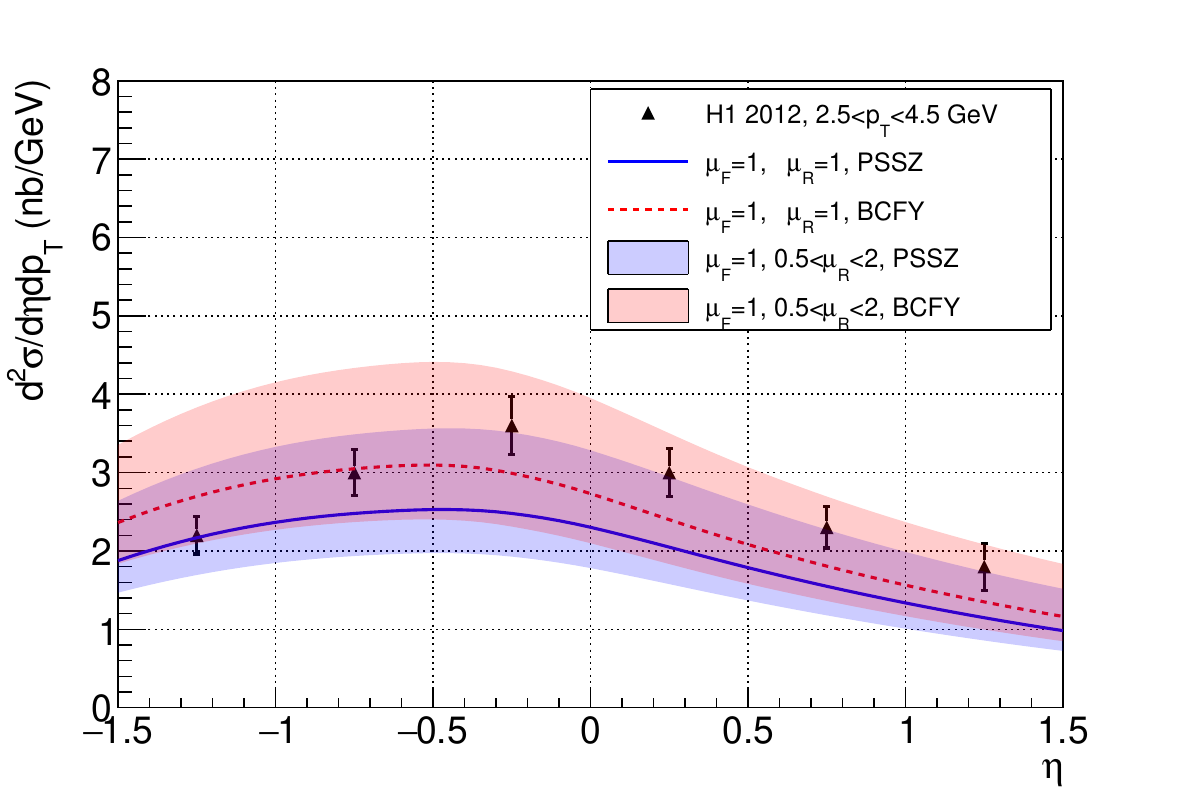}
    \label{fig:h1eta2_bcfy}
\end{subfigure}
\hfill
\begin{subfigure}{0.49\textwidth}
    \centering
    \includegraphics[width=\textwidth]{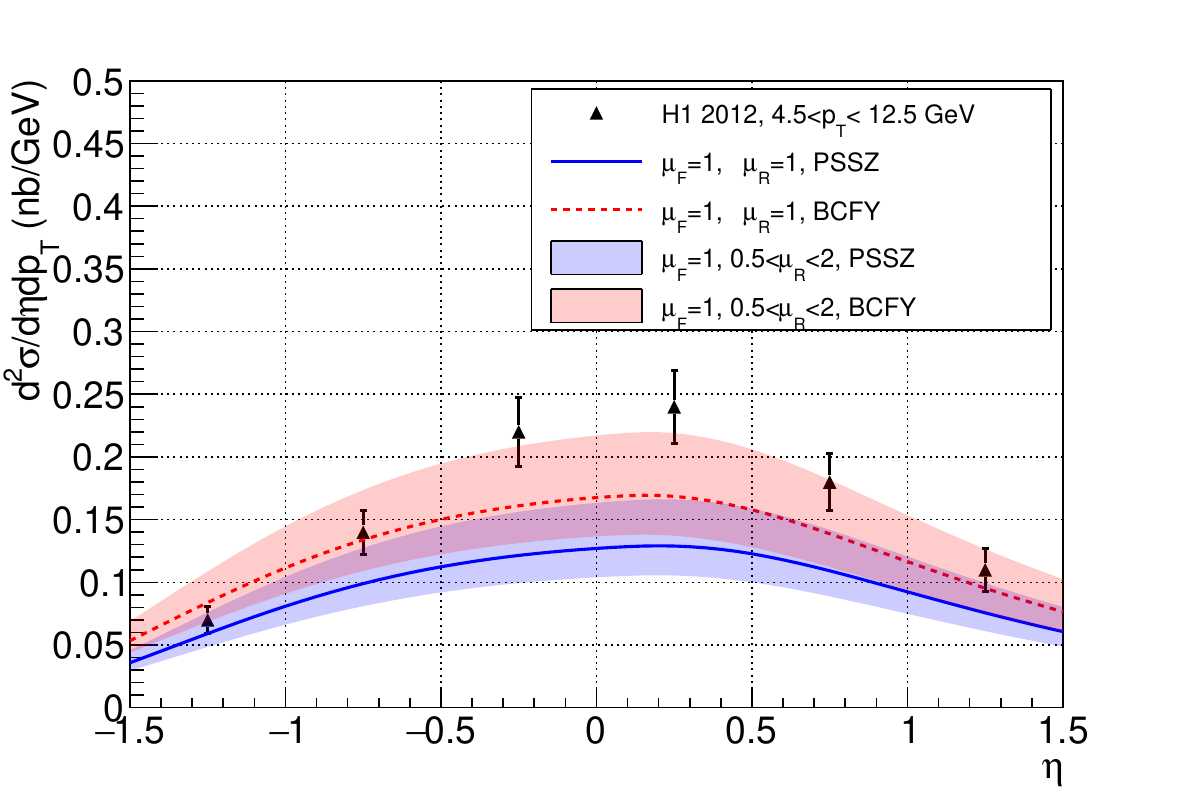}
    \label{fig:h1eta3_bcfy}
\end{subfigure}
 \caption{Pseudorapidity distribution of $D^*$ mesons in photoproduction in electron-proton collisions at HERA from FONLL calculation \cite{Frixione:2002zv}. Bands denote renormalization scale variation  $0.5<\mu_R<2$ while $\mu_F=1$ is fixed. Blue band: PSSZ fragmentation function with $\varepsilon=0.02$, red band BCFY fragmentation function with $r=0.1$.  Compared with data from H1  \cite{H1:2011myz}, for different $p_T$ bins. Note that data and results in three $p_T$ bins $(1.8,2.5), (2.5,4.5), (4.5,12.5)$ are presented as double differential cross sections. Positive pseudorapidity is proton direction.}
 \label{fig:h1datay2011_bcfy}
\end{figure}

\noindent
In Fig.~\ref{fig:heradata} all four data sets from HERA are compared with the theoretical calculations based  on the FONLL  with PSSZ fragmentation function. The plots show transverse momentum distribution, while the rapidity (or pseudorapidity) is integrated out in the kinematic regions defined in Table \ref{table:names_numbers}, corresponding to each data set. The proton PDF used in all cases is CT18ANLO. The blue solid curve corresponds to FONLL and blue bands indicate the renormalization scale dependence while the factorization scale is fixed. There are also two cases shown, which indicate the factorization scale dependence (dotted and dashed magenta). As demonstrated in \cite{Frixione:2002zv}, and observed also here,  the renormalization scale dependence is the dominant one, except for lower values of transverse momenta ($p_T \lesssim 2$ GeV) where  factorization scale variation is sizeable. For some choices of renormalization and factorization scale the results become unreliable at  low transverse momentum (see \cite{Frixione:2002zv}). Thus for one of such scenario  we only show the curve down to about $2 \rm \, GeV$. The charm quark mass is set to \(m_c=1.3 \, \rm GeV\). The description of older H1 data is very good, for the ZEUS and  later H1  data the FONLL prediction is somewhat too soft, the description is very good in the low $p_T$ region and underestimates the data at higher $p_T$. Also the calculation based on newer PDFs is  consistent with the original  FONLL calculations \cite{Frixione:2002zv}  based on  CTEQ5M PDF set\footnote{For consistency check, we also used CTEQ5M PDF set in FONLL code and reproduced the calculations of \cite{Frixione:2002zv}.} For comparison we also show the calculation based on the fixed order (FO) NLO calculation, black solid with grey bands. As mentioned above, for this calculation we take $\varepsilon=0.035$, consistently with parameters used in FMNR calculation in \cite{H1:2011myz}. The distribution is harder in that case, and the description of the data is slightly better, especially at higher $p_T$ and this is particularly visible  for ZEUS \cite{ZEUS:1998wxs} and H1 2012 \cite{H1:2011myz} data. 
\\[4pt] 
\noindent
In Fig.~\ref{fig:heradata_bcfy} we compare the transverse-momentum distributions of $\Dzero$ mesons obtained with the PSSZ (blue band and curve) and BCFY (red band and curve) fragmentation functions. Relative to PSSZ, the BCFY parametrization yields a slightly harder $p_T$ spectrum and provides a better description of the HERA data, particularly at high $p_T$.
\\[4pt] 
\noindent
\noindent
\begin{figure}
\centering
\begin{subfigure}{0.49\textwidth}
    \centering
    \includegraphics[width=\textwidth]{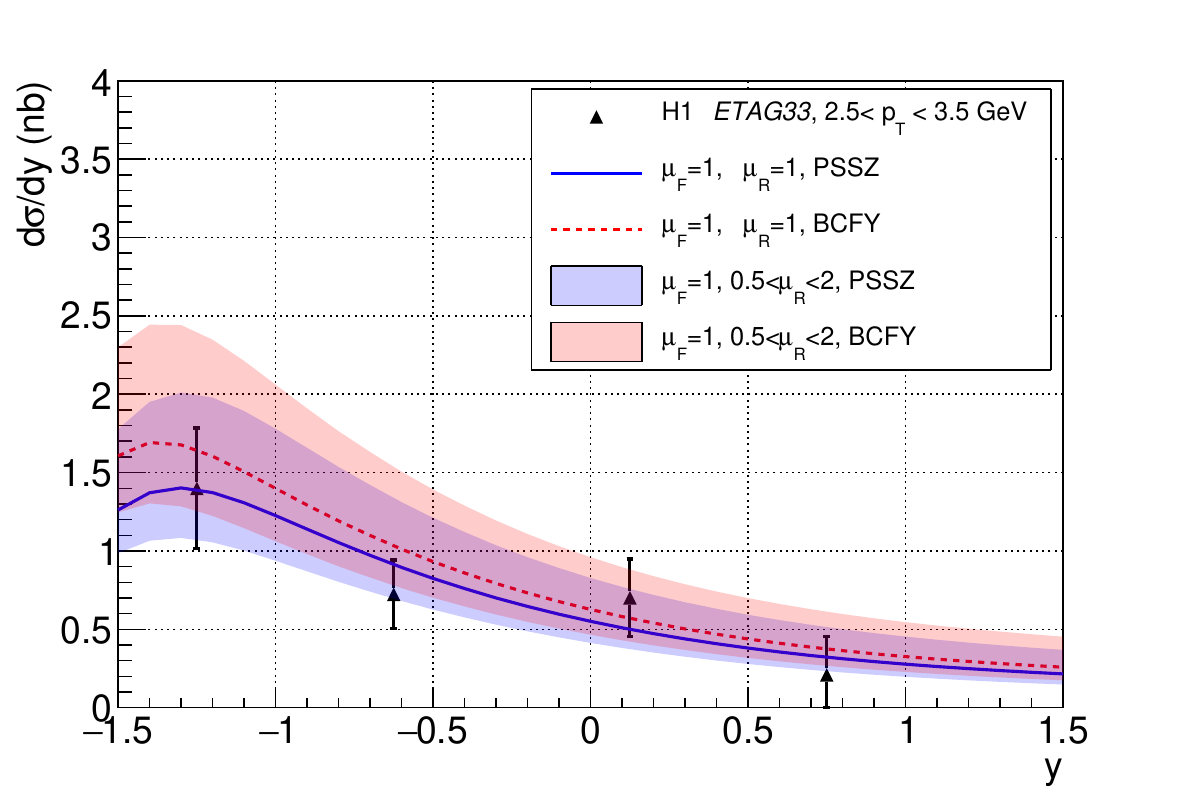}
    \label{fig:h1331y_bcfy}
\end{subfigure}
\hfill
\begin{subfigure}{0.49\textwidth}
    \centering
    \includegraphics[width=\textwidth]{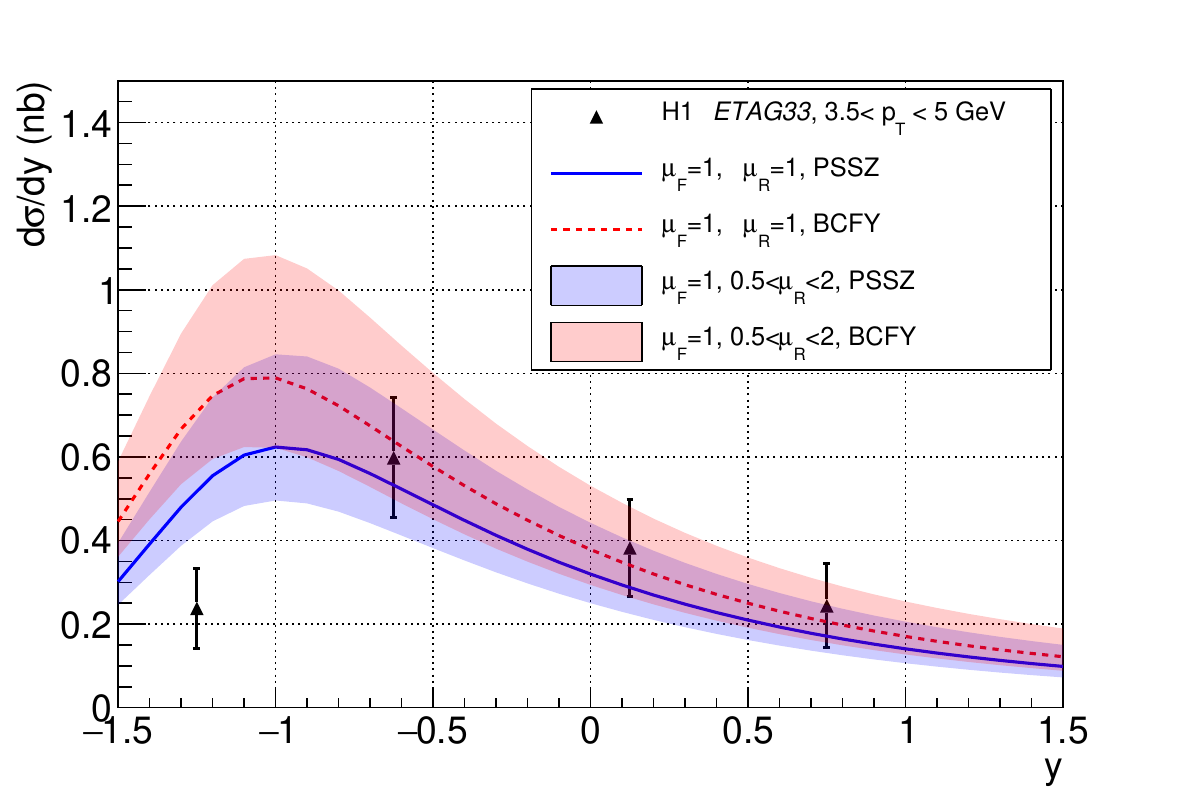}
    \label{fig:h1332y_bcfy}
\end{subfigure} 
\begin{subfigure}{0.49\textwidth}
    \centering
    \includegraphics[width=\textwidth]{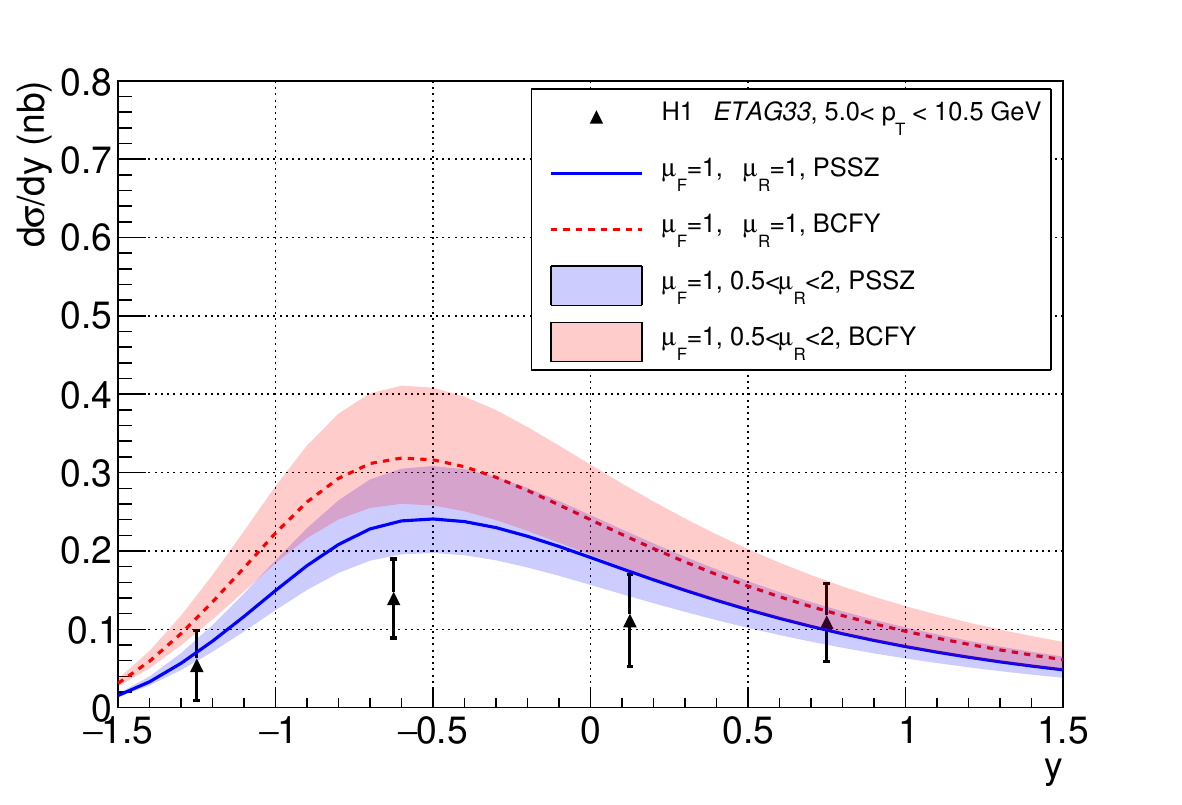}
    \label{fig:h1333y_bcfy}
\end{subfigure}
\hfill
\begin{subfigure}{0.49\textwidth}
    \centering
    \includegraphics[width=\textwidth]{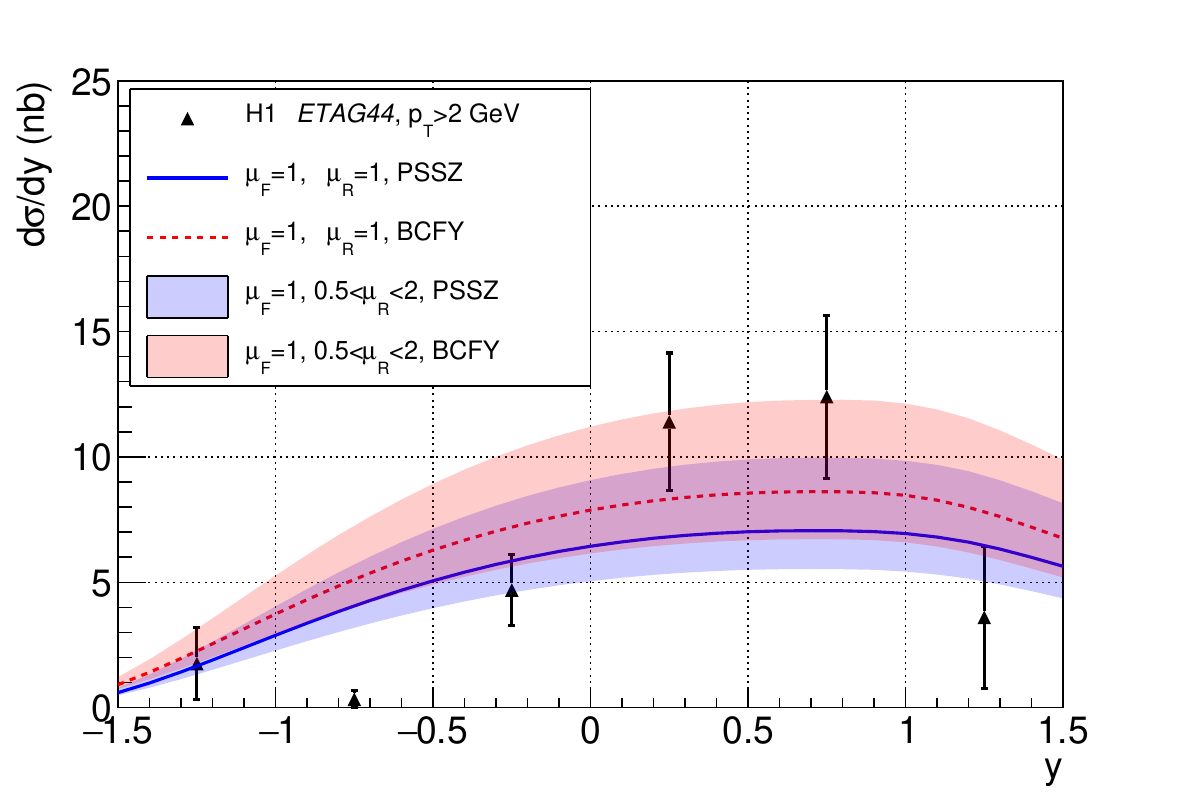}
    \label{fig:h144y_bcfy}
\end{subfigure}
 \caption{Rapidity distribution of $D^*$ mesons in photoproduction in electron-proton collisions at HERA from FONLL calculation \cite{Frixione:2002zv}. Bands denote renormalization scale variation  $0.5<\mu_R<2$ while $\mu_F=1$ is fixed. Blue band: PSSZ fragmentation function with $\varepsilon=0.02$, red band BCFY fragmentation function with $r=0.1$. Compared with data from H1  \cite{H1:1998csb}, for different $p_T$ bins. Note different vertical scales. Positive rapidity is proton going direction.}
 \label{fig:h1datay_bcfy}
\end{figure}
\begin{figure}
\centering
\begin{subfigure}{0.49\textwidth}
    \centering
    \includegraphics[width=\textwidth]{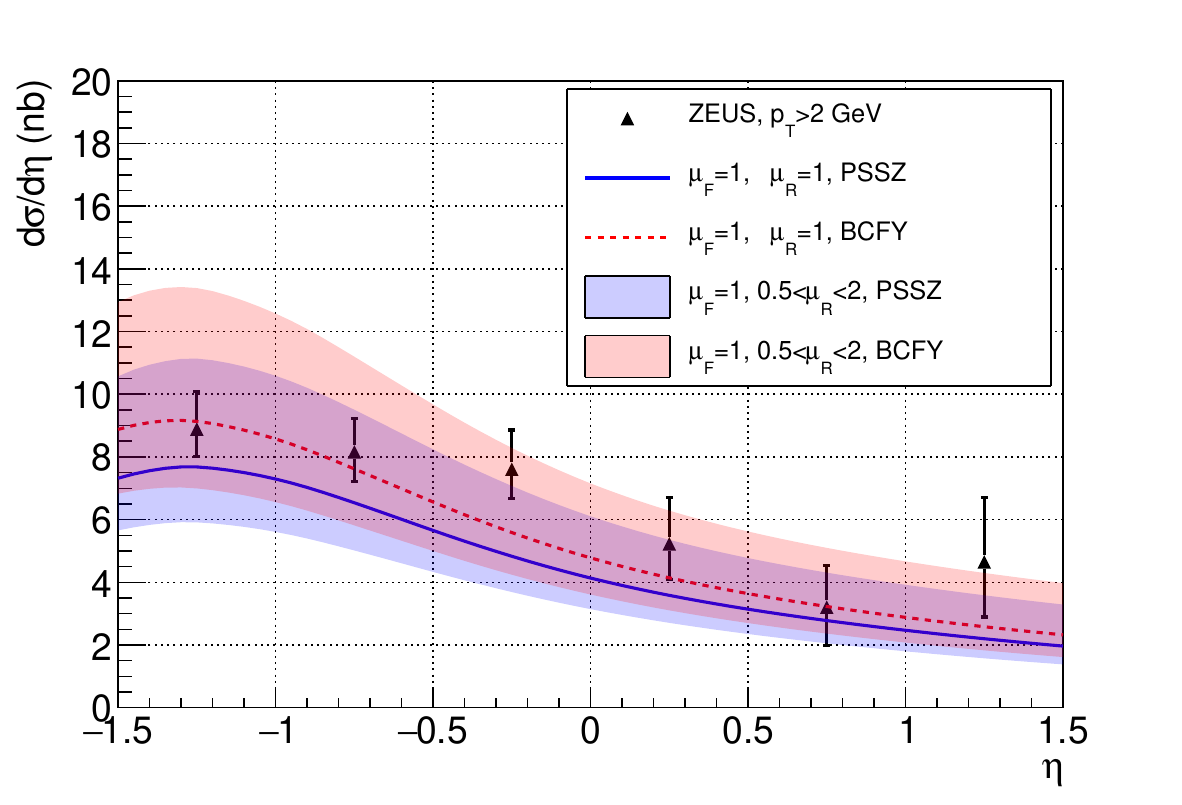}
    \label{fig:zeusy2_bcfy}
\end{subfigure}
\hfill
\begin{subfigure}{0.49\textwidth}
    \centering
    \includegraphics[width=\textwidth]{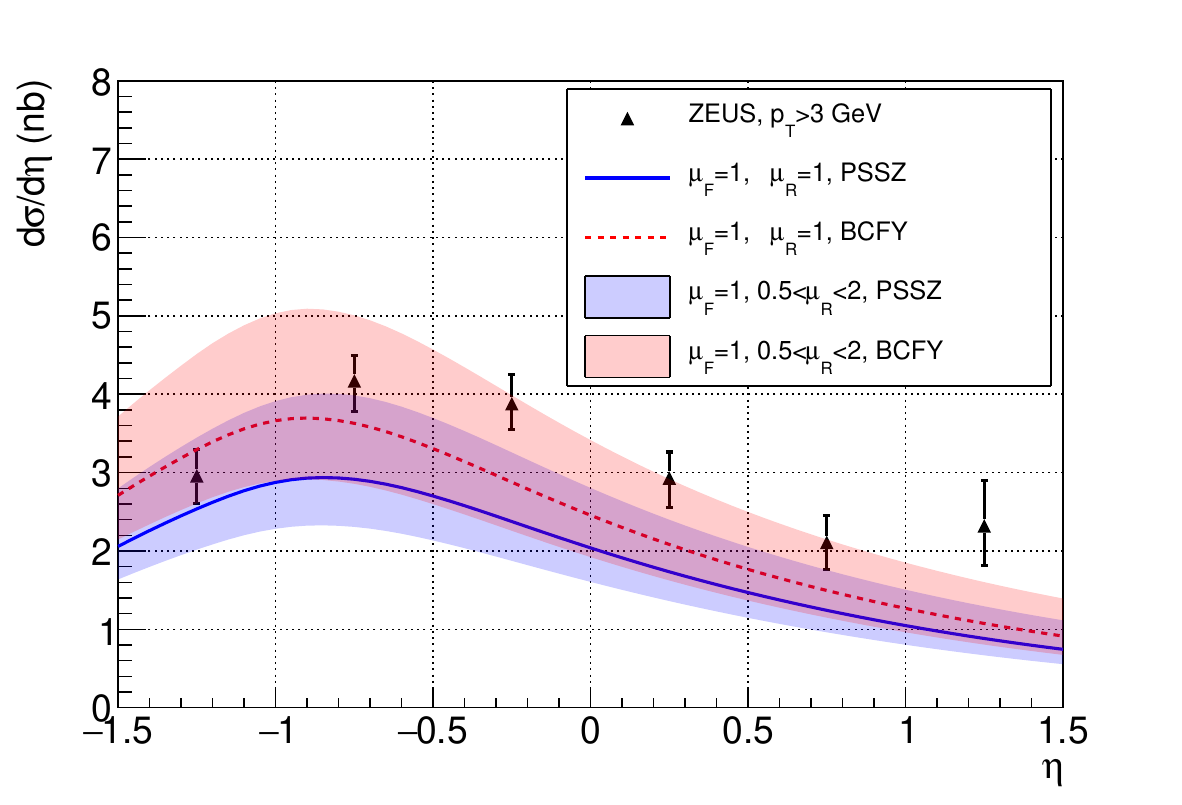}
    \label{fig:zeusy3_bcfy}
\end{subfigure} 
\begin{subfigure}{0.49\textwidth}
    \centering
    \includegraphics[width=\textwidth]{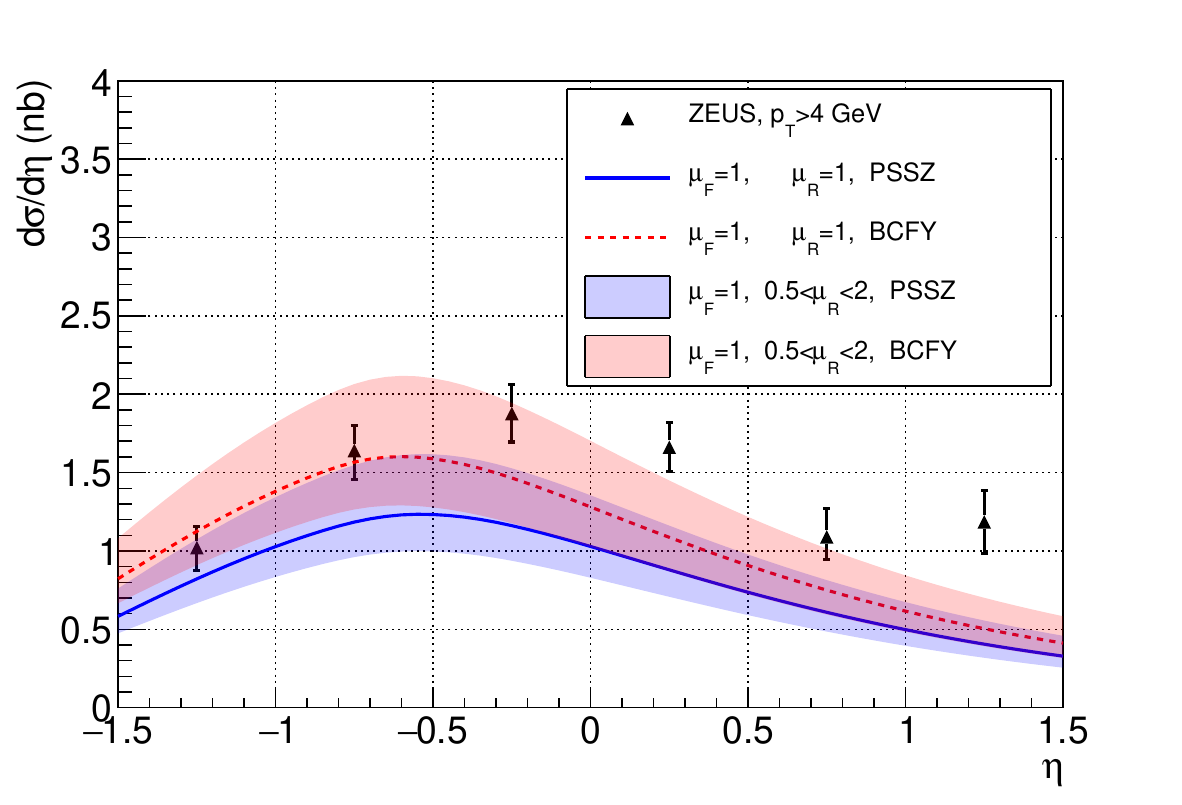}
    \label{fig:zeusy4_bcfy}
\end{subfigure}
\hfill
\begin{subfigure}{0.49\textwidth}
    \centering
    \includegraphics[width=\textwidth]{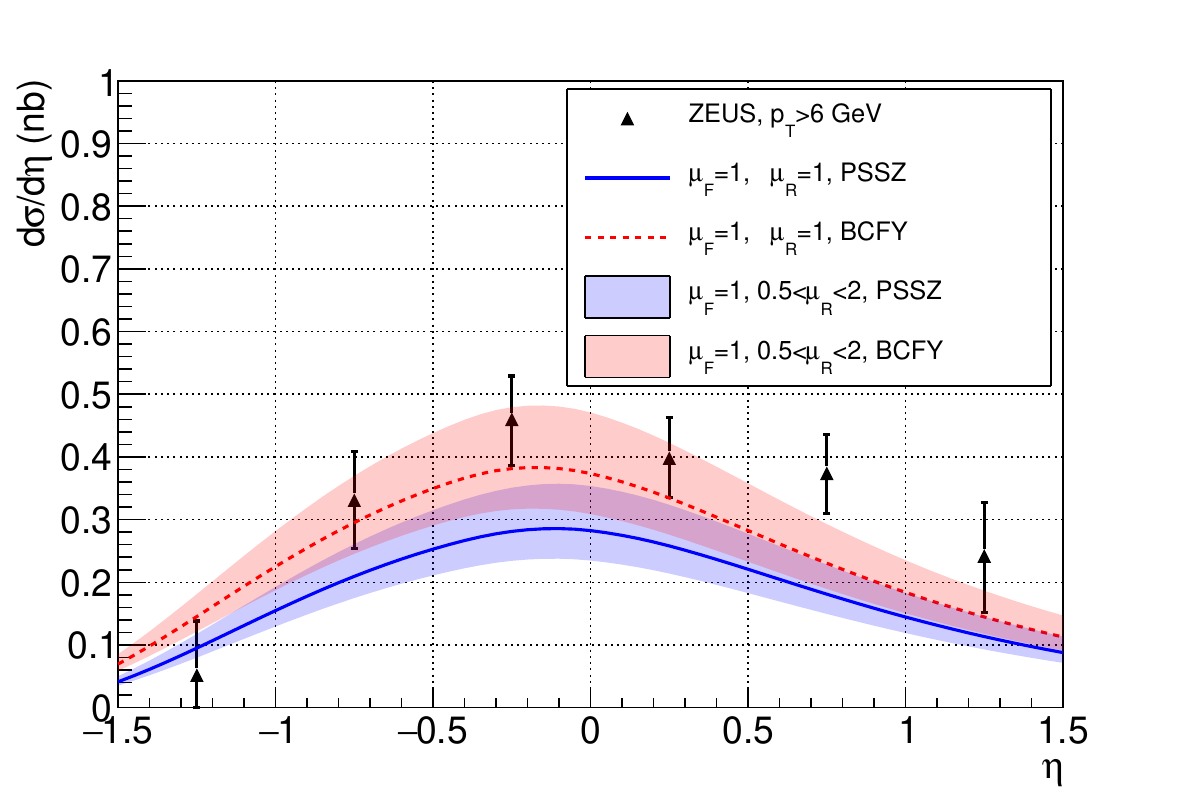}
    \label{fig:zeusy6_bcfy}
\end{subfigure}
 \caption{Pseudorapidity distribution of $D^*$ mesons in photoproduction in electron-proton collisions at HERA from FONLL calculation \cite{Frixione:2002zv}. Bands denote renormalization scale variation  $0.5<\mu_R<2$ while $\mu_F=1$ is fixed. Blue band: PSSZ fragmentation function with $\varepsilon=0.02$, red band BCFY fragmentation function with $r=0.1$.  Compared with data from ZEUS  \cite{ZEUS:1998wxs}, for different minimum $p_T$ cuts: $2,3,4,6$ GeV. Note different vertical scales. Positive pseudorapidity is proton direction.}
 \label{fig:zeusdatay_bcfy}
\end{figure}
\noindent
In Figs.~\ref{fig:h1datay_bcfy}, \ref{fig:zeusdatay_bcfy}, and \ref{fig:h1datay2011_bcfy}, we show FONLL rapidity and pseudorapidity distributions for various $p_T$ cuts across multiple data sets. The rapidity convention matches that of HERA, with the proton beam taken as the positive direction.
The blue bands represent the renormalization-scale uncertainty for the calculation employing the PSSZ fragmentation function, whereas the red bands correspond to the computation using BCFY. The solid blue and dashed red curves denote the respective central predictions. The factorization scale is fixed at $\mu_F/\mu_0 = 1$. Overall, the shape  of the distributions are very well reproduced by both calculations.
The calculation with BFCY fragmentation seems to be closer in normalization to data in majority of  $p_T$ bins. \\[6pt]
Detailed comparison between FONLL-based calculations and predictions obtained using fixed-order calculations are shown in Appendix A.
\\[4pt]
\noindent
We also studied  the impact of changes in the photon and proton PDFs on the calculation. We checked that using AFG photon PDF \cite{Aurenche:1994in} has negligible impact on the cross sections in the HERA kinematics. Similarly, we have checked that the differences between the calculations using CT18ANLO, HERAPDF2.0 and nNNPDF3.0\_p  are of the order of few percent for $p_T>2$ GeV, and only in the very low $p_T$ region ($<1$ GeV) the differences become larger. We have also tested the sensitivity of the calculation to the different value of charm mass. Increasing mass of the heavy quark leads to the harder $p_T$ spectrum. At low values of $p_T$ the calculation with higher mass is suppressed, while at higher values of $p_T$ it is enhanced since the effect of the logarithmic resummation is smaller for this case (see discussion in \cite{Frixione:2002zv}).
\\[4pt]
\noindent
Summarizing, we have updated FONLL calculations with newer proton PDFs and compared with the HERA data on $D^*$ photoproduction. The change due to the proton PDF is small with respect to the original calculations presented  in \cite{Frixione:2002zv}. 
Overall the data are well described by the FONLL calculation with BCFY fragmentation function. The calculation with PSSZ fragmentation function describes older H1 \cite{H1:1998csb} data reasonably well, while it leads to softer $p_T$ spectrum than the  ZEUS \cite{ZEUS:1998wxs}
 and H1 data from 2012 \cite{H1:2015ubc}.
 The calculation with BCFY fragmentation function leads to a better description of the HERA data, especially at high values of transverse momenta.
\section{Predictions for $D^0$ production in PbPb ultraperipheral collisions at the LHC}
\label{sec:upcfinebin}

\begin{figure}[H]
\centering
\begin{subfigure}{0.49\textwidth}
    \centering
    \includegraphics[width=\textwidth]{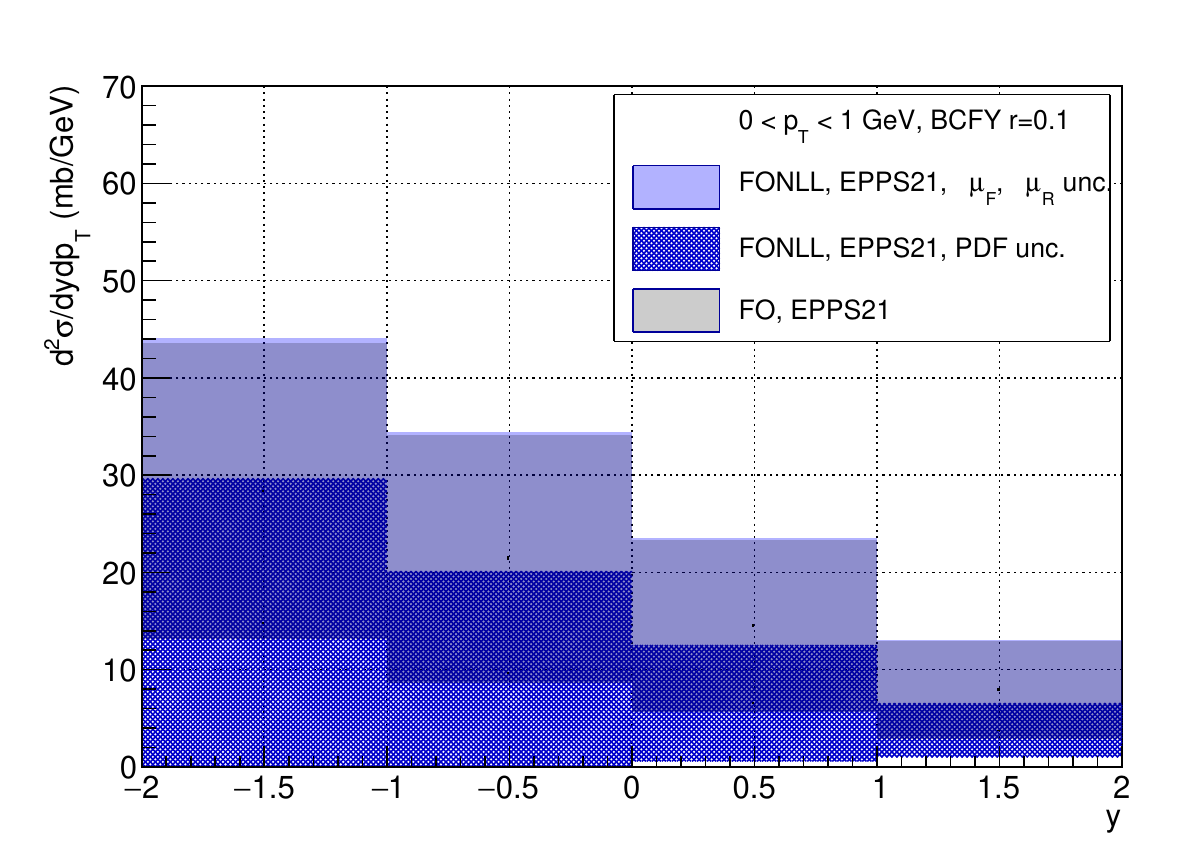}
    \label{fig:ydpt0_fine_epps21bcfy}
\end{subfigure}
\hfill
\begin{subfigure}{0.49\textwidth}
    \centering
    \includegraphics[width=\textwidth]{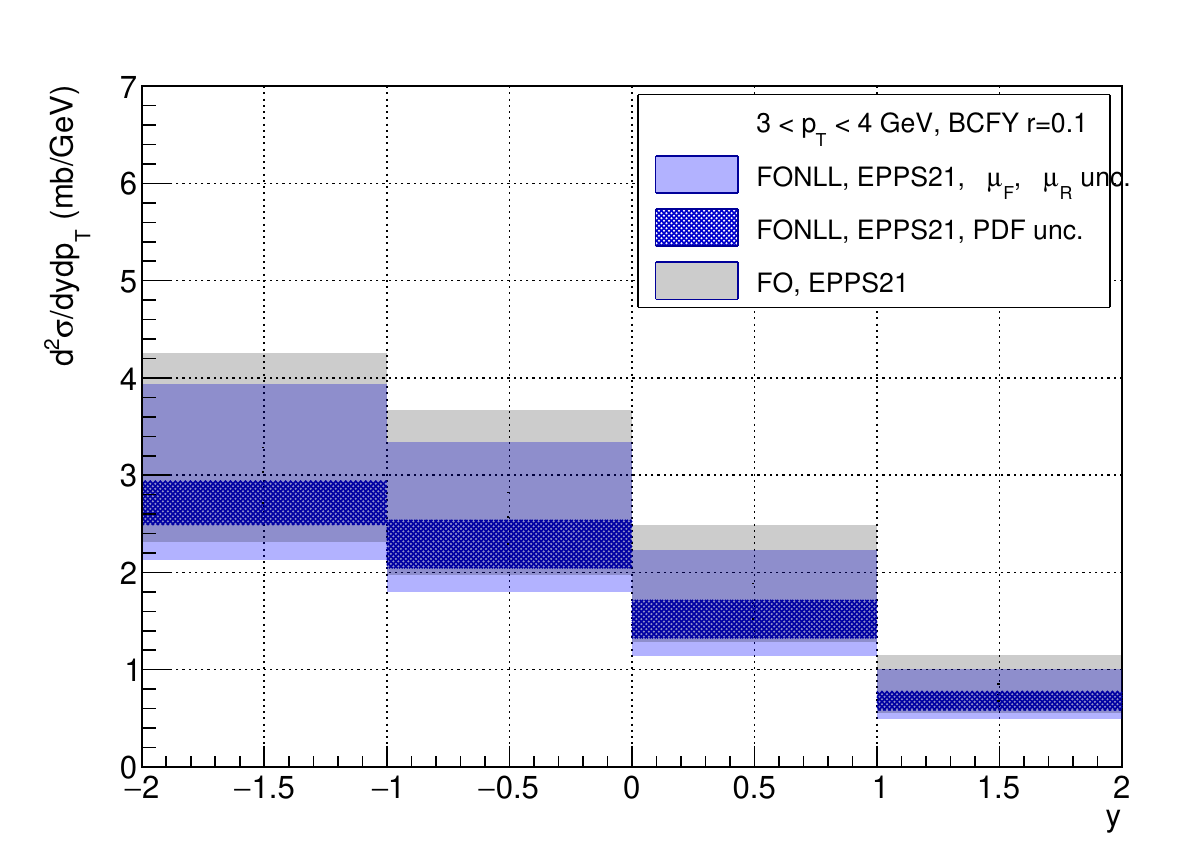}
    \label{fig:ydpt1_fine_epps21bcfy}
\end{subfigure} 
\begin{subfigure}{0.49\textwidth}
    \centering
    \includegraphics[width=\textwidth]{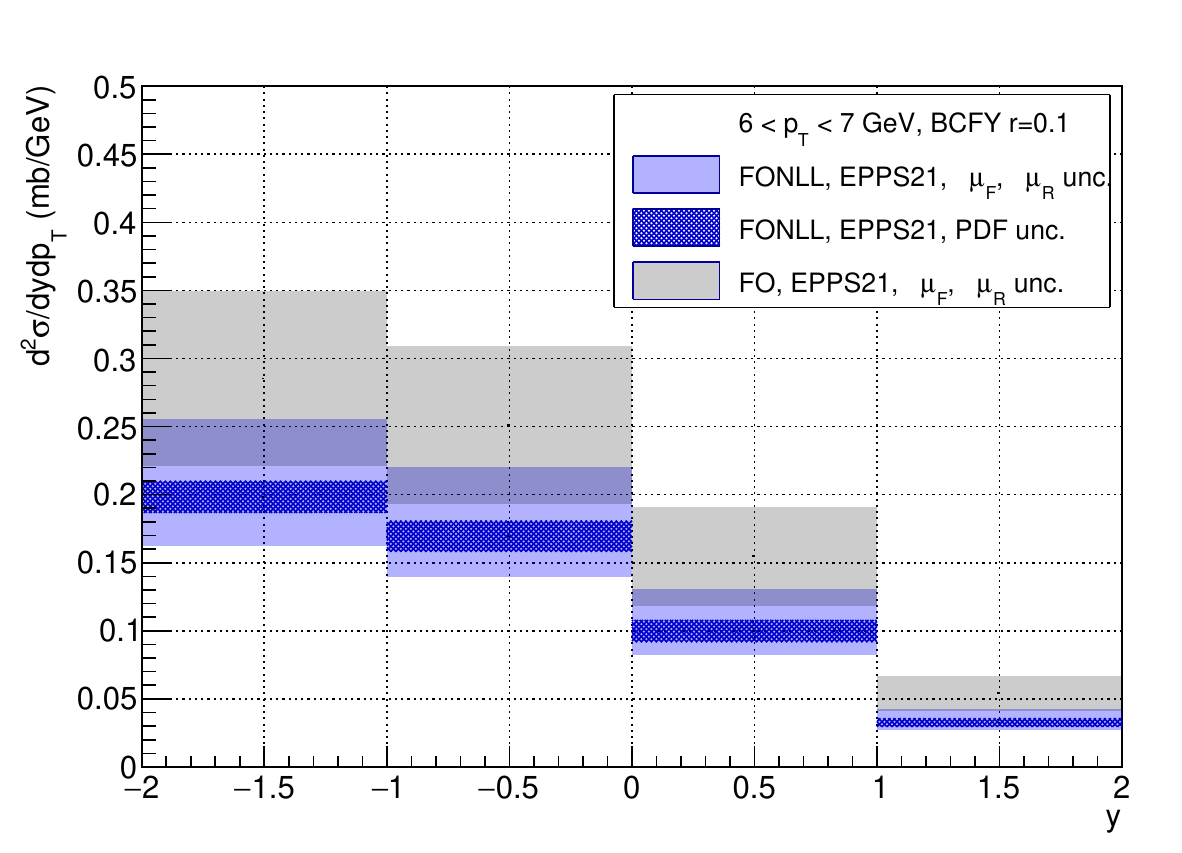}
    \label{fig:ydpt2_fine_epps21bcfy}
\end{subfigure}
\hfill
\begin{subfigure}{0.49\textwidth}
    \centering
    \includegraphics[width=\textwidth]{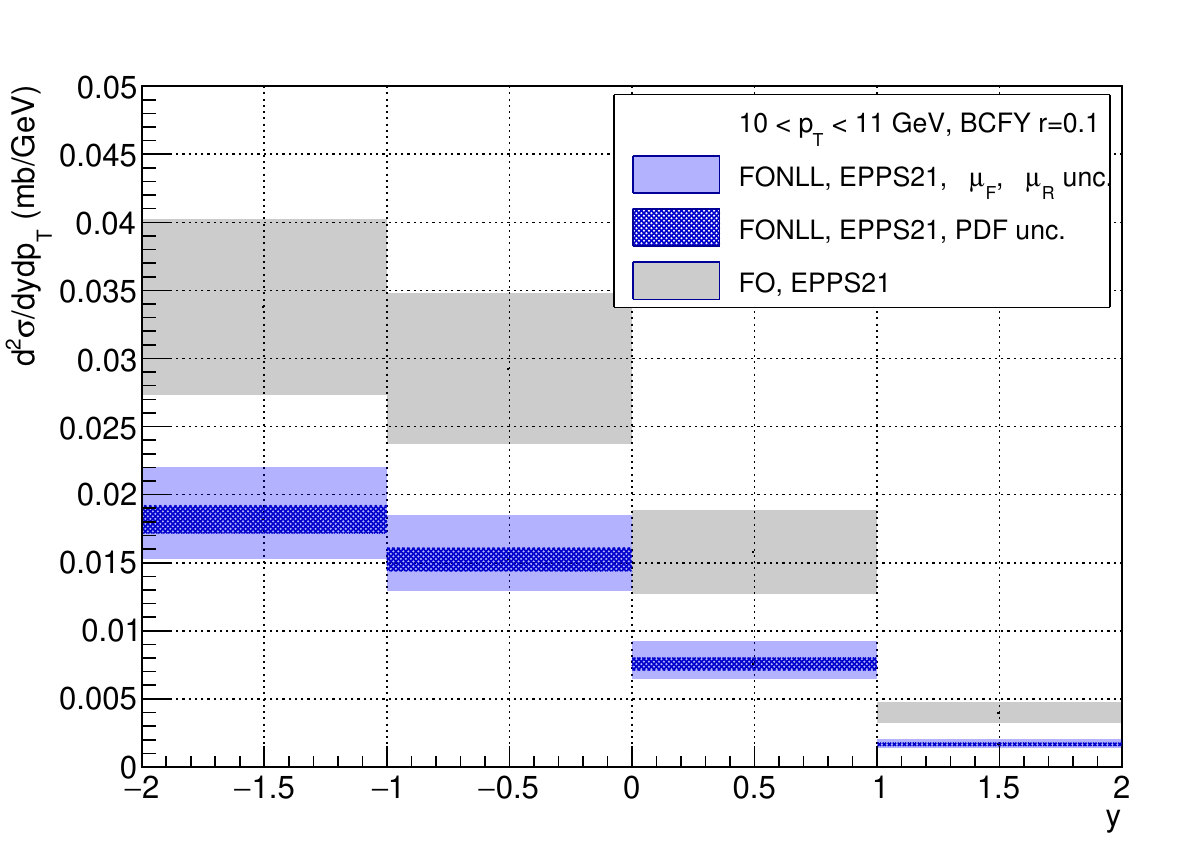}
    \label{fig:ydpt3_fine_epps21bcfy}
\end{subfigure}
 \caption{Rapidity distribution  for the $D^0$  production in UPC PbPb collisions at $\sqrt{\rm s_{\scriptscriptstyle{NN}}}=5.36$ TeV.  Four panels correspond to $p_{T}$ bins $(0-1), (3-4), (6-7), (10-11)$ GeV. Light blue band: FONLL  calculation with factorization and renormalization scale variation, grey band: FO  calculation with factorization and renormalization scale variation, dark blue band: FONLL EPPS21 \cite{Eskola:2021nhw} PDF uncertainty. Both FONLL and FO calculation done with BCFY fragmentation function \cite{Braaten:1994bz,Cacciari:2003zu} with parameter $r=0.1$. Photon-emitting nucleus is moving in the positive rapidity direction.}
 \label{fig:ydpt_fine_epps21bcfy}
\end{figure}

\begin{figure}[H]
\centering
\begin{subfigure}{0.49\textwidth}
    \centering
    \includegraphics[width=\textwidth]{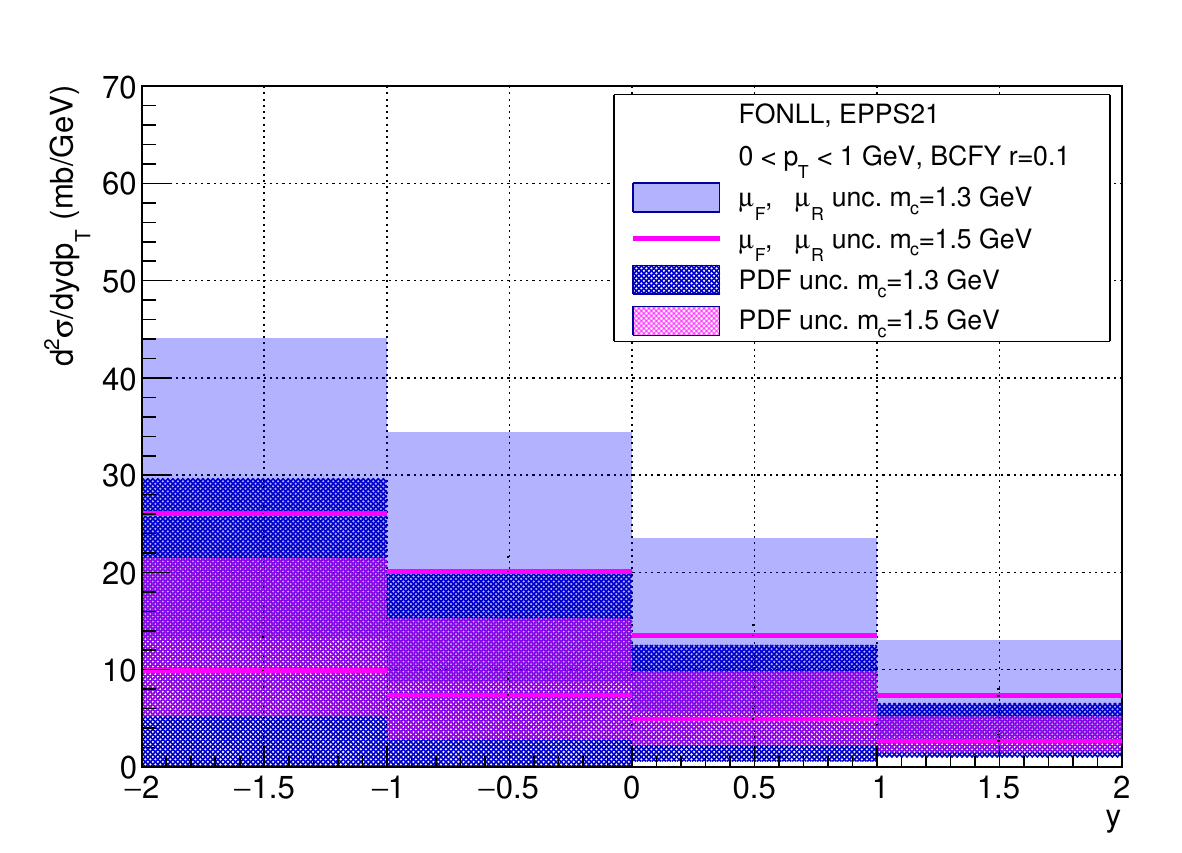}
    \label{fig:ydpt0mc}
\end{subfigure}
\hfill
\begin{subfigure}{0.49\textwidth}
    \centering
    \includegraphics[width=\textwidth]{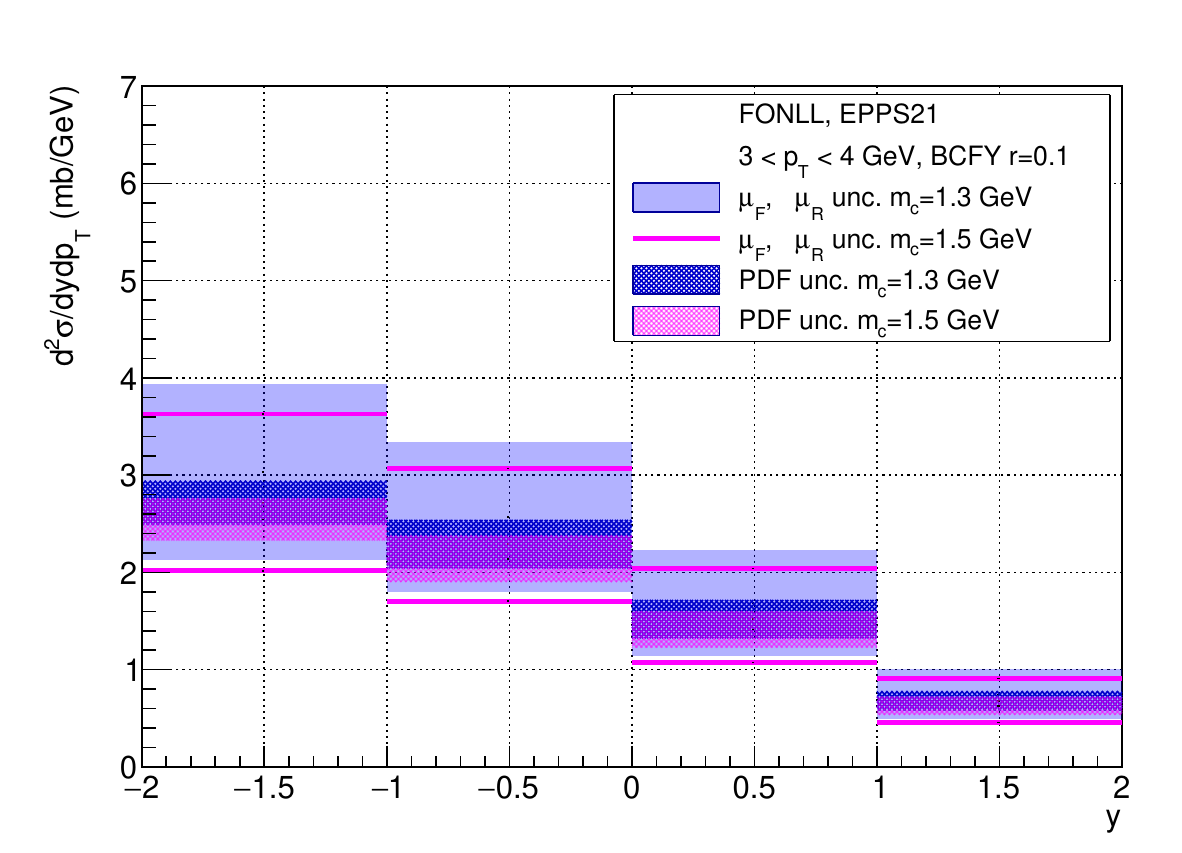}
    \label{fig:ydpt1mc}
\end{subfigure} 
\begin{subfigure}{0.49\textwidth}
    \centering
    \includegraphics[width=\textwidth]{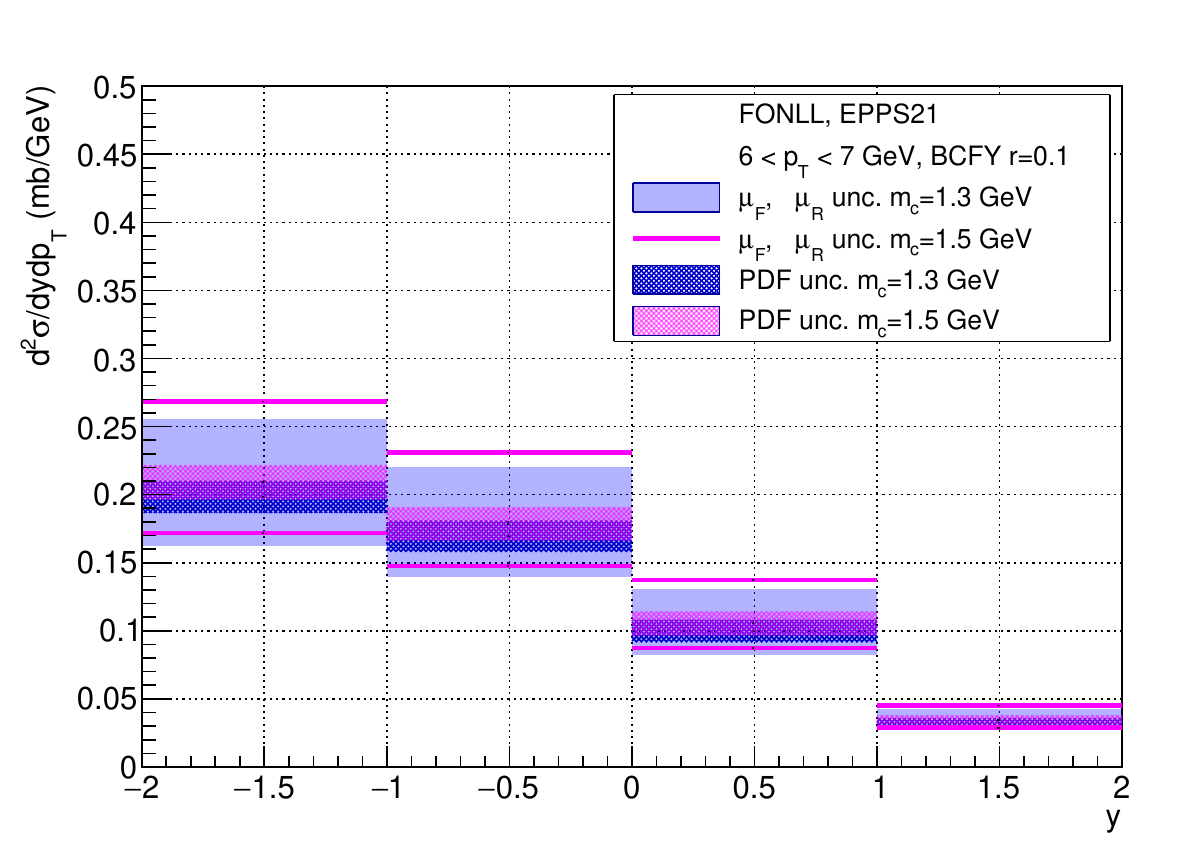}
    \label{fig:ydpt2mc}
\end{subfigure}
\hfill
\begin{subfigure}{0.49\textwidth}
    \centering
    \includegraphics[width=\textwidth]{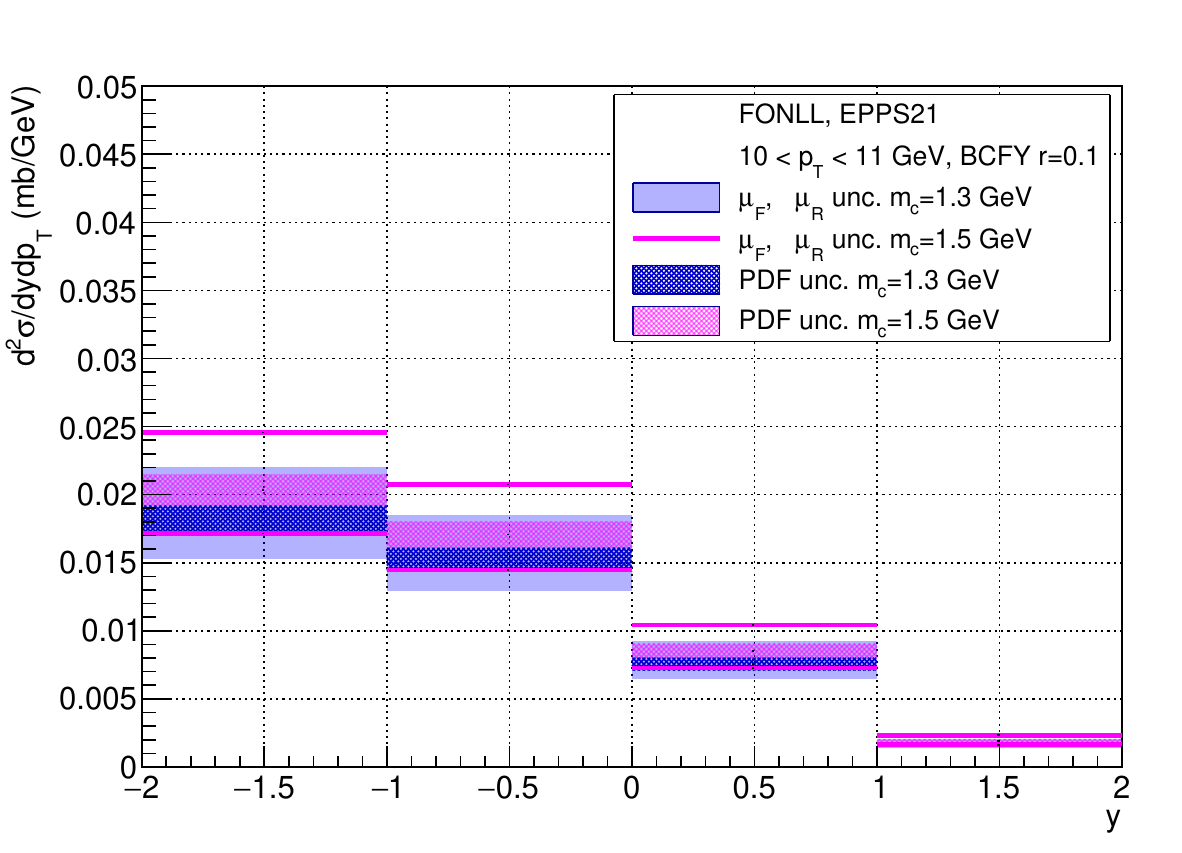}
    \label{fig:ydpt3mc}
\end{subfigure}
 \caption{Rapidity distribution  for the $D^0$  production in UPC PbPb collisions at $\sqrt{\rm s_{\scriptscriptstyle{NN}}}=5.36$ TeV with EPPS21 nuclear PDF.  Four panels correspond to $p_{T}$ bins $(0-1), (3-4), (6-7), (10-11)$ GeV. Light blue band: FONLL  calculation with factorization and renormalization scale variation for $m_c=1.3 \, \rm GeV$, magenta lines band: FONLL  calculation with factorization and renormalization scale variation and $m_c=1.5 \, \rm GeV$, dark blue band: FONLL EPPS21 \cite{Eskola:2021nhw} PDF uncertainty, $m_c=1.3 \, \rm GeV$. Magenta  band: FONLL EPPS21 \cite{Eskola:2021nhw} PDF uncertainty, $m_c=1.5 \, \rm GeV$.  BCFY fragmentation function \cite{Braaten:1994bz,Cacciari:2003zu} with parameter $r=0.1$. Photon-emitting nucleus is moving in the positive rapidity direction.}
 \label{fig:ydpt_fine_epps21bcfy_mc}
\end{figure}

\begin{figure}[H]
\centering
\begin{subfigure}{0.49\textwidth}
    \centering
    \includegraphics[width=\textwidth]{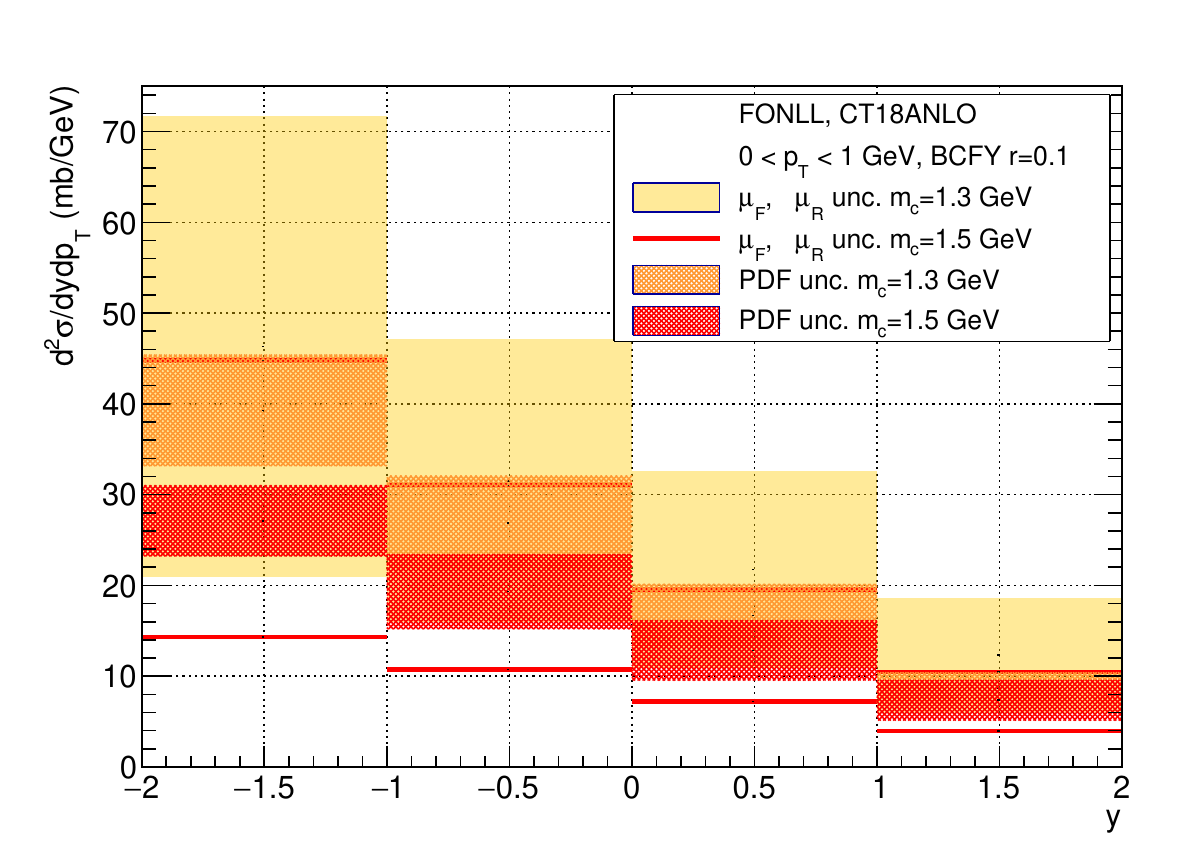}
    \label{fig:ydpt0protonmc}
\end{subfigure}
\hfill
\begin{subfigure}{0.49\textwidth}
    \centering
    \includegraphics[width=\textwidth]{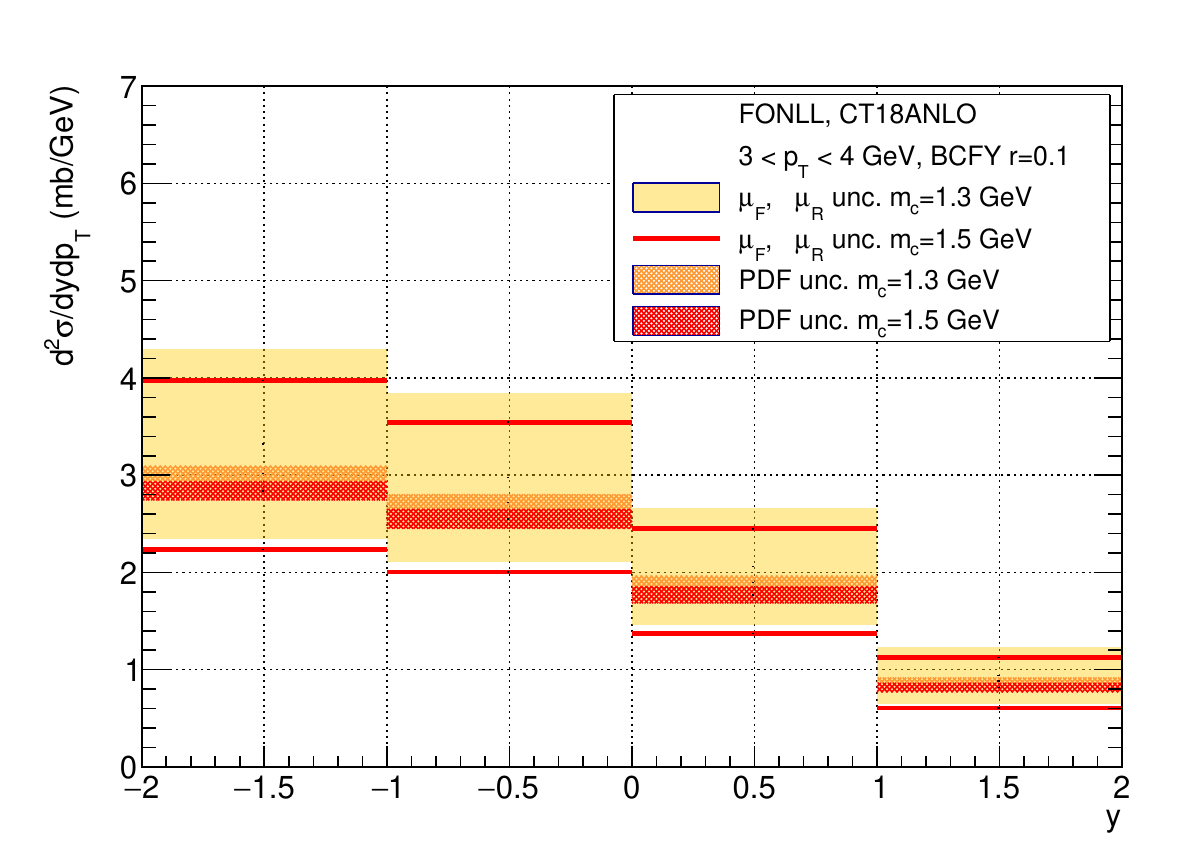}
    \label{fig:ydpt1protonmc}
\end{subfigure} 
\begin{subfigure}{0.49\textwidth}
    \centering
    \includegraphics[width=\textwidth]{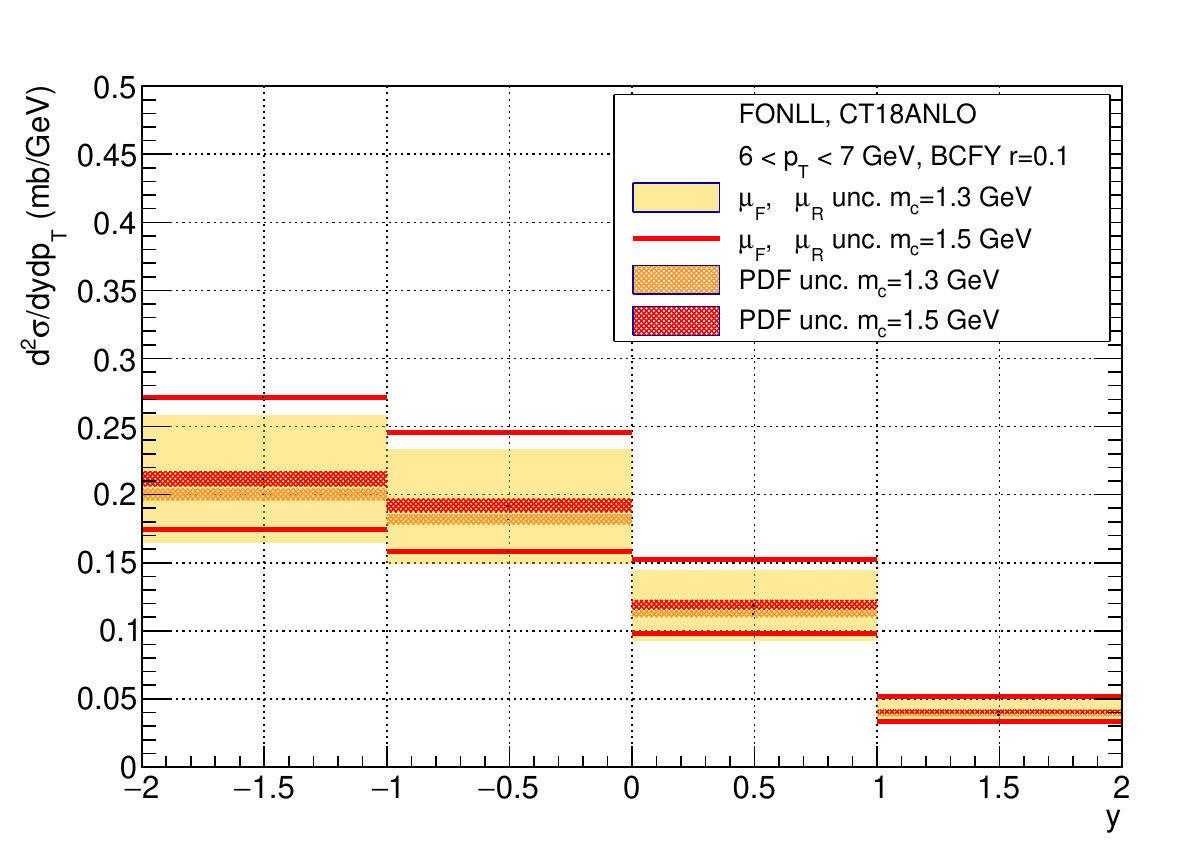}
    \label{fig:ydpt2protonmc}
\end{subfigure}
\hfill
\begin{subfigure}{0.49\textwidth}
    \centering
    \includegraphics[width=\textwidth]{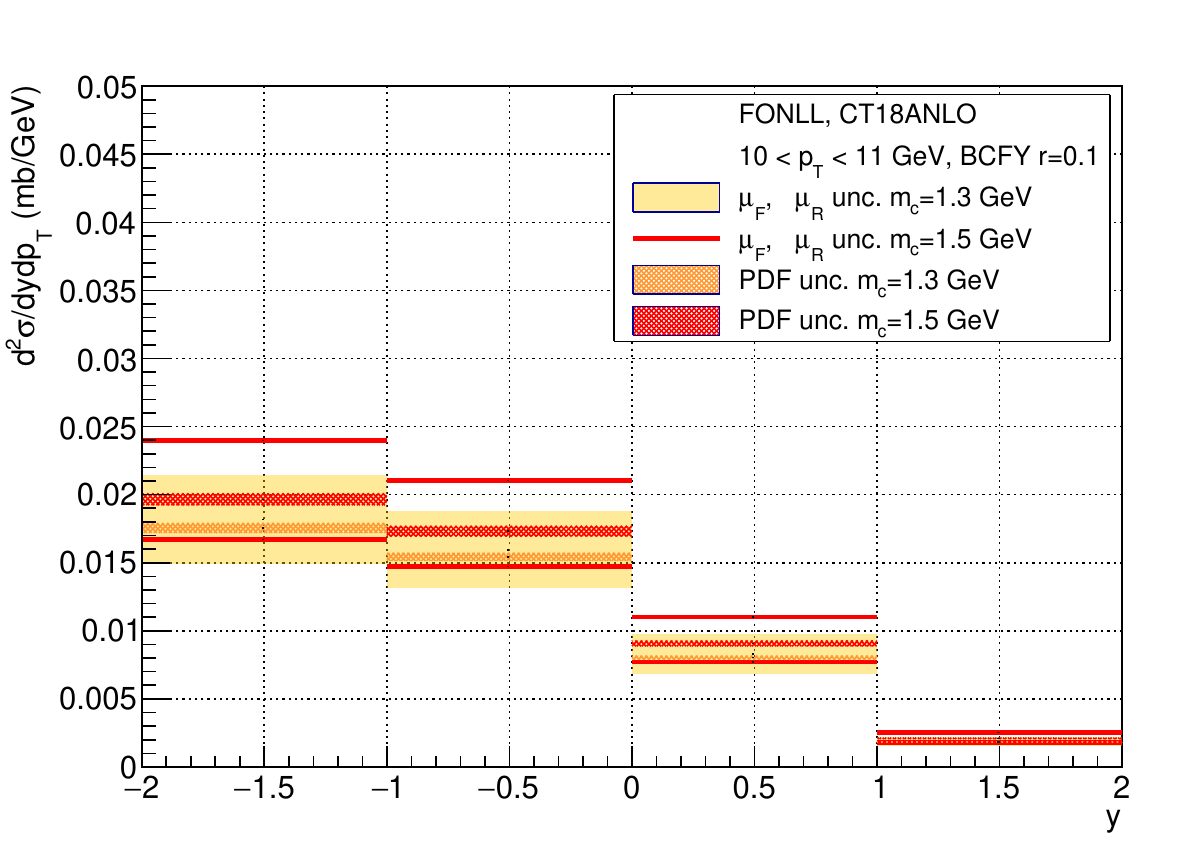}
    \label{fig:ydpt3protonmc}
\end{subfigure}
 \caption{Rapidity distribution  for the $D^0$  production in UPC PbPb collisions at $\sqrt{\rm s_{\scriptscriptstyle{NN}}}=5.36$ TeV with CT18ANLO proton PDF.  Four panels correspond to $p_{T}$ bins $(0-1), (3-4), (6-7), (10-11)$ GeV. Yellow band: FONLL  calculation with factorization and renormalization scale variation for $m_c=1.3 \, \rm GeV$, red lines band: FONLL  calculation with factorization and renormalization scale variation and $m_c=1.5 \, \rm GeV$, orange band: FONLL CT18ANLO \cite{Eskola:2021nhw} PDF uncertainty, $m_c=1.3 \, \rm GeV$. Red  band: FONLL CT18ANLO \cite{Eskola:2021nhw} PDF uncertainty, $m_c=1.5 \, \rm GeV$.  BCFY fragmentation function \cite{Braaten:1994bz,Cacciari:2003zu} with parameter $r=0.1$. Photon-emitting nucleus is moving in the positive rapidity direction.}
 \label{fig:ydpt_fine_ct18anlobcfy_mc}
\end{figure}

\begin{figure}[H]
\centering
\begin{subfigure}{0.49\textwidth}
    \centering
    \includegraphics[width=\textwidth]{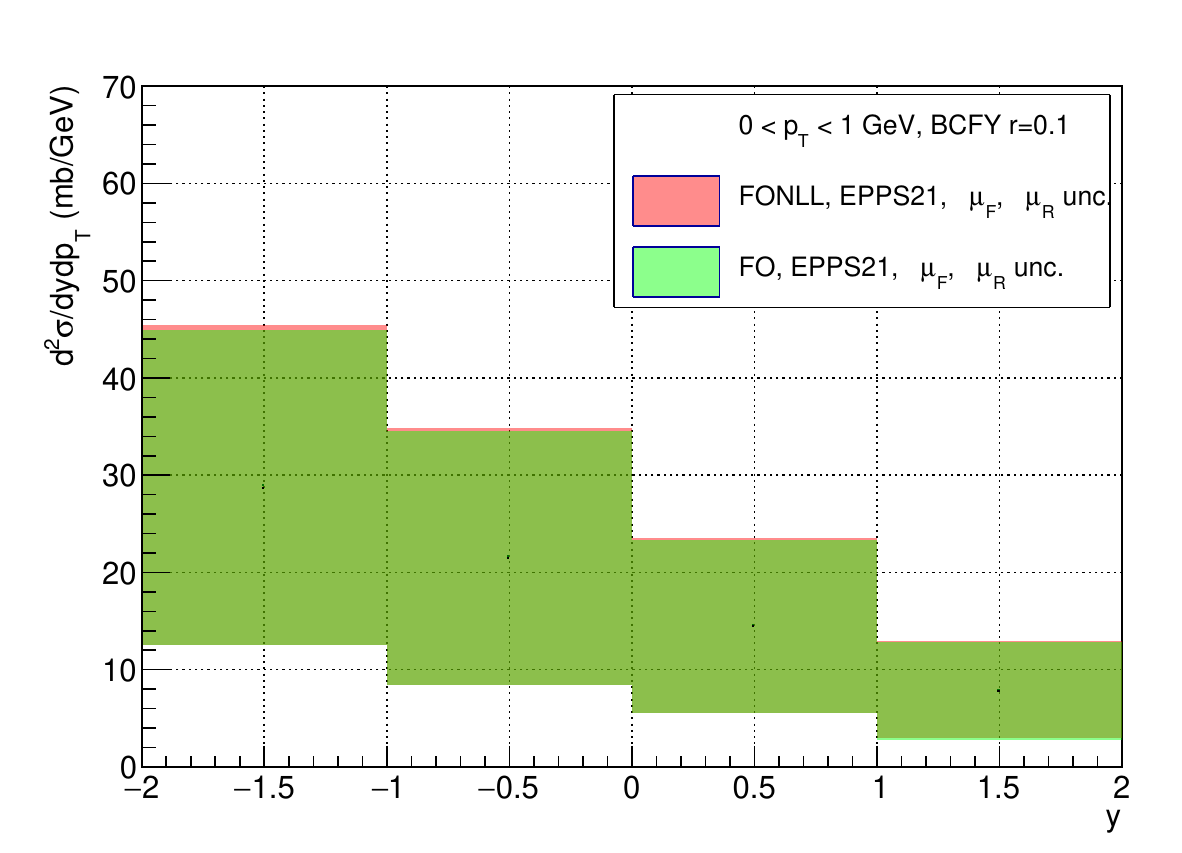}
    \label{fig:ydpt0_afg}
\end{subfigure}
\hfill
\begin{subfigure}{0.49\textwidth}
    \centering
    \includegraphics[width=\textwidth]{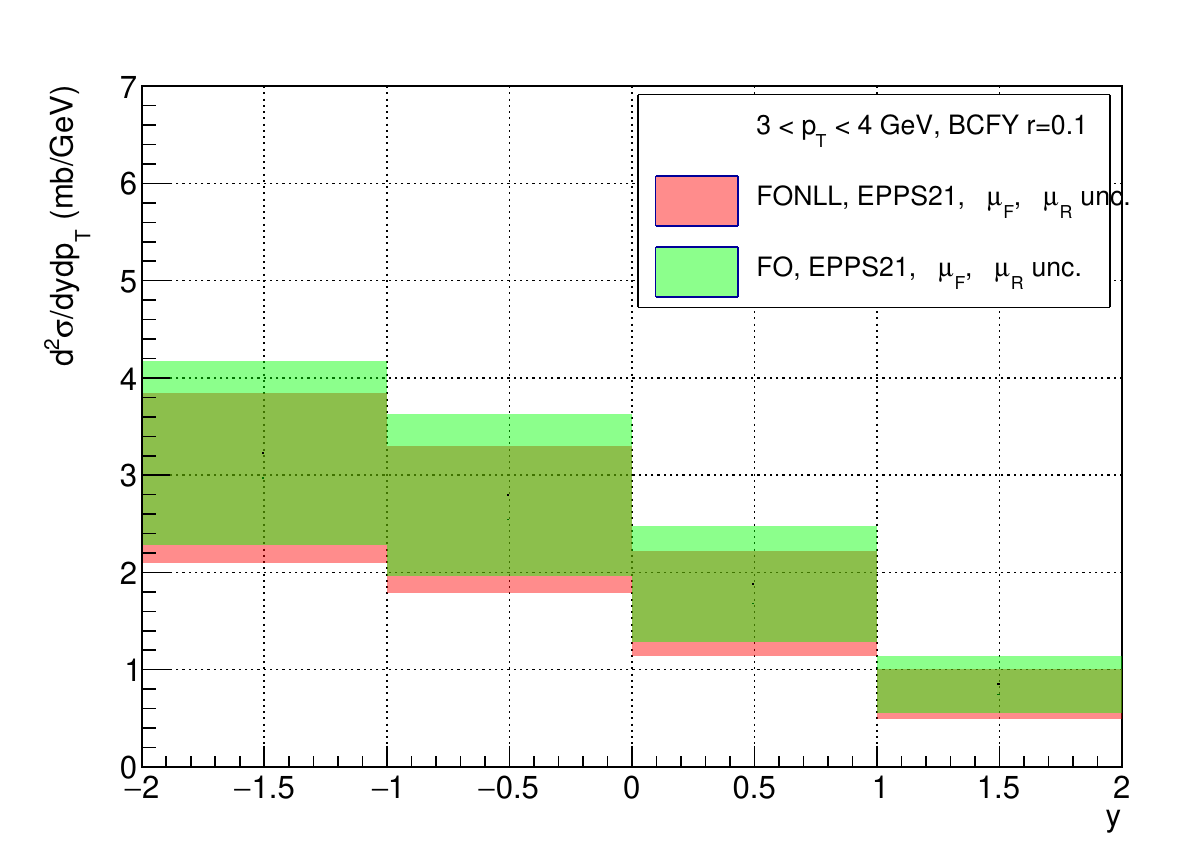}
    \label{fig:ydpt1_afg}
\end{subfigure} 
\begin{subfigure}{0.49\textwidth}
    \centering
    \includegraphics[width=\textwidth]{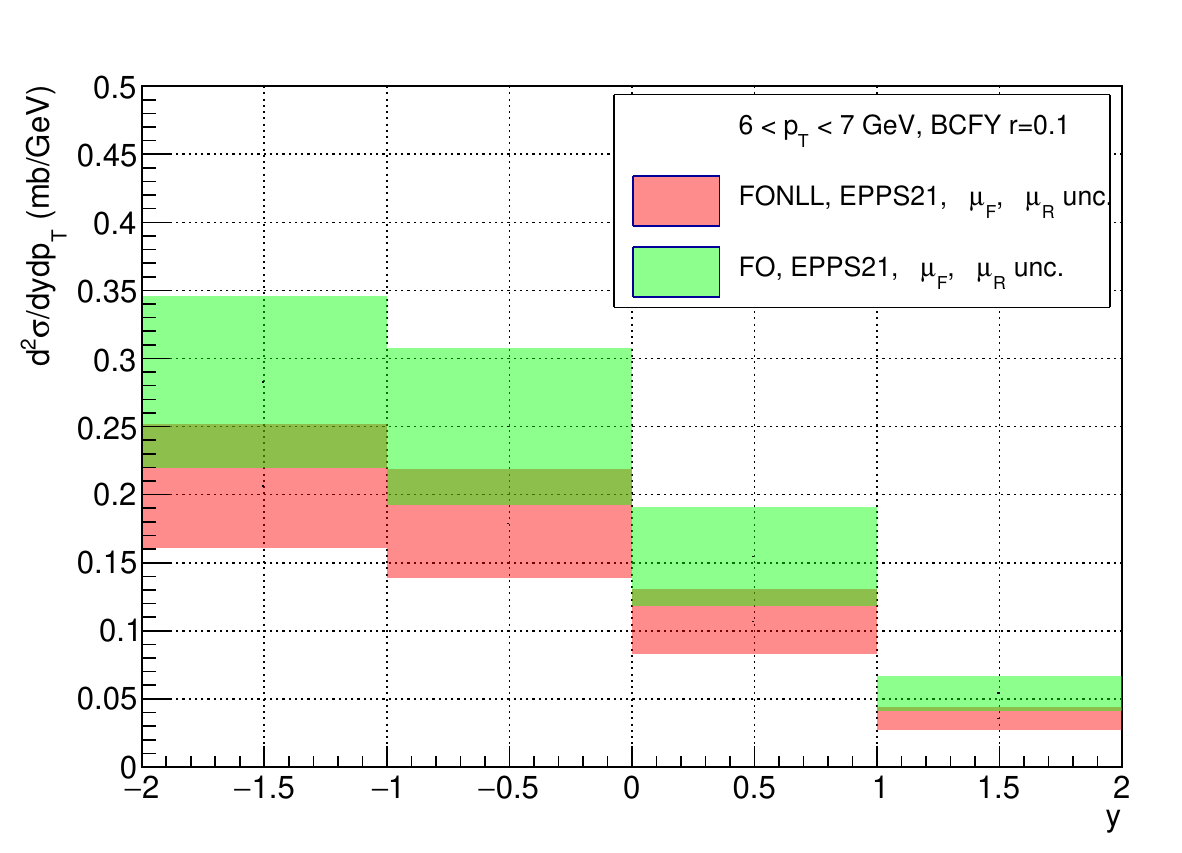}
    \label{fig:ydpt2_afg}
\end{subfigure}
\hfill
\begin{subfigure}{0.49\textwidth}
    \centering
    \includegraphics[width=\textwidth]{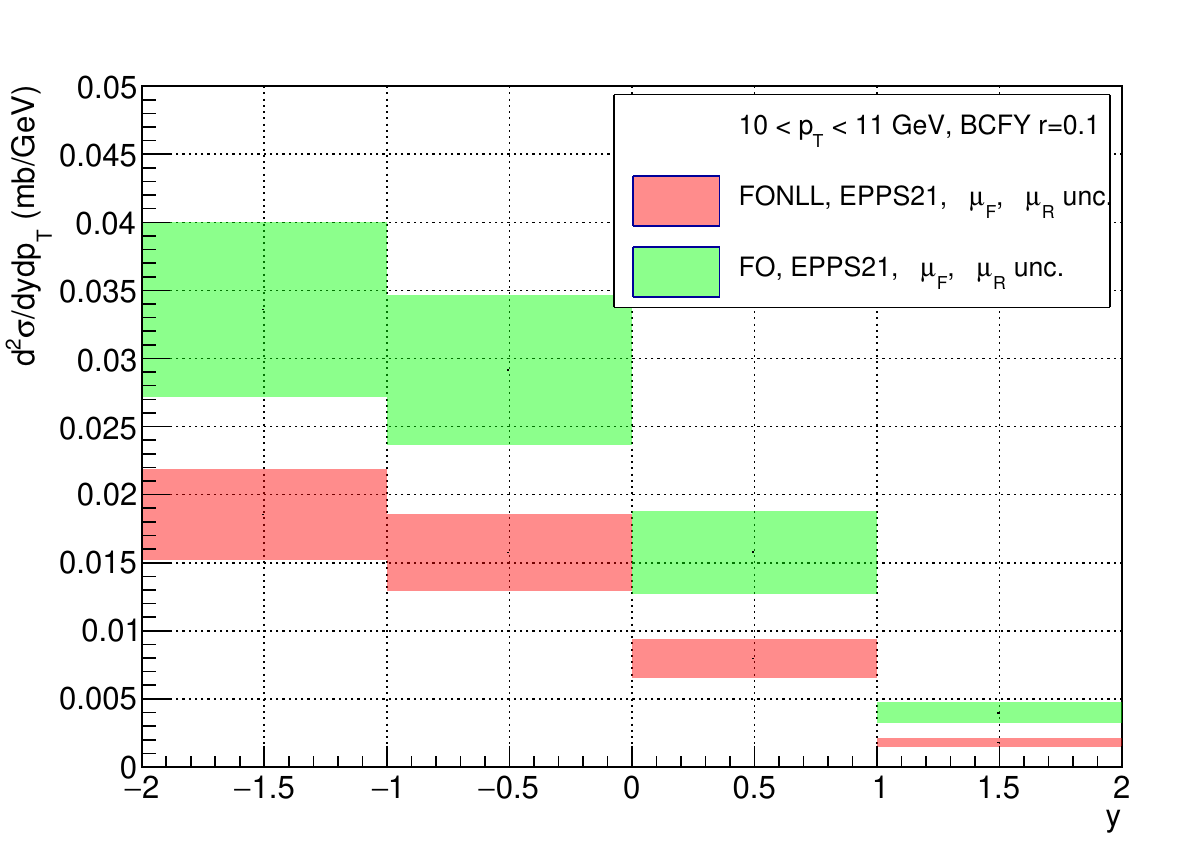}
    \label{fig:ydpt3_afg}
\end{subfigure}
 \caption{Rapidity distribution  for the $D^0$  production in UPC PbPb collisions at $\sqrt{\rm s_{\scriptscriptstyle{NN}}}=5.36
 $ TeV. Calculations with AFG photon PDF. Four panels correspond to $p_{T}$ bins $(0-1), (3-4), (6-7), (10-11)$ GeV. Red band: FONLL  calculation with factorization and renormalization scale variation, green band: FO  calculation with factorization and renormalization scale variation.  FONLL and FO calculation done with BCFY fragmentation function with parameter $r=0.1$. Photon-emitting nucleus is moving in the positive rapidity direction.}
 \label{fig:ydpt_fine_epps21afg}
\end{figure}

\begin{figure}[H]
\centering
\begin{subfigure}{0.49\textwidth}
    \centering
    \includegraphics[width=\textwidth]{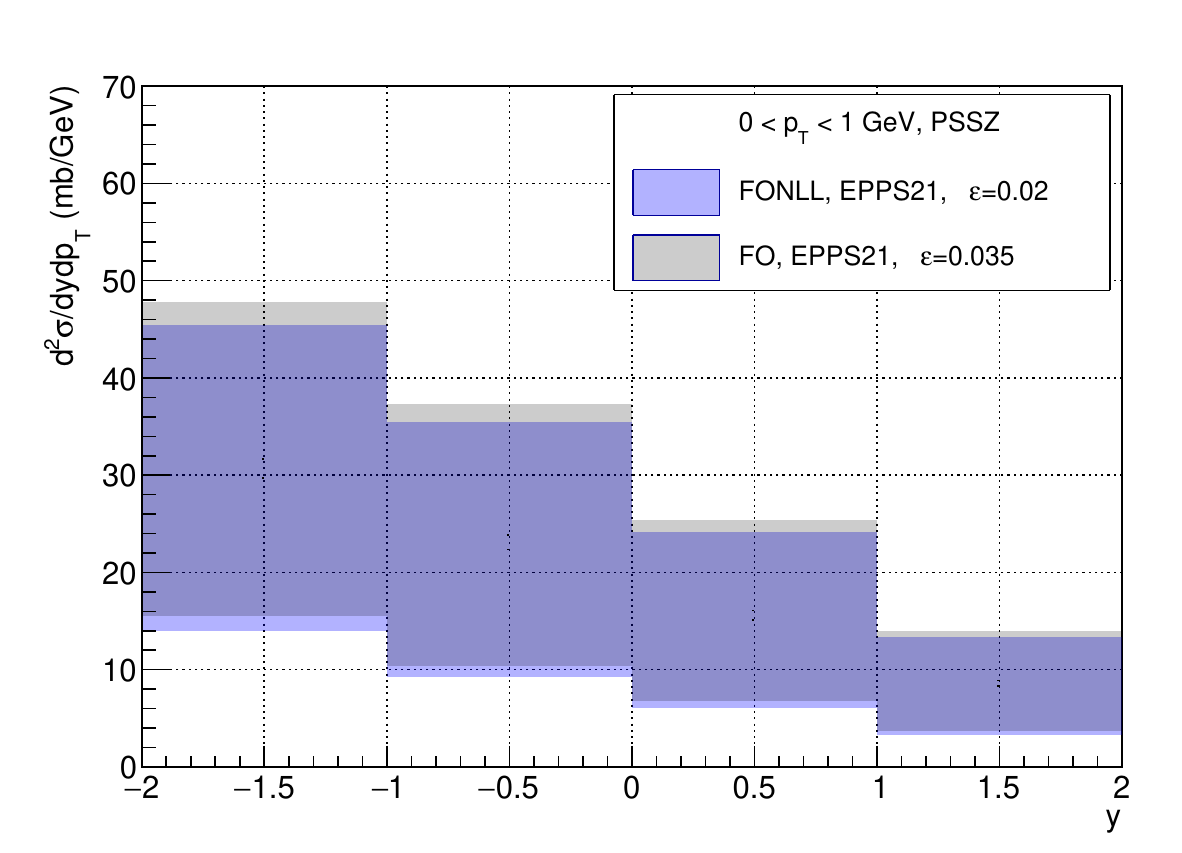}
    \label{fig:ydpt0finepet}
\end{subfigure}
\hfill
\begin{subfigure}{0.49\textwidth}
    \centering
    \includegraphics[width=\textwidth]{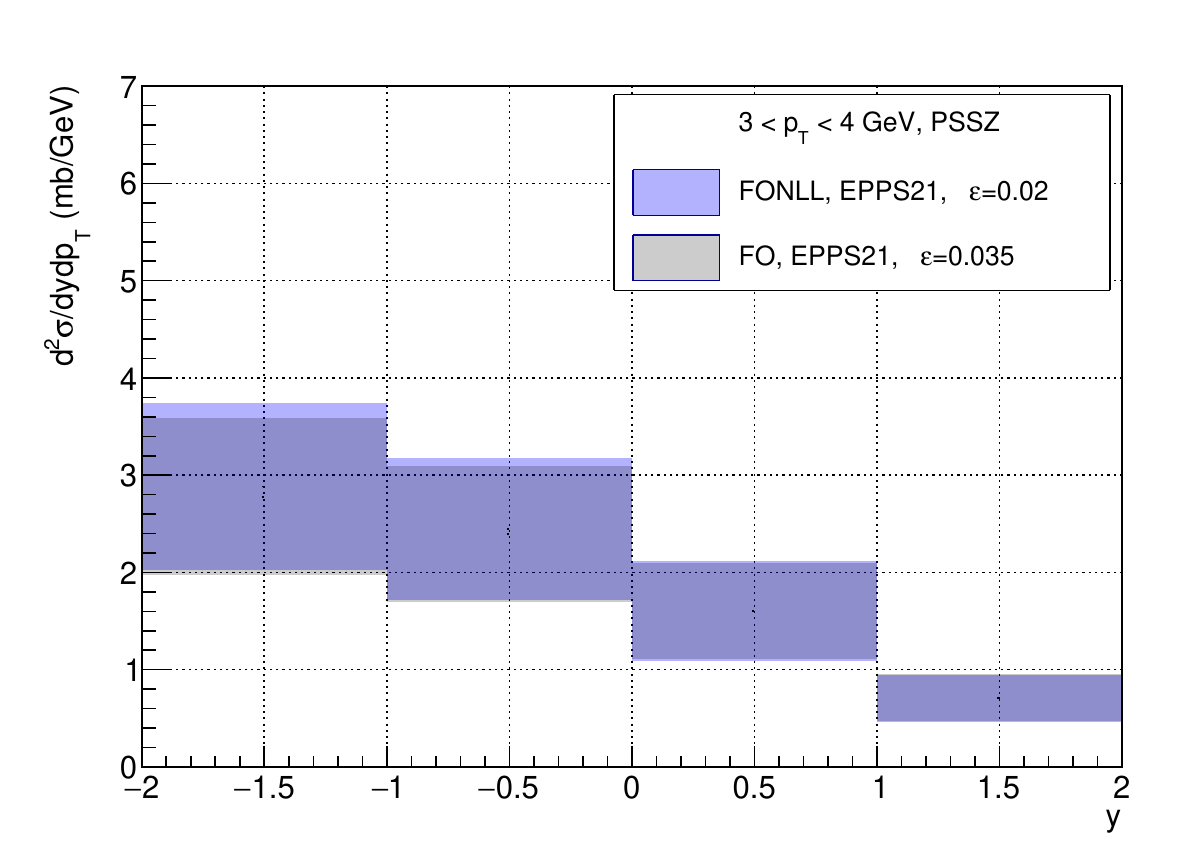}
    \label{fig:ydpt1finepet}
\end{subfigure} 
\begin{subfigure}{0.49\textwidth}
    \centering
    \includegraphics[width=\textwidth]{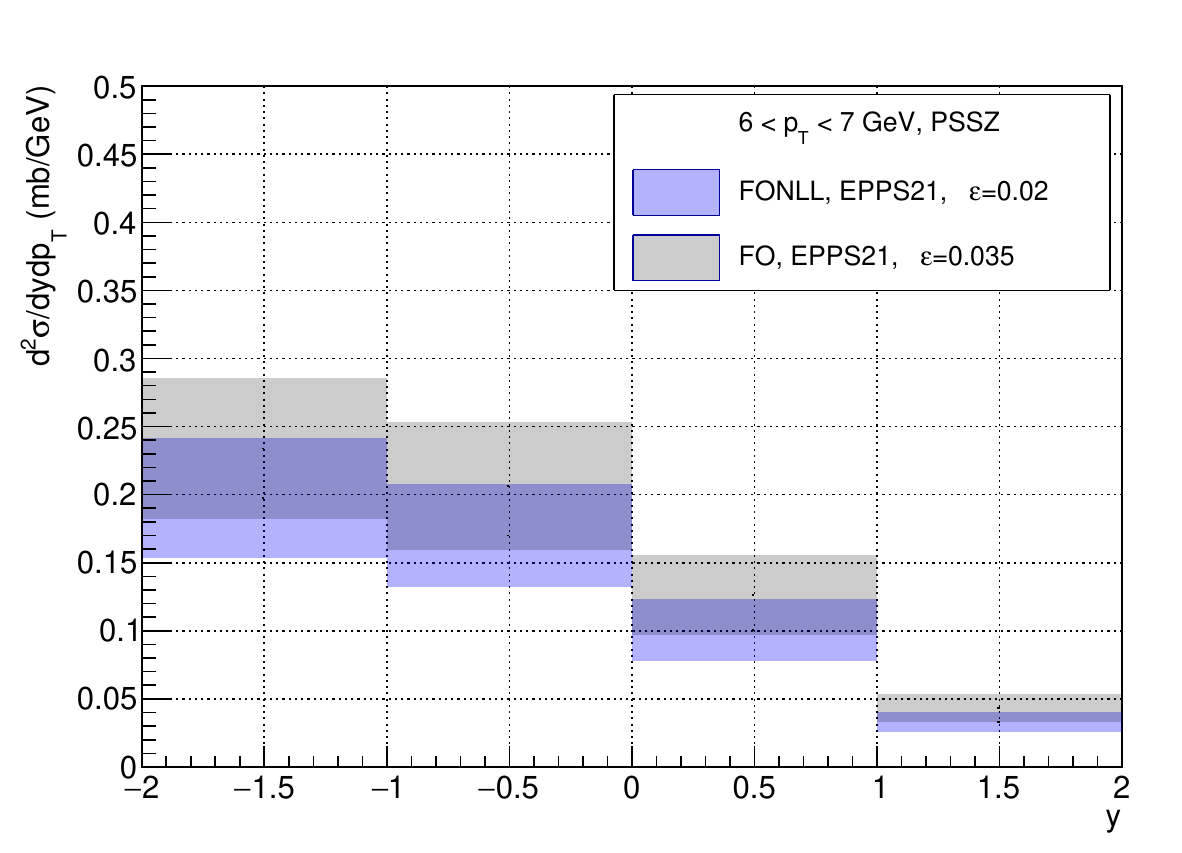}
    \label{fig:ydpt2finepet}
\end{subfigure}
\hfill
\begin{subfigure}{0.49\textwidth}
    \centering
    \includegraphics[width=\textwidth]{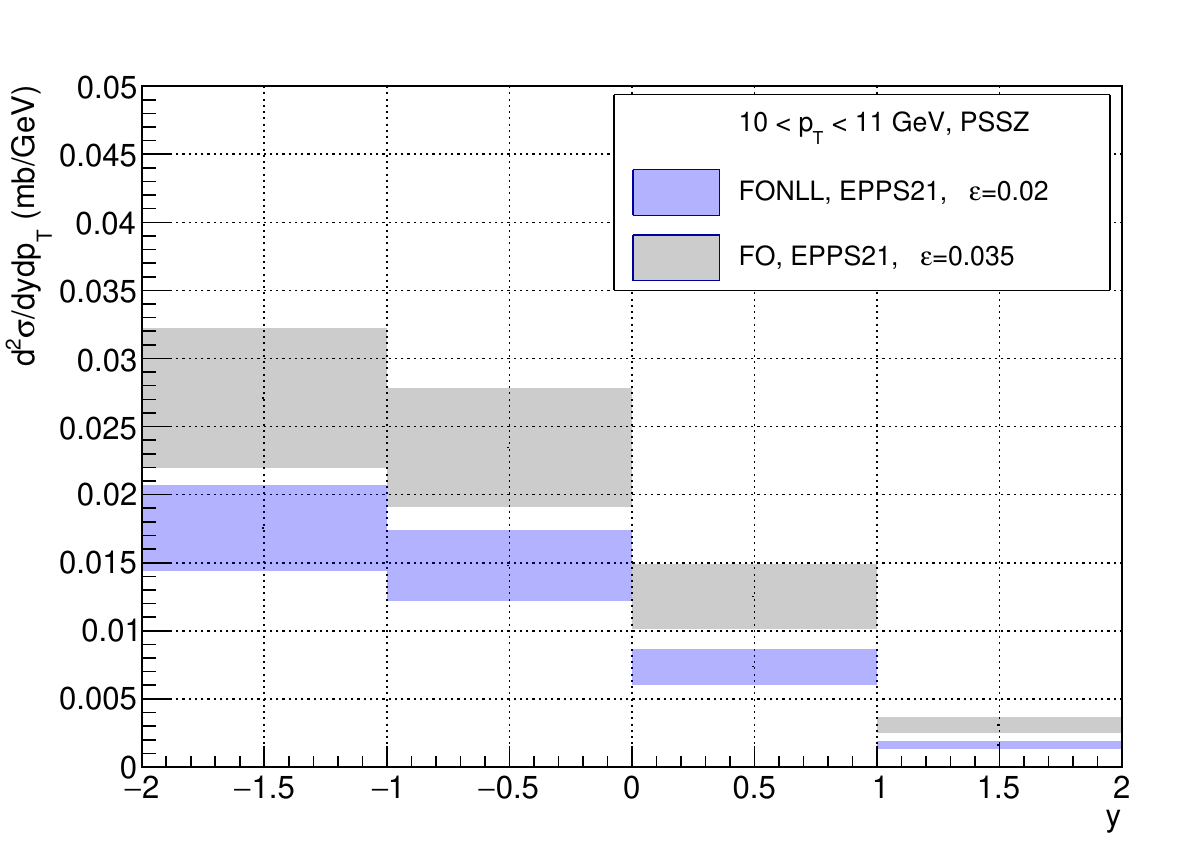}
    \label{fig:ydpt3finepet}
\end{subfigure}
 \caption{Rapidity distribution  for the $D^0$  production in UPC PbPb collisions at $\sqrt{\rm s_{\scriptscriptstyle{NN}}}=5.36$ TeV. Four panels correspond to $p_{T}$ bins $(0-1), (3-4), (6-7), (10-11)$ GeV. Light blue band: FONLL  calculation with factorization and renormalization scale variation, grey band: FO  calculation with factorization and renormalization scale variation.  FONLL and FO calculation done with PSSZ fragmentation function with parameter $\varepsilon=0.02$ and $0.035$ respectively. Photon-emitting nucleus is moving in the positive rapidity direction.}
 \label{fig:ydpt_fine_epps21pet}
\end{figure}

Having validated the FONLL calculations against the $D^*$ photoproduction data at HERA, we now address the calculations of the $D^0$ inclusive production in ultraperipheral collisions at the LHC. Since the nucleus, which is the source of the photons, is not detected, one should in principle integrate the flux  over the entire range of photon energies. In practice, we assume that the fractional energy of the photon is limited by 
\begin{equation}
    z \in [z_{\rm min},z_{\rm max}] \; .
\end{equation}
We take $z_{\rm max} = 0.1$ based on the fact that the photon flux from the nucleus is negligible beyond there, as evident from  Fig.~\ref{fig:fluxes}. The minimal value $z_{\rm min}$ is given by
\begin{equation}
z_{\rm min} = \frac{M_{c\bar{c}}^2}{s} \; , 
\end{equation}
where the invariant mass of the charm pair is $M_{c\bar{c}}^2=2 m_T^2(1+\cosh(y_1-y_2))$. In practice we take the lower limit $z_{\rm min}=\frac{4m_c^2}{s}$ (we checked that moving the cutoff to $z_{\rm min}=10^{-5}$ does not affect the calculation within the rapidity range of $[-2,2]$). \\[4pt] 
\noindent
As discussed in Subsec.~\ref{sec:nPDF}, for the nuclear PDF we used the EPPS21 \cite{Eskola:2021nhw} set as well as the   nNNPDF3.0 \cite{AbdulKhalek:2022fyi} set. In addition, we also perform the calculations with the EPPS21 proton baseline PDF CT18ANLO \cite{Hou:2019efy}. This allows  to quantify the size of the nuclear effects in the kinematics for $D^0$ production at CMS. \\[4pt]
For the fragmentation, we adopt the BCFY parametrization, as described in Eq.\eqref{eq:bcfyadd}, with $r=0.1$ \cite{Braaten:1994bz}. For comparison we also used  PSSZ function with $\varepsilon=0.2$ and $\varepsilon=0.035$ \cite{Peterson:1982ak}. Finally, the results have been corrected by the EMD factor, described in Sec.~\ref{sec:emd} and shown in the right plot of Fig.~\ref{fig:ratiofluxEMD},  to account for the survival probability of the nucleus due to the photon interactions.
\\[4pt] 
\noindent
We first present the results of theoretical calculations in the finer bins in $\pt$, studying the sensitivity due to the variation of different parameters. In the following section we show the comparison of calculations in the bins used by CMS in their measurement \cite{CMS:2025jjx}.
\\[4pt] 
\noindent
In Fig.~\ref{fig:ydpt_fine_epps21bcfy} distributions in rapidity for $D^0$ production are shown as a function of $\pt$ and rapidity in four bins of $\pt$ : (0-1), (3-4), (6-7), (10-11) GeV and rapidity range $-2<y<2$. 
The convention in these plots is such that the positive rapidity is the photon going direction, thus it is reverse with respect to HERA.
FONLL calculation is shown (blue band) for the BCFY choice of the fragmentation function with parameter $r=0.1$, and EPPS21 nPDF is used. The bands correspond to the variation of the factorization $\mu_F/\mu_0=0.5,1.0,2.0$ and renormalization scales $\mu_R/\mu_0=0.5,1.0,2.0$, (with constraint $1/2 \le \mu_F/\mu_R\le 2$) where $\mu_0=\sqrt{p_T^2+m_c^2}$ and charm mass was set to $m_c=1.3$ GeV (consistent with the charm mass value used in the EPPS21 PDF set). The inner, dark blue band in FONLL calculation is the EPPS21 PDF uncertainty. We see that, for moderate and high $p_T$, the PDF uncertainty is much smaller than the variation of the renormalization and factorization scales. It is in the  lowest $\pt$ bin, that the nPDF uncertainty is quite large and comparable to the scale uncertainty. It is understood, since the PDF uncertainty will also directly influence the magnitude of the scale uncertainty.  For comparison we also show FO calculation (grey bands) in Fig.~\ref{fig:ydpt_fine_epps21bcfy}. As  observed for the HERA case, the FO is higher than FONLL at large $p_T$. 
\\[4pt] 
\noindent
In Fig.~\ref{fig:ydpt_fine_epps21bcfy_mc} the  FONLL results are plotted, but using  two different mass choices for the charm quark $m_c=1.3$ and $m_c=1.5$ GeV for the EPPS21 nuclear PDF. Similar to the case of the HERA data, the higher value of the charm mass leads to slightly lower results for low $p_T$ but higher values for highest $p_T$ bins. The high $p_T$ trend is related to the FONLL resummation. The scale and PDF uncertainties are slightly smaller for $m_c=1.5$ GeV.
\\[4pt] 
\noindent
In Fig.~\ref{fig:ydpt_fine_ct18anlobcfy_mc} a similar study on the charm mass dependence is shown, but the nuclear PDF is taken to be equal to A (mass number) times the proton PDF. For the proton PDF CT18ANLO is selected. Therefore this calculation is without any effects of nuclear modification. We observe that the calculations are comparable to the ones with nuclear PDFs for two higher $p_T$ bins. For the lowest $p_T$ bin the  theoretical calculations are slightly higher than for EPPS21 case, as expected due to the increased role of the shadowing at small $x$ and $p_T$. More precisely, at the largest values of rapidity, which corresponds to smallest values of $x$, the nuclear shadowing  leads to a reduction of cross section at the level of  about 30-40\% at $\pt<2$ GeV, $15-20\%$ at $3<\pt<7$ GeV, and less than 10\% for $\pt>9$ GeV. 
\\[4pt] 
\noindent
In Fig.~\ref{fig:ydpt_fine_epps21afg} the same distributions in rapidity are shown for FONLL and FO but for the case of the AFG photon PDF \cite{Aurenche:1994in}, with just variation in renormalization and factorization scale, using nuclear EPPS21 PDF set. Comparing with Fig.~\ref{fig:ydpt_fine_epps21bcfy} we see that the differences between GRV and AFG photon PDFs are completely negligible in the kinematics studied. 
\\[4pt] 
\noindent
In Fig.~\ref{fig:ydpt_fine_epps21pet} the same distributions in rapidity are shown but for different choices of the fragmentation function (nuclear PDF is still EPPS21). Here we show the FONLL with PSSZ function and parameter $\varepsilon=0.02$ and FO calculation with PSSZ function and parameter $\varepsilon=0.035$. This is the same setup as presented in Fig.~\ref{fig:heradata} for the HERA data. We see that the FONLL and FO are somewhat closer for the highest $p_T$ bin, than in the case of the BCFY fragmentation function. This is understandable, since the fragmentation with $\varepsilon=0.035$ leads to a softer spectrum than with $\varepsilon=0.02$. 
Overall, the differences between the two fragmentation schemes BCFY and PSSZ are very small in the studied range of $p_T$ and $y$, with BCFY fragmentation function resulting in slightly higher cross section at large values of $p_T$ as was found for ep case in Sec.~\ref{sec:hera}. 
\\[4pt] 
\noindent
For the sake of completeness, and despite the two separate contributions being unphysical (e.g. \cite{Cacciari:2001td} and \cite{Kramer:2003jw}) and dependent on what theoretical approach and parameters are used, we have also explored the relative sizes of the direct and resolved cross sections. Within the approach that we have chosen (FONLL, with a massive NLO calculation matched to a massless NLL-resummed one, with the introduction of a damping function $G(m,p_T)$ (see eq.~(\ref{eq:fonllmatching}), with its free parameter $c$ set to 5, for suppressing spurious higher orders terms), we observe that the direct contribution tends to dominate at positive rapidities and medium transverse momentum values ($p_T < 10$~GeV), whereas the resolved contribution can reach up to about 50\% in two kinematical regions, namely a small region of negative rapidities ($y < 1$) and small transverse momentum ($p_T < 1$~GeV), and a broader region of transverse momentum larger than about 10~GeV.
\\[4pt] 
\noindent
Appendix B presents supplementary plots of the calculations versus transverse momentum, binned in rapidity, for the EPPS21, nNNPDF3.0, and CT18ANLO PDF sets.
\\[4pt]

\clearpage
\section{Comparison with CMS  measurement}
\label{sec:cms}

In this section, the G$\gamma$A-FONLL predictions are compared to the first measurement of photonuclear D$^0$ meson production in ultraperipheral heavy-ion collisions, recently performed by the CMS Collaboration at the LHC~\cite{CMS:2025jjx}. The measurement uses 1.38~$\rm nb^{-1}$ of lead–lead data collected at $\rm \sqrt{s_{\scriptscriptstyle NN}} = 5.36$~TeV. Photonuclear events were selected by requiring that at least one of the two colliding nuclei breaks (0nXn), using the Zero Degree Calorimeters, and by imposing the presence of a large rapidity gap in the direction of the photon-emitting nucleus.
The measurement was performed as a function of the transverse momentum and rapidity of the $D^0$ meson, in the ranges $2 < p_T < 12$~GeV and $-2 < y < 2$. Specifically, the $2 < p_T < 5$~GeV interval was measured in one rapidity bin, $|y| < 1$, while the $5 < p_T < 8$~GeV and $8 < p_T < 12$~GeV intervals were measured in four equal-size rapidity bins spanning $-2 < y < 2$. \\[4pt] 

\begin{figure}[H]
\centering
\begin{subfigure}{0.49\textwidth}
    \centering
    \includegraphics[width=\textwidth]{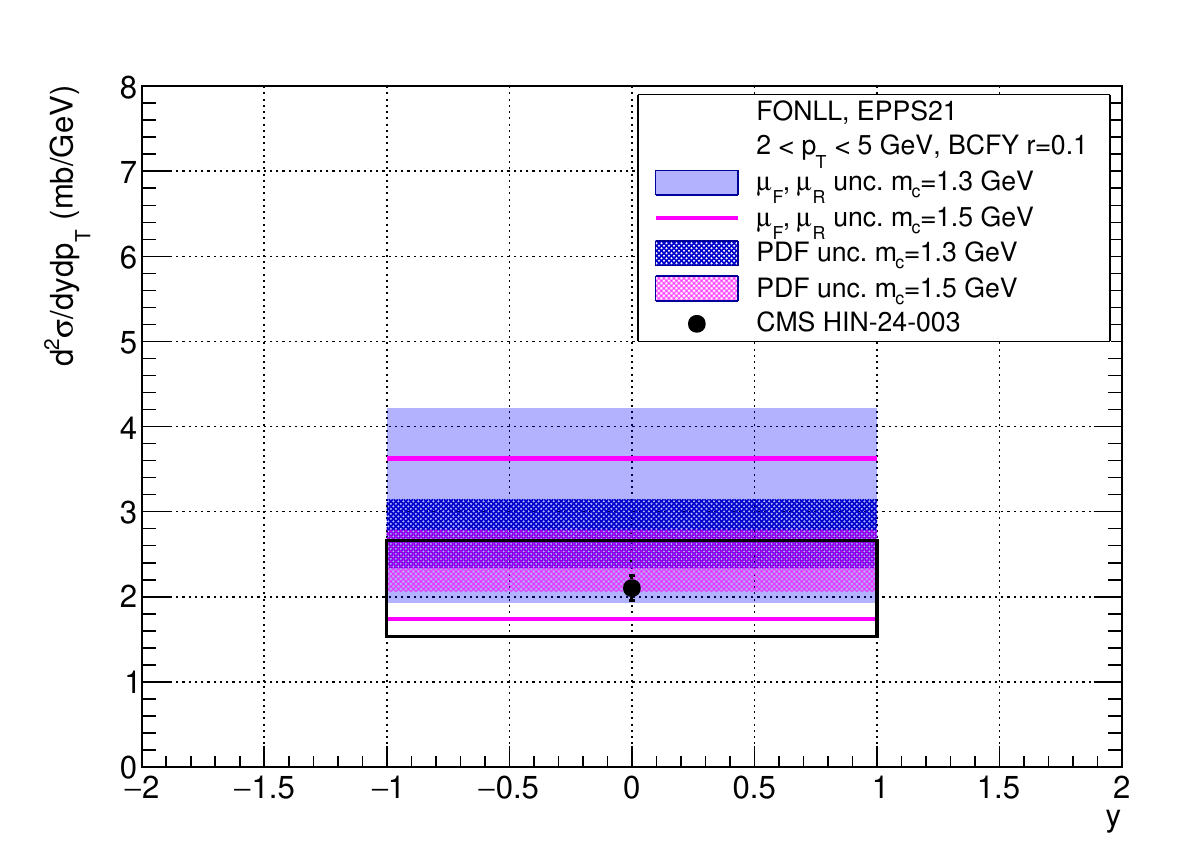}
    \label{fig:ydpt0mc_cms}
\end{subfigure}
\begin{subfigure}{0.49\textwidth}
    \centering
    \includegraphics[width=\textwidth]{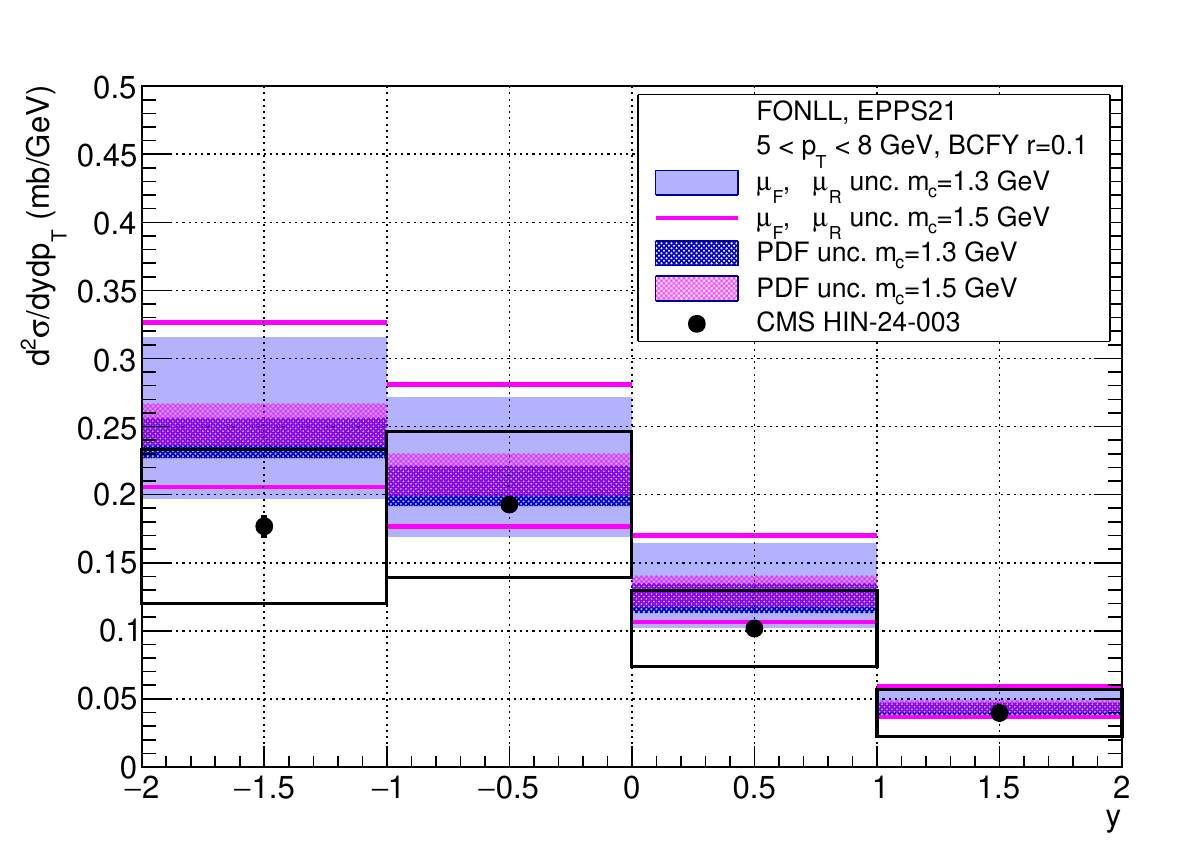}
    \label{fig:ydpt2mc_cms}
\end{subfigure}
\hfill
\begin{subfigure}{0.49\textwidth}
    \centering
    \includegraphics[width=\textwidth]{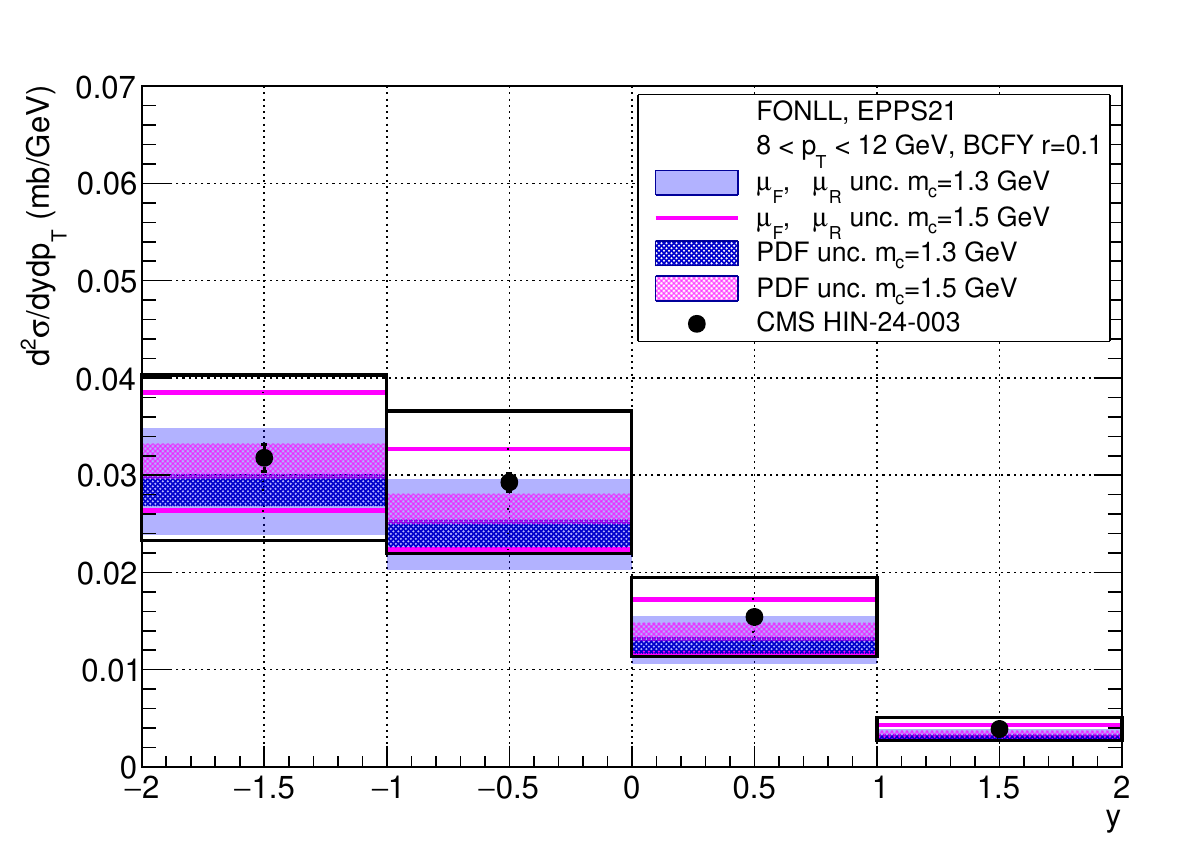}
    \label{fig:ydpt3mc_cms}
\end{subfigure}
 \caption{Rapidity distribution  for the $D^0$  production in UPC PbPb collisions at ${\rm\sqrt{s_{\scriptscriptstyle NN}} }=5.36$ TeV with EPPS21 nuclear PDF in  $p_{T}$ bins $(2-5), (5-8), (8-12)$ GeV. Light blue band: FONLL  calculation with factorization and renormalization scale variation for $m_c=1.3 \, \rm GeV$, magenta lines band: FONLL  calculation with factorization and renormalization scale variation and $m_c=1.5 \, \rm GeV$, dark blue band: FONLL EPPS21 \cite{Eskola:2021nhw} PDF uncertainty, $m_c=1.3 \, \rm GeV$. Magenta  band: FONLL EPPS21 \cite{Eskola:2021nhw} PDF uncertainty, $m_c=1.5 \, \rm GeV$.  BCFY fragmentation function \cite{Braaten:1994bz,Cacciari:2003zu} with parameter $r=0.1$. Data are from CMS \cite{CMS:2025jjx}.}
 \label{fig:ydpt_epps21bcfy_mc_cms}
\end{figure}
\begin{figure}
\centering
\begin{subfigure}{0.49\textwidth}
    \centering    \includegraphics[width=\textwidth]{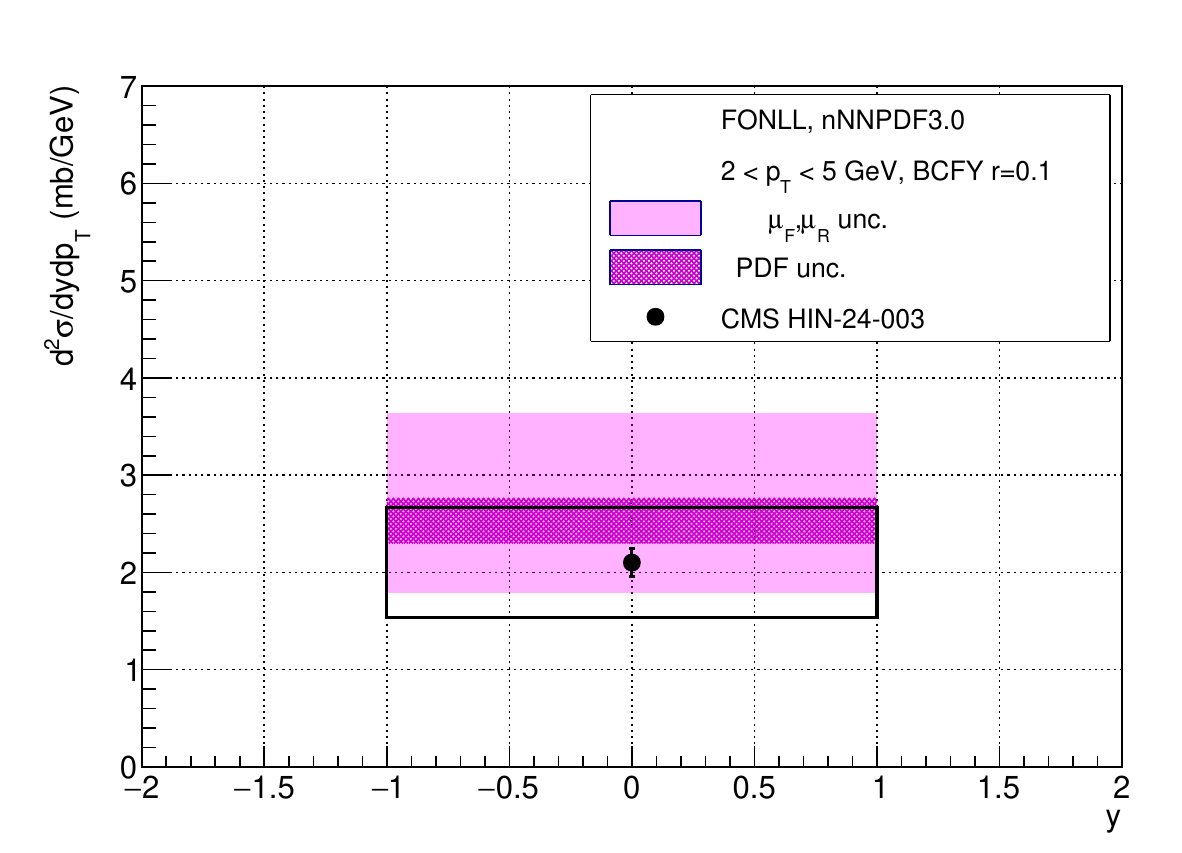}
    \label{fig:ydpt0nnpdfmc_cms}
\end{subfigure}
\hfill
\begin{subfigure}{0.49\textwidth}
    \centering
    \includegraphics[width=\textwidth]{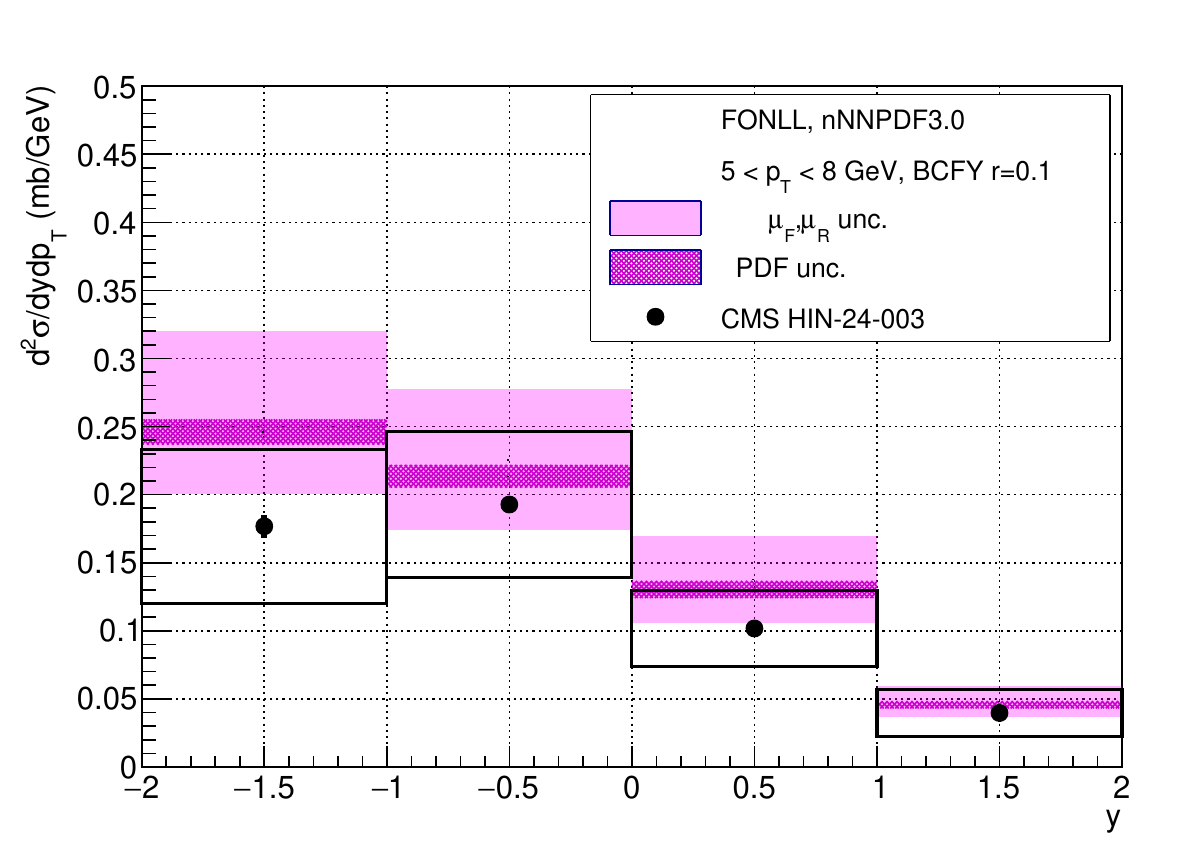}
    \label{fig:ydpt2nnpdfmc_cms}
\end{subfigure}
\hfill
\begin{subfigure}{0.49\textwidth}
    \centering
    \includegraphics[width=\textwidth]{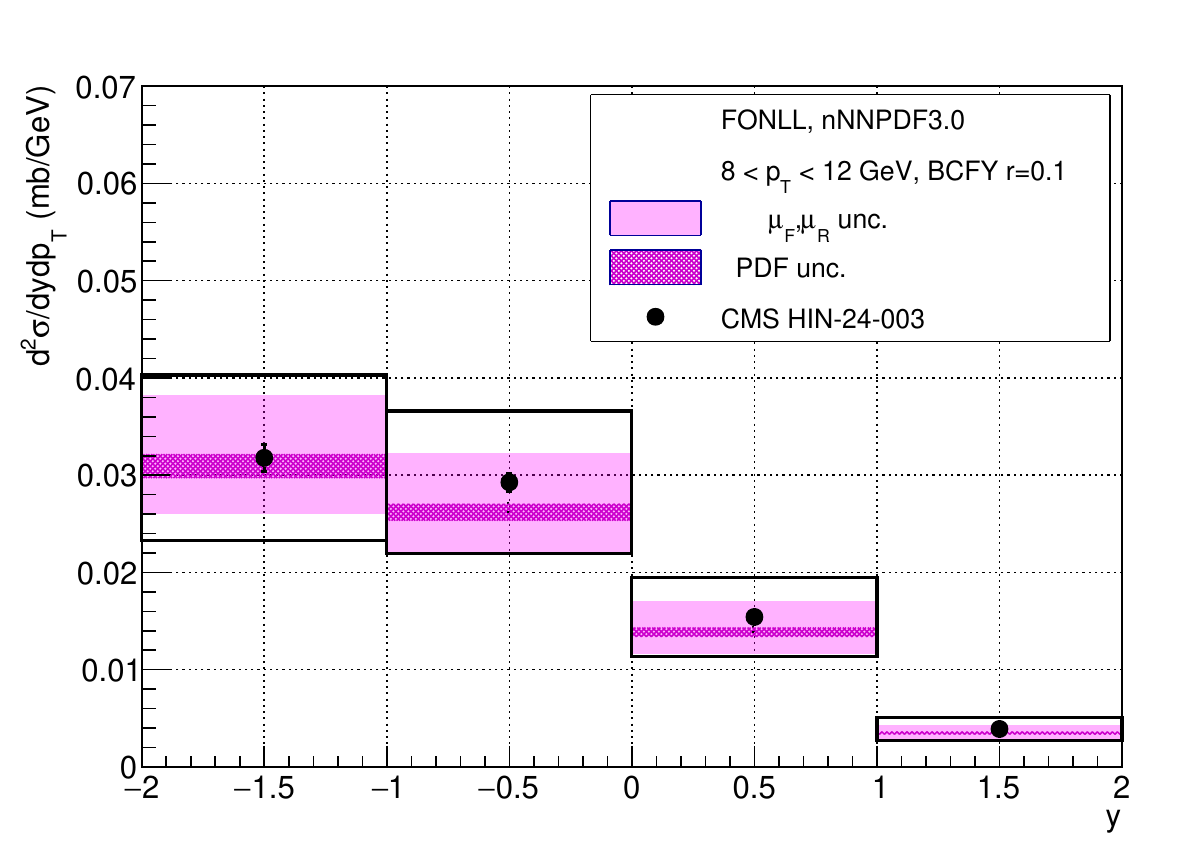}
    \label{fig:ydpt3nnpdfmc_cms}
\end{subfigure}
 \caption{Rapidity distribution  for the $D^0$  production in UPC PbPb collisions at ${\rm\sqrt{s_{\scriptscriptstyle NN}} }=5.36$ TeV with nNNPDF3.0 nuclear PDF in  $p_{T}$ bins $(2-5), (5-8), (8-12)$ GeV. Light magenta band: FONLL  calculation with factorization and renormalization scale variation for $m_c=1.3 \, \rm GeV$, darker magenta band PDF uncertainty, $m_c=1.5 \, \rm GeV$.  BCFY fragmentation function \cite{Braaten:1994bz,Cacciari:2003zu} with parameter $r=0.1$. Data are from CMS \cite{CMS:2025jjx}.}
 \label{fig:ydpt_nndpfbcfy_cms}
\end{figure}

\noindent
In Fig.~\ref{fig:ydpt_epps21bcfy_mc_cms}, the G$\gamma$A-FONLL predictions obtained with EPPS21 nuclear PDFs are compared to the  CMS data. Two calculations are shown: one with $m_c = 1.3$~GeV, which is the charm quark mass used in the extraction of the EPPS21 set, and another with $m_c = 1.5$~GeV to test the sensitivity of the prediction to the charm mass. Figure~\ref{fig:ydpt_nndpfbcfy_cms} shows analogous predictions obtained using the nNNPDF3.0 set ($m_c = 1.5$~GeV). 

\begin{figure}
\centering
\begin{subfigure}{0.49\textwidth}
    \centering
    \includegraphics[width=\textwidth]{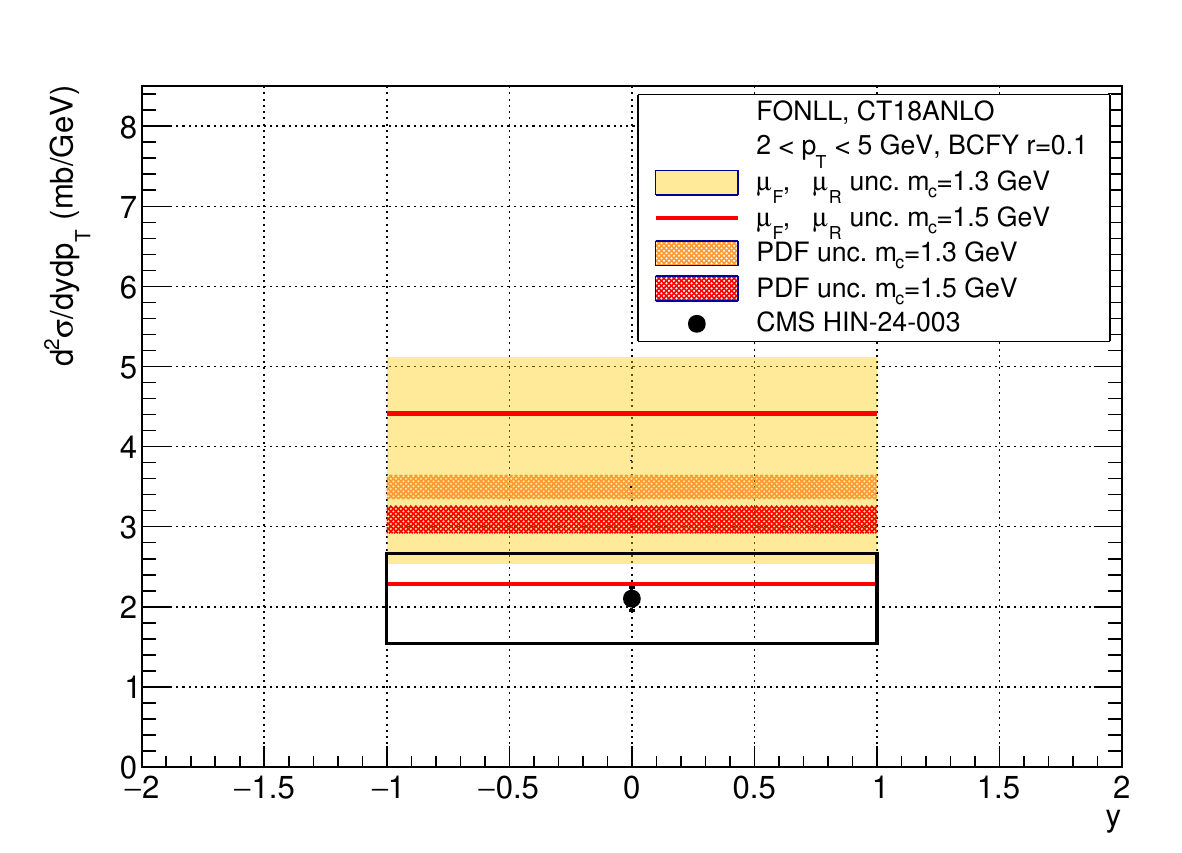}
    \label{fig:ydpt0protonmc_cms}
\end{subfigure}
\hfill
\begin{subfigure}{0.49\textwidth}
    \centering
    \includegraphics[width=\textwidth]{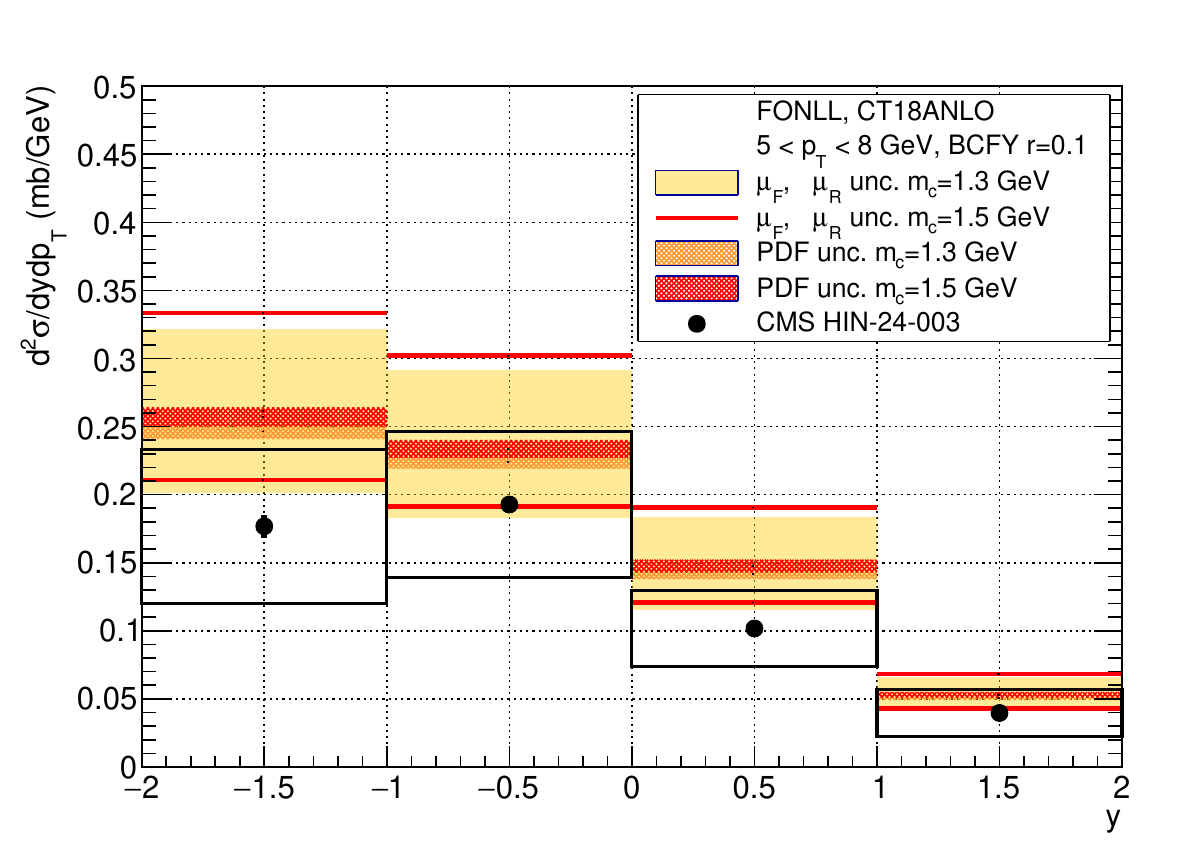}
    \label{fig:ydpt2protonmc_cms}
\end{subfigure}
\hfill
\begin{subfigure}{0.49\textwidth}
    \centering
    \includegraphics[width=\textwidth]{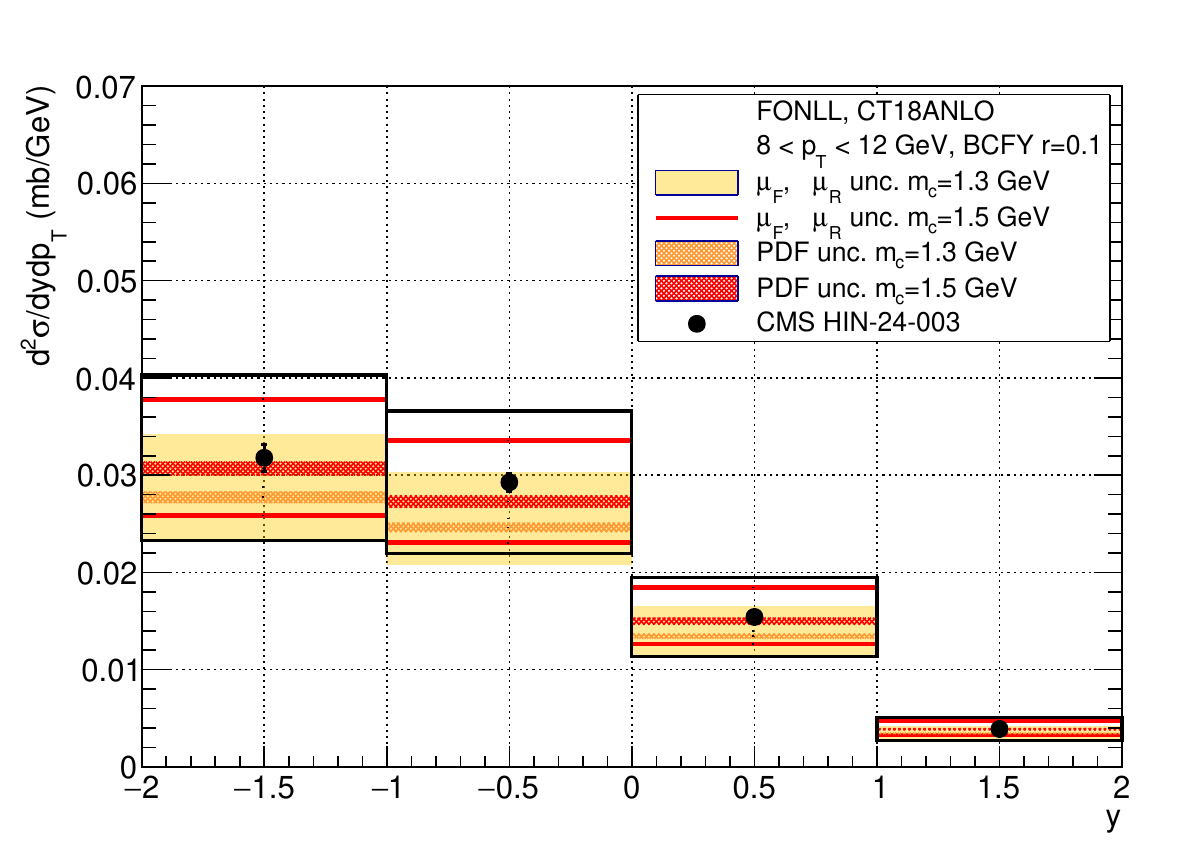}
    \label{fig:ydpt3protonmc_cms}
\end{subfigure}
 \caption{Rapidity distribution  for the $D^0$  production in UPC PbPb collisions at $\rm\sqrt{s_{\scriptscriptstyle NN}}=5.36$ TeV with CT18ANLO proton PDFin  $p_{T}$ bins $(2-5), (5-8), (8-12)$ GeV. Yellow band: FONLL  calculation with factorization and renormalization scale variation for $m_c=1.3 \, \rm GeV$, red lines band: FONLL  calculation with factorization and renormalization scale variation and $m_c=1.5 \, \rm GeV$, orange band: FONLL CT18ANLO \cite{Eskola:2021nhw} PDF uncertainty, $m_c=1.3 \, \rm GeV$. Red  band: FONLL CT18ANLO \cite{Eskola:2021nhw} PDF uncertainty, $m_c=1.5 \, \rm GeV$.  BCFY fragmentation function \cite{Braaten:1994bz,Cacciari:2003zu} with parameter $r=0.1$. Data are from CMS \cite{CMS:2025jjx}.}
 \label{fig:ydpt_ct18anlobcfy_mc_cms}
\end{figure}

\noindent
Within the still sizeable experimental uncertainties, the data are well described by the G$\gamma$A-FONLL predictions obtained using both EPPS21 and nNNPDF3.0 across all $p_T$ and rapidity bins. The central values of the data appear to favor calculations with a larger charm quark mass. \\[4pt]
\noindent
In Fig.~\ref{fig:ydpt_ct18anlobcfy_mc_cms}, the experimental data are compared to the G$\gamma$A-FONLL predictions obtained using the CT18ANLO proton PDF for two values of the charm quark mass, $m_c = 1.3$ and $1.5$ GeV. In the two lower $p_T$ intervals of Fig.\ref{fig:ydpt_ct18anlobcfy_mc_cms}, the calculations using proton PDFs systematically overshoot the data. The nuclear modifications in the $2 < p_T < 5$~GeV interval amount to approximately 20$\%$, while in the intermediate $5 < \pt < 8$~GeV range they lie between 5$\%$ and 15$\%$, with stronger suppression observed at higher rapidities. The comparison with predictions using proton PDFs suggests that the absence of nuclear modifications significantly reduces the agreement with the experimental data. The need for sizable nuclear effects is particularly evident for low-$p_T$ $\Dzero$ mesons at forward rapidity, where low-$x$ effects play a more prominent role. 
This effect certainly  deserves a further detailed analysis. We only tested here the parton densities obtained using standard linear DGLAP evolution but it is well known \cite{Gribov:1983ivg} that it is precisely the low \(p_T\), low \(x\) regime where gluon recombination effects are expected to be important and can lead to saturation.  They can be accounted for by including in the evolution  the non-linear terms in parton density. One can then potentially expect sizeable modifications to the cross section, as demonstrated  in  work on nuclear PDFs and nonlinear GLR-MQ equation \cite{Rausch:2022nkv}. The impact of nonlinear  evolution for parton densities used in the charm photoproduction process, while being beyond the scope of the present work,  certainly deserves to be investigated further, and we postpone it for the future study. \\[4pt]
\noindent
Finally, we note that in the highest $p_T$ bin, where nuclear effects are expected to be negligible, the central FONLL predictions, obtained with both proton and nuclear PDFs, lie slightly below the data. Interestingly, a similar trend was observed at HERA, where the FONLL prediction yielded a slightly softer $p_T$ spectrum than the experimental data (see Sec.~\ref{sec:hera}).

\section{Conclusions}
In this paper we presented calculations of charm photoproduction in electron–proton collisions at HERA and in ultraperipheral heavy-ion collisions (UPCs) at the LHC with the new G$\gamma$A–FONLL framework. \\[4pt] \noindent
We performed detailed comparison with HERA measurements of $D^*$ production in electron–proton collisions, employing  resummed FONLL  calculations as functions of the $D^*$ transverse momentum and rapidity. We examined how the predictions vary with different photon and proton PDF parametrizations and assessed the impact of alternative fragmentation functions. The calculations agree with earlier FONLL predictions and with the $ep$ measurements at HERA. As previously observed in earlier FONLL studies, the predictions tend to be marginally softer than the measurements at higher transverse momentum. Among the fragmentation models, the BCFY parametrization provides a better description of the experimental data than the PSSZ one, thanks to its harder spectrum at large $p_T$. Theoretical uncertainties remain sizable, with variations in the renormalization scale constituting the dominant contribution. \\[4pt]
After validating the framework with HERA data, we generated predictions for $D^{0}$-meson production in UPCs at the LHC. The effect of electromagnetic dissociation is incorporated in G$\gamma$A-FONLL to model the survival probability of the photon-emitting nucleus and enable direct comparison with the the UPC data. Various predictions for $\Dzero$ photoproduction in UPCs at the LHC, performed using the latest proton and nuclear PDF parametrizations and adopting different charm-quark mass hypotheses and fragmentation functions, were presented and discussed. The G$\gamma$A-FONLL predictions obtained with both EPPS21 and nNNPDF3.0 nuclear PDFs provide a good description of the recent CMS measurement of the $\Dzero$ photonuclear cross section in UPCs across all $p_T$ and rapidity intervals. In contrast, calculations based on the CT18ANLO proton PDF systematically overshoot the data, underscoring the importance of nuclear modifications. These effects are most pronounced for low-$p_T$ $\Dzero$ mesons at forward rapidity, where low-$x$ effects are expected to become more relevant. 
At high $p_T$, where nuclear effects are expected to be negligible, the G$\gamma$A--FONLL predictions appear slightly softer than the data. The predictions presented in this paper also highlights the sensitivity of the predictions to the assumed charm-quark mass. While confirming the strong experimental and theoretical constraining power of charm production in UPCs, this study also underscores the need for further work to quantify competing effects, namely electromagnetic dissociation, possible modifications to charm-meson fragmentation, and uncertainties in the charm-quark mass. If left unaccounted for, these factors could bias the interpretation of low-$x$ nuclear modifications. \\[6pt]
In summary, the G$\gamma$A–FONLL framework introduced here establishes a unified theoretical baseline for high-accuracy charm-photoproduction studies in both UPCs at the LHC, and in electron–ion collisions at the future Electron-Ion Collider. High-precision measurements across complementary regions of $x$ and $Q^{2}$, together with comparisons to saturation-based approaches such as the Color Glass Condensate~\cite{Iancu:2003xm,Gimeno-Estivill:2025rbw}, promise unprecedented insight into low-$x$ gluon dynamics and the onset of gluon saturation.
\label{sec:conclu}

\section*{Acknowledgments}
We thank Vadim Guzey, Mark Strikman, Petja Paakkinen, Ilkka Helenius, Cristian Baldenegro, Chris McGinn, Balázs Kovács, Spencer Klein for the fruitful discussions. GMI is supported by the U.S. Department of Energy grant No. DOE-SC0011088. AMS is  supported by the U.S. Department of Energy grant No. DE-SC-0002145 and within the framework of the of the Saturated Glue (SURGE) Topical Theory Collaboration.

\clearpage
\appendix
\section*{Appendix A)~Fixed-Order calculations for HERA}
\label{sec:appendixA}

\begin{figure}
\centering
\begin{subfigure}{0.49\textwidth}
    \centering
    \includegraphics[width=\textwidth]{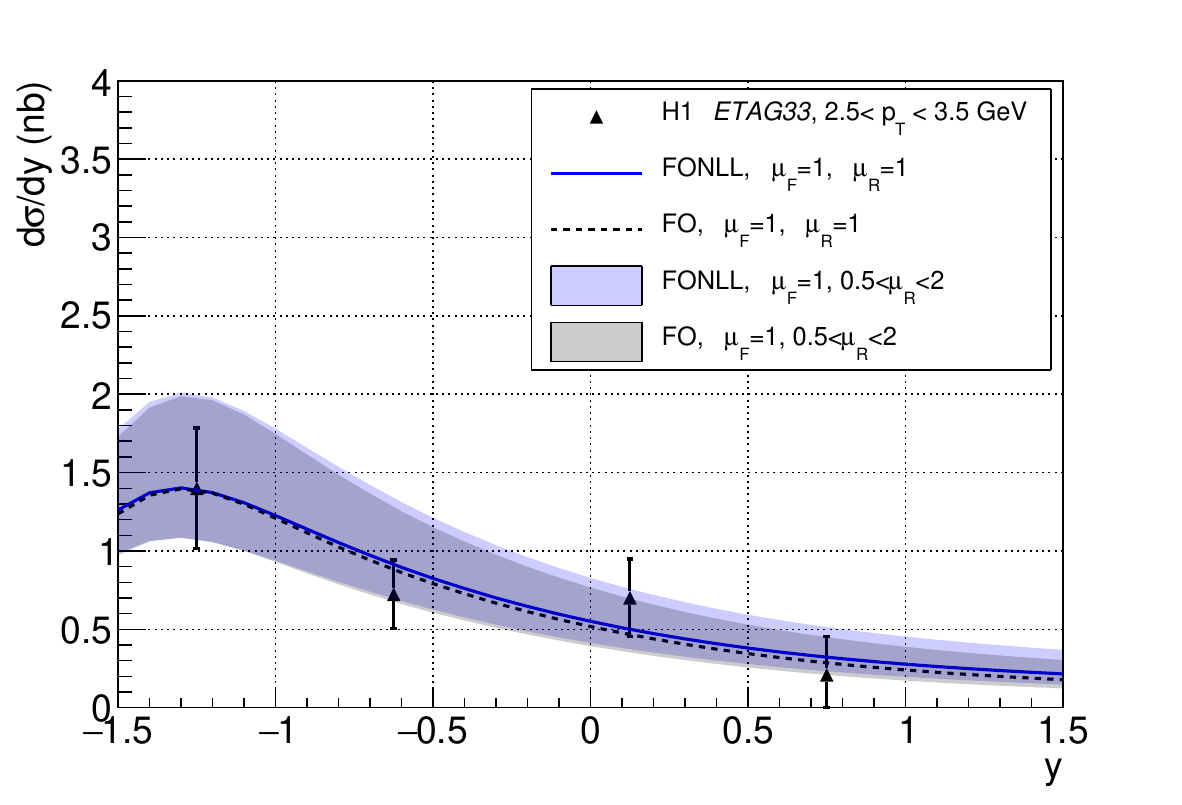}
    \label{fig:h1331y}
\end{subfigure}
\hfill
\begin{subfigure}{0.49\textwidth}
    \centering
    \includegraphics[width=\textwidth]{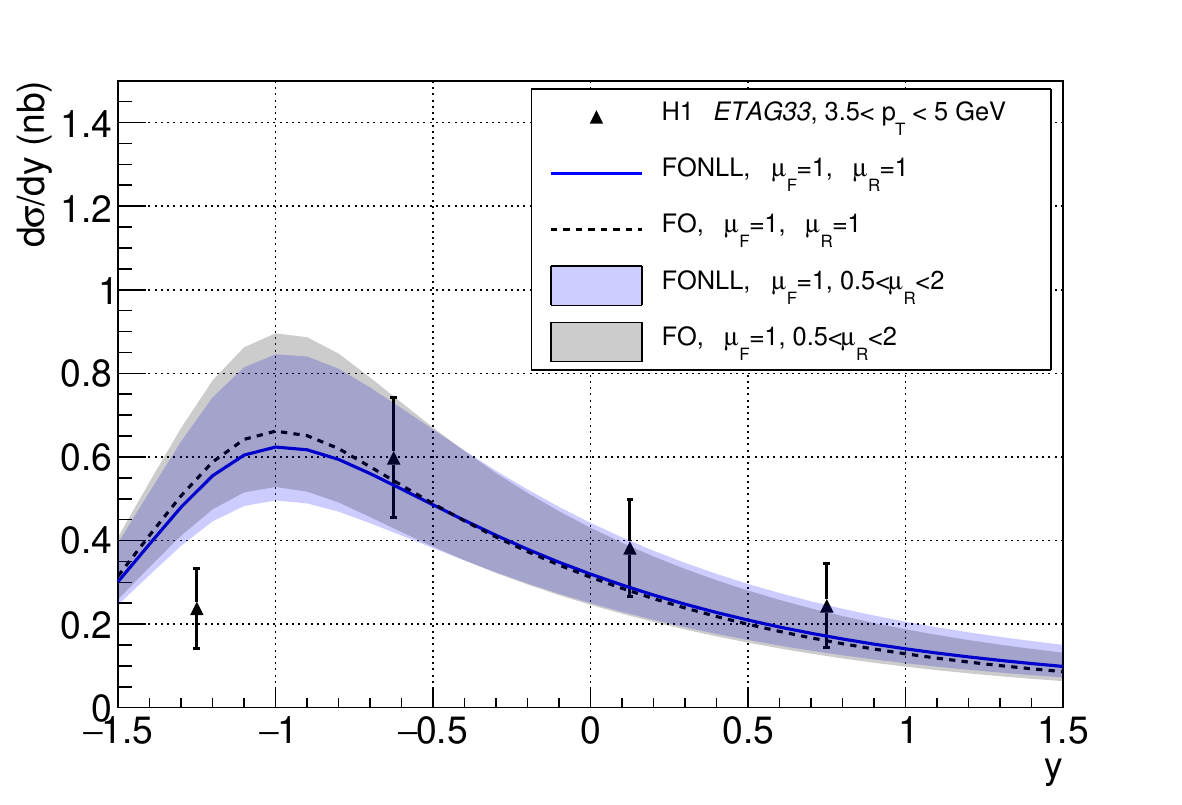}
    \label{fig:h1332y}
\end{subfigure} 
\begin{subfigure}{0.49\textwidth}
    \centering
    \includegraphics[width=\textwidth]{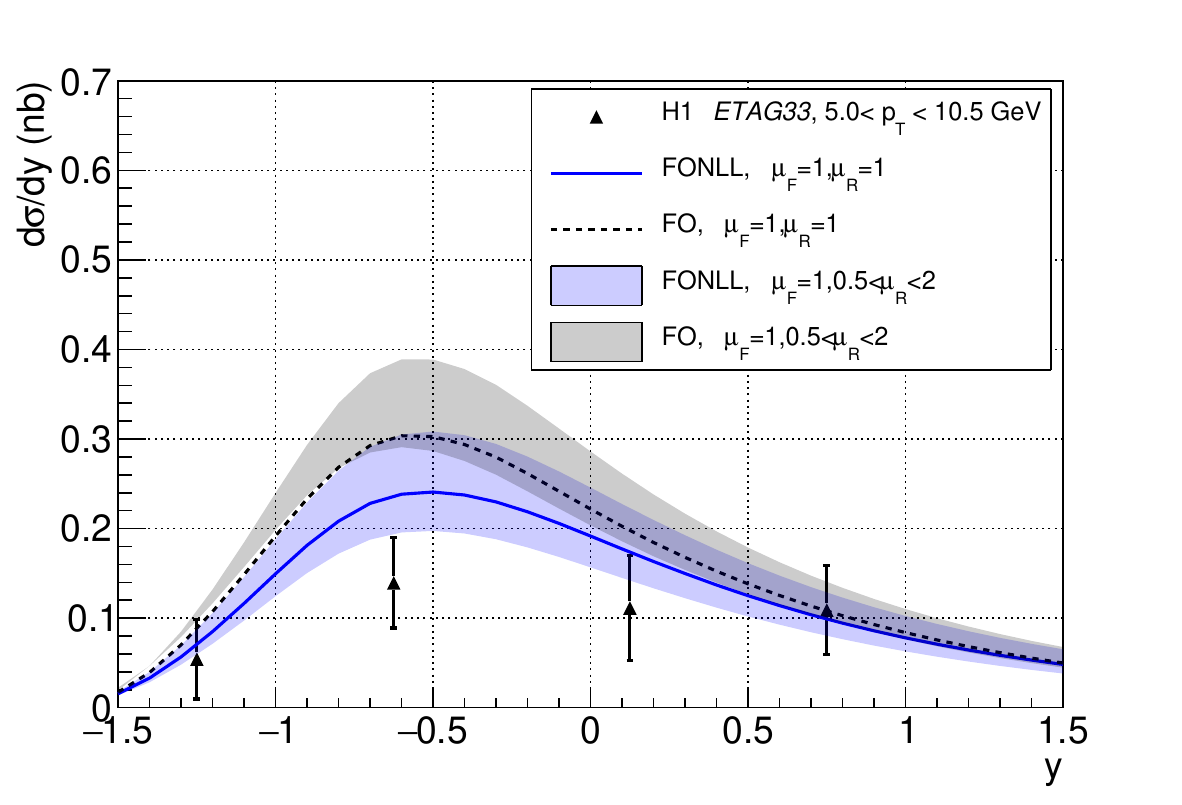}
    \label{fig:h1333y}
\end{subfigure}
\hfill
\begin{subfigure}{0.49\textwidth}
    \centering
    \includegraphics[width=\textwidth]{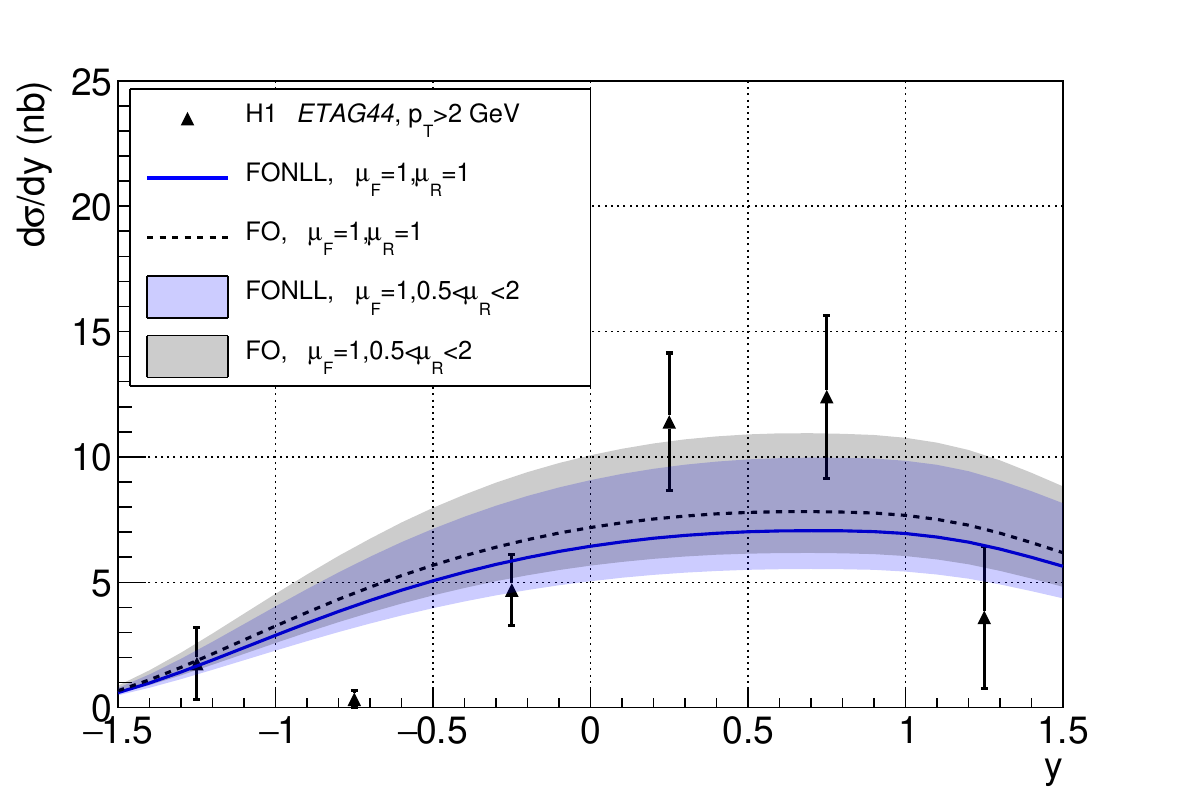}
    \label{fig:h144y}
\end{subfigure}
 \caption{Rapidity distribution of $D^*$ mesons in photoproduction in electron-proton collisions at HERA.  FONLL calculation \cite{Frixione:2002zv} $\mu_F=\mu_R=1$ (blue solid), FO calculation (black dashed). Shaded blue indicates variation of $0.5<\mu_R<2$ while $\mu_F=1$ is fixed; shaded grey indicates renormalization scale variation for FO. PSSZ fragmentation function is used with $\varepsilon=0.02$ for FONLL and $\varepsilon=0.035$ for FO. Compared with data from H1  \cite{H1:1998csb}, for different $p_T$ bins. Note different vertical scales. Positive rapidity is proton going direction.}
 \label{fig:h1datay}
\end{figure}

\begin{figure}
\centering
\begin{subfigure}{0.49\textwidth}
    \centering
    \includegraphics[width=\textwidth]{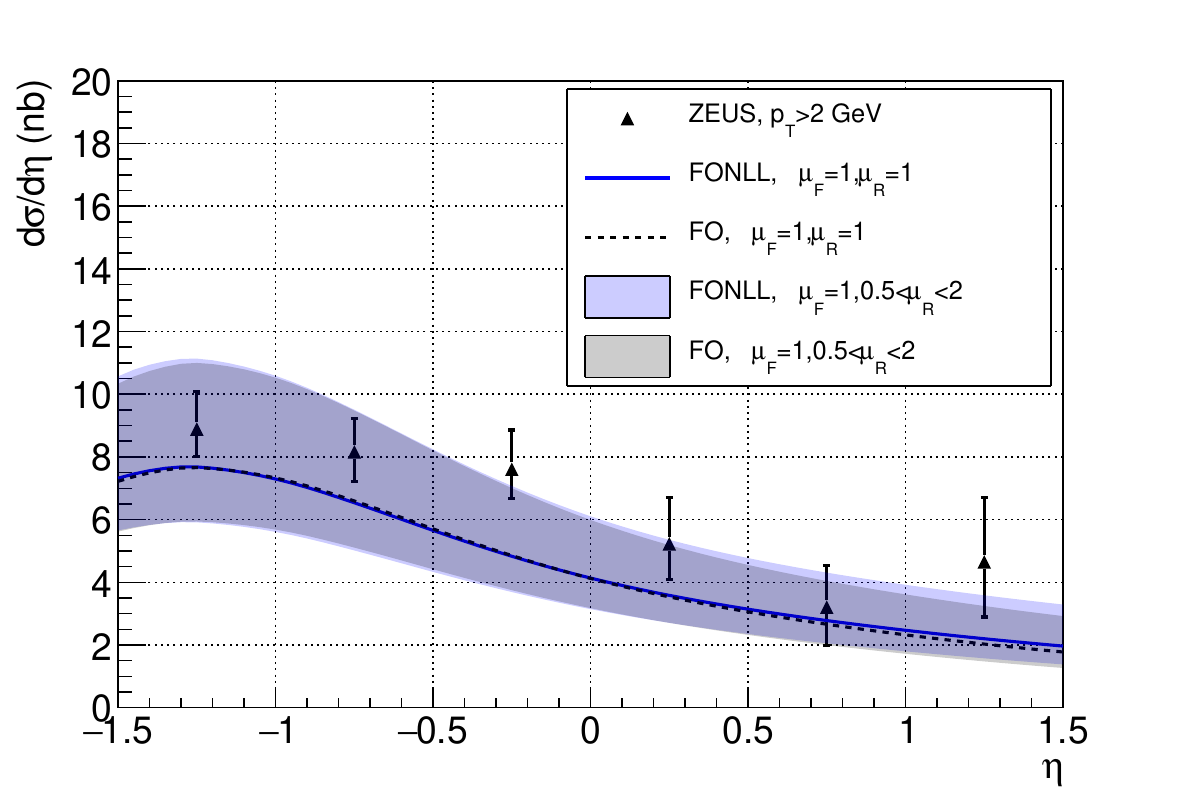}
    \label{fig:zeusy2}
\end{subfigure}
\hfill
\begin{subfigure}{0.49\textwidth}
    \centering
    \includegraphics[width=\textwidth]{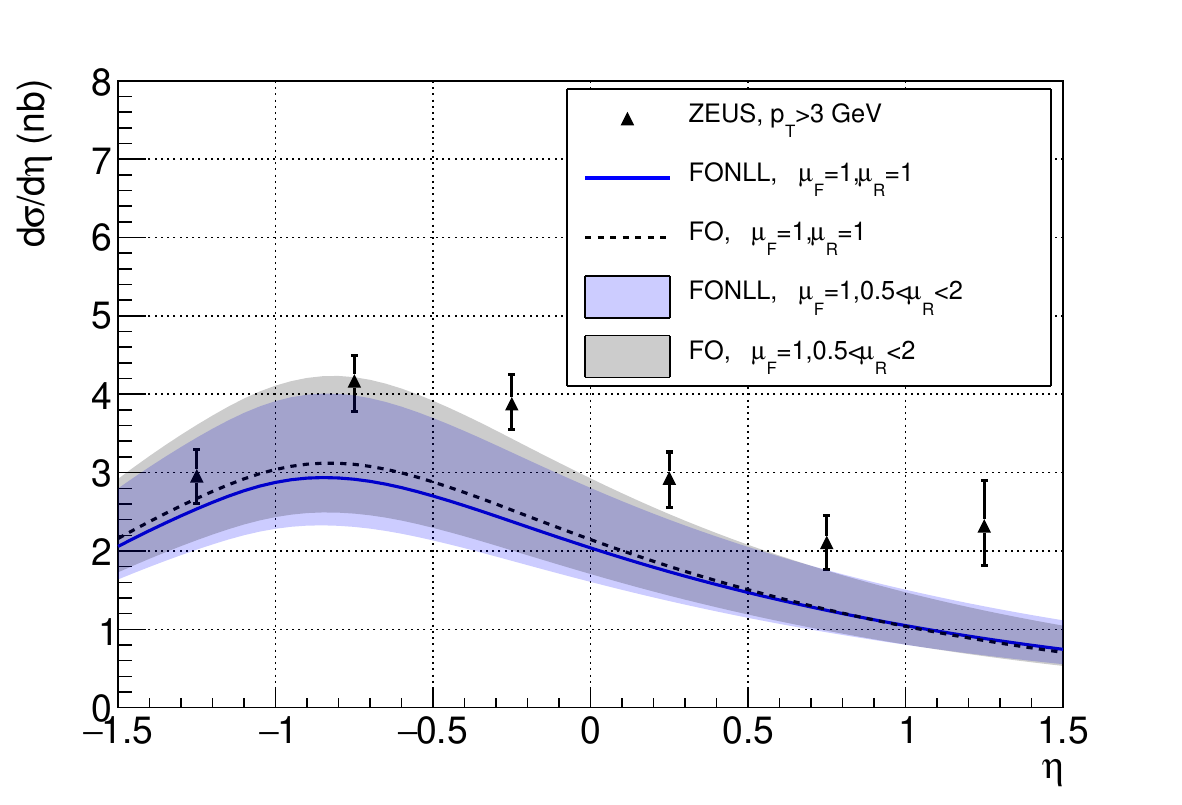}
    \label{fig:zeusy3}
\end{subfigure} 
\begin{subfigure}{0.49\textwidth}
    \centering
    \includegraphics[width=\textwidth]{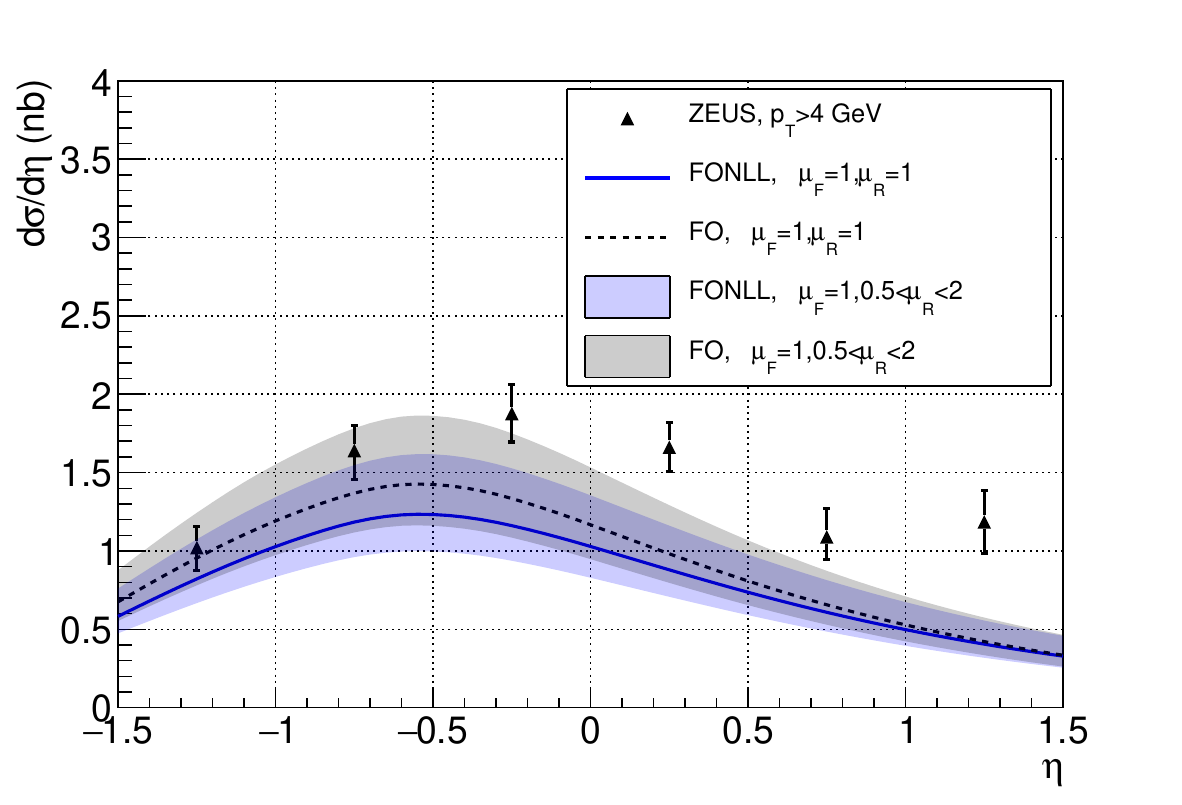}
    \label{fig:zeusy4}
\end{subfigure}
\hfill
\begin{subfigure}{0.49\textwidth}
    \centering
    \includegraphics[width=\textwidth]{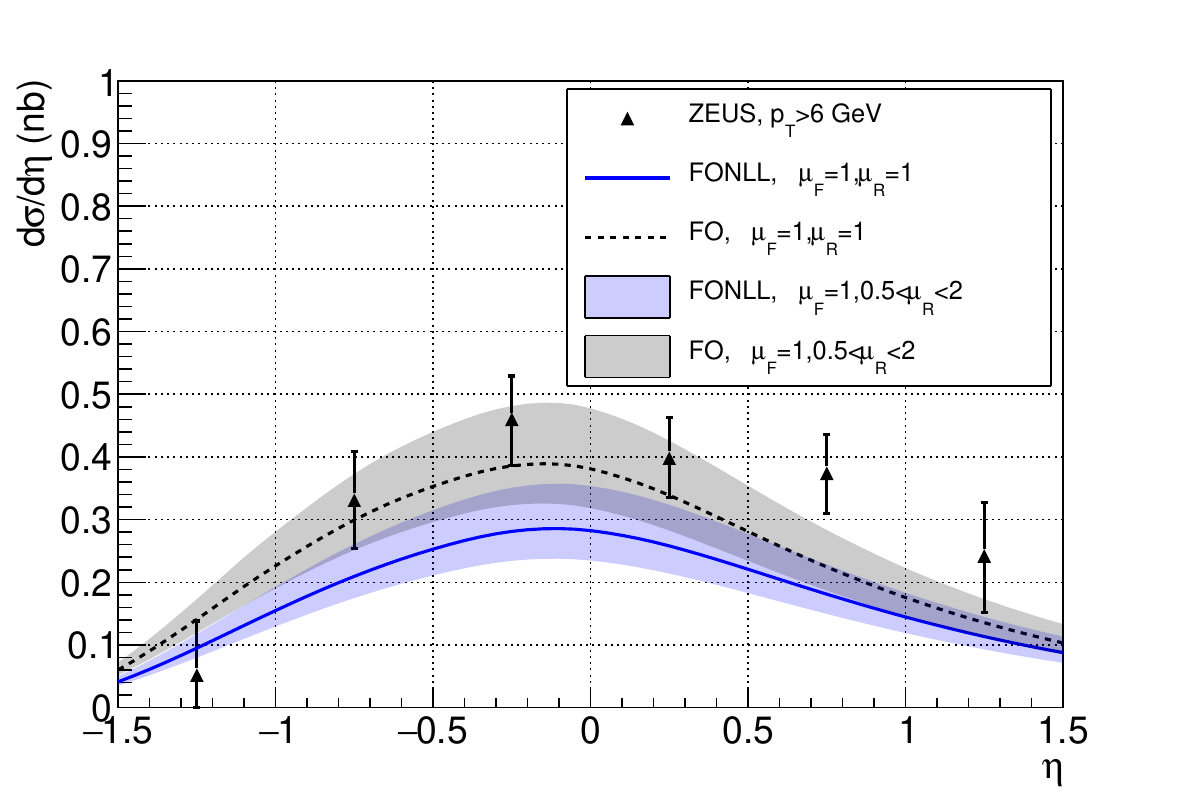}
    \label{fig:zeusy6}
\end{subfigure}
 \caption{Pseudorapidity distribution of $D^*$ mesons in photoproduction in electron-proton collisions at HERA. FONLL calculation \cite{Frixione:2002zv} $\mu_F=\mu_R=1$ (blue solid), FO calculation (black dashed). Shaded blue indicates variation of $0.5<\mu_R<2$ while $\mu_F=1$ is fixed; shaded grey indicates renormalization scale variation for FO. PSSZ fragmentation function is used with $\varepsilon=0.02$ for FONLL and $\varepsilon=0.035$ for FO. Compared with data from ZEUS  \cite{ZEUS:1998wxs}, for different minimum $p_T$ cuts: $2,3,4,6$ GeV. Note different vertical scales. Positive pseudorapidity is proton direction.}
 \label{fig:zeusdatay}
\end{figure}

\begin{figure}
\centering
\begin{subfigure}{0.49\textwidth}
    \centering
    \includegraphics[width=\textwidth]{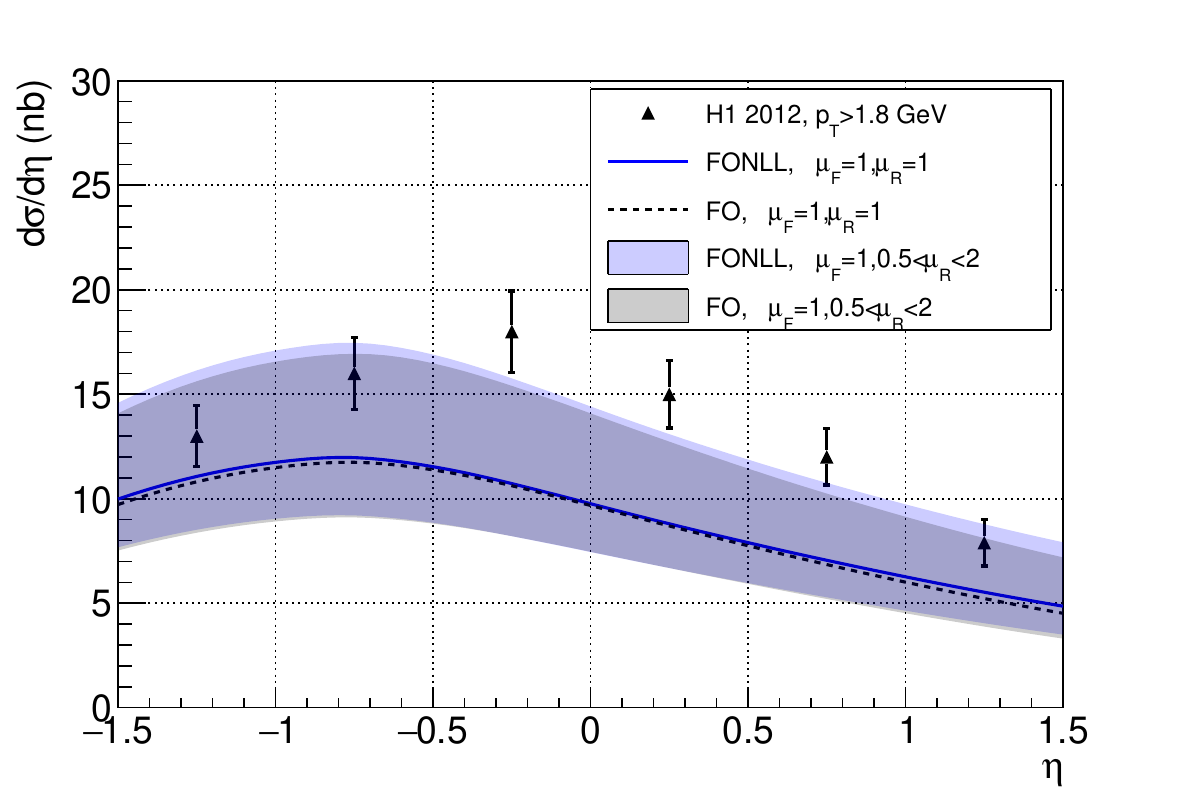}
    \label{fig:h1eta}
\end{subfigure}
\hfill
\begin{subfigure}{0.49\textwidth}
    \centering
    \includegraphics[width=\textwidth]{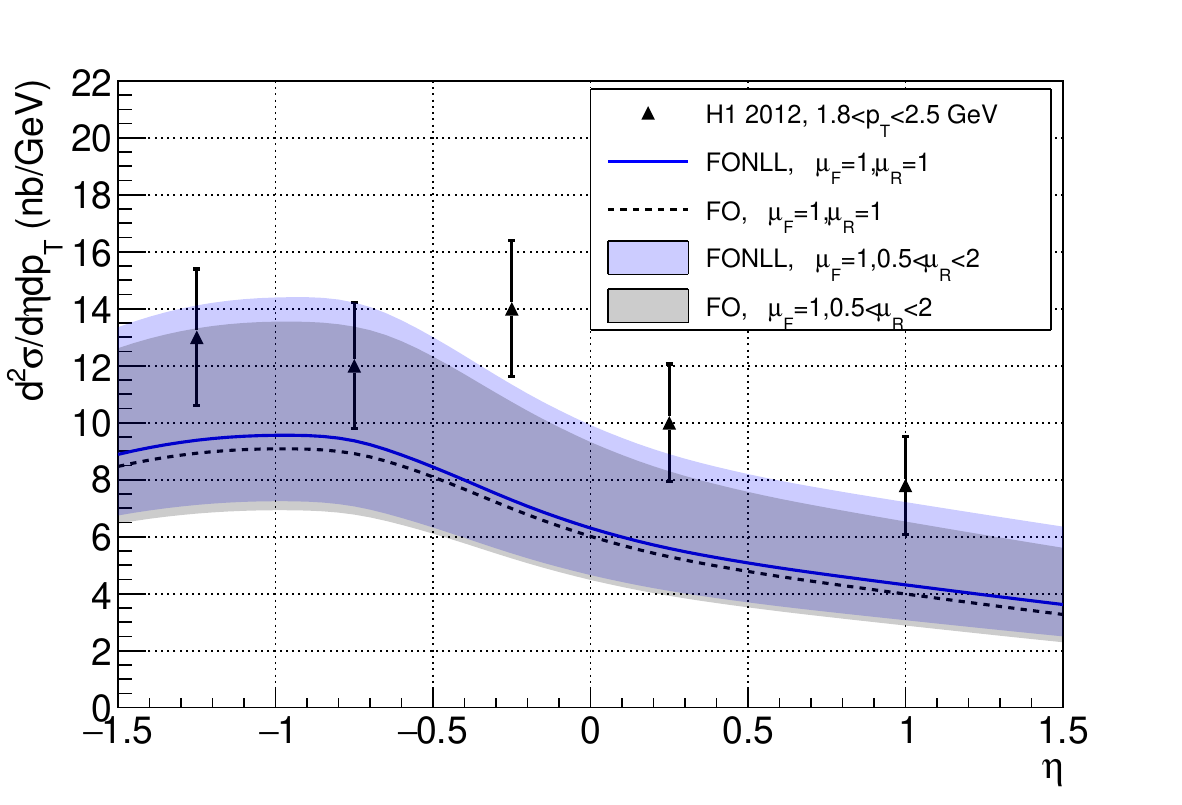}
    \label{fig:h1eta1}
\end{subfigure} 
\begin{subfigure}{0.49\textwidth}
    \centering
    \includegraphics[width=\textwidth]{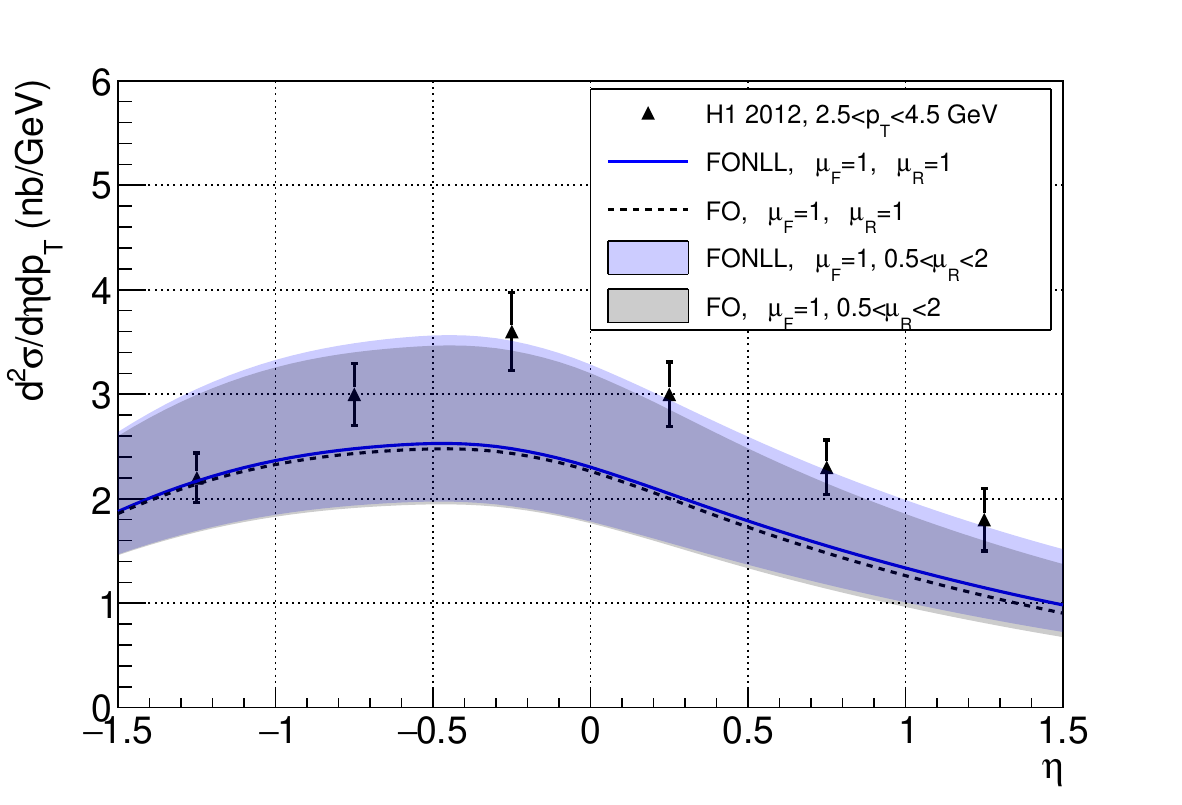}
    \label{fig:h1eta2}
\end{subfigure}
\hfill
\begin{subfigure}{0.49\textwidth}
    \centering
    \includegraphics[width=\textwidth]{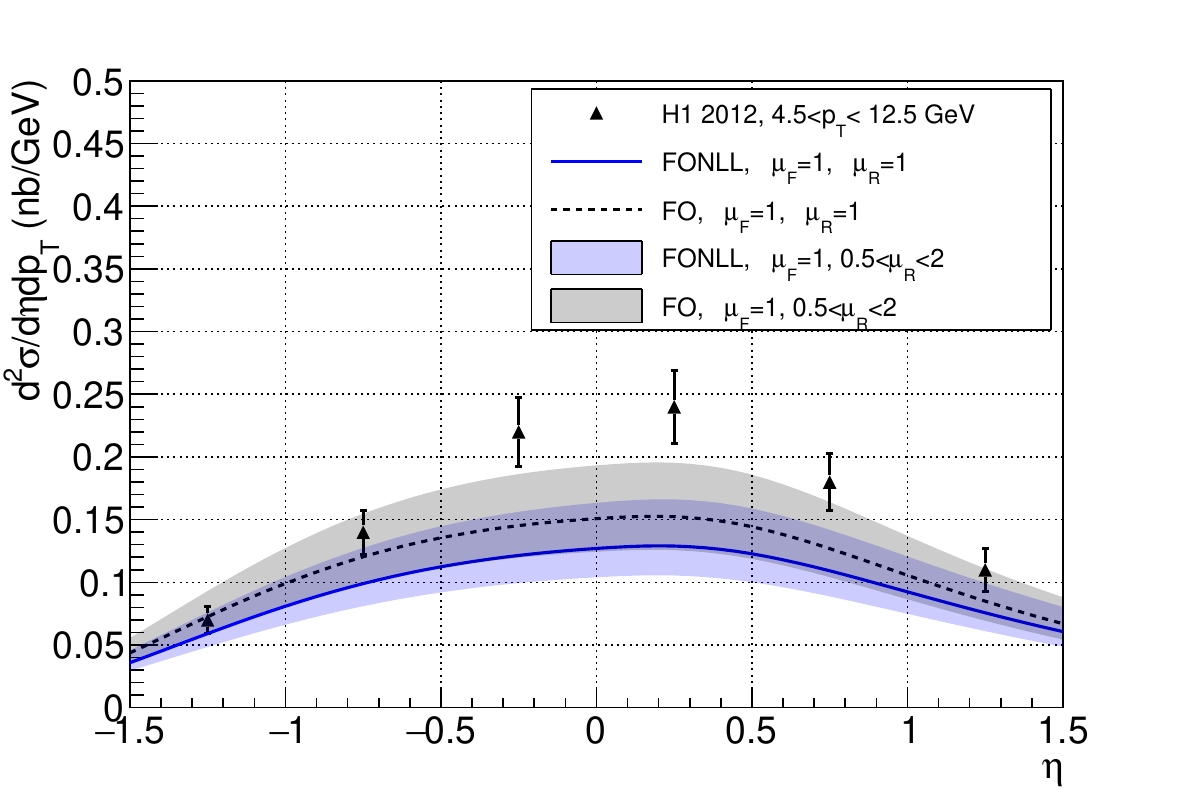}
    \label{fig:h1eta3}
\end{subfigure}
 \caption{Pseudorapidity distribution of $D^*$ mesons in photoproduction in electron-proton collisions at HERA.  FONLL calculation \cite{Frixione:2002zv} $\mu_F=\mu_R=1$ (blue solid), FO calculation (black dashed). Shaded blue indicates variation of $0.5<\mu_R<2$ while $\mu_F=1$ is fixed; shaded grey indicates renormalization scale variation for FO. PSSZ fragmentation function is used with $\varepsilon=0.02$ for FONLL and $\varepsilon=0.035$ for FO. Compared with data from H1  \cite{H1:2011myz}, for different $p_T$ bins. Note that data and results in three $p_T$ bins $(1.8,2.5), (2.5,4.5), (4.5,12.5)$ are presented as double differential cross sections. Positive pseudorapidity is proton direction. \\ \\}
 \label{fig:h1datay2011}
\end{figure}

In Figs.~\ref{fig:h1datay}, \ref{fig:zeusdatay}, \ref{fig:h1datay2011} the fixed-order (FO) predictions for the rapidity and pseudorapidity distributions of $D^{*+}$ mesons in electron-proton collisions at HERA are compared to the G$\gamma$A--FONLL predictions and to the experimental data.
The blue bands represent the renormalization-scale uncertainty of the FONLL calculation, whereas the grey bands correspond to the fixed-order (FO) result; the solid blue and dashed black curves denote their respective central predictions. The factorization scale is fixed at $\mu_F/\mu_0 = 1$. The PSSZ fragmentation parameter is set to $\varepsilon = 0.02$ for FONLL and $\varepsilon = 0.35$ for the FO calculation.\\[4pt]

\clearpage

\clearpage

\section*{Appendix B) Supplementary predictions for UPCs with nuclear and proton PDFs}
\begin{figure}
    \centering
    \includegraphics[width=0.6\linewidth]{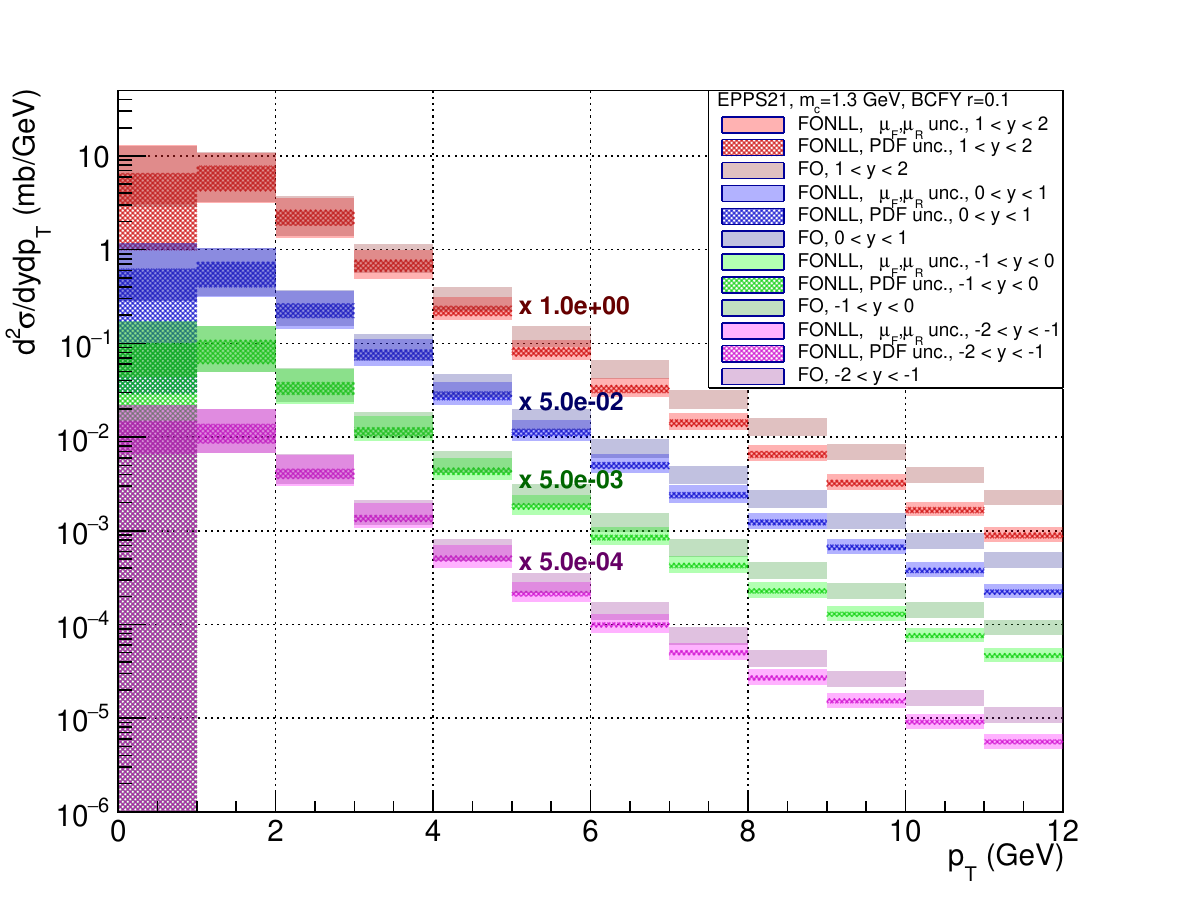}
    \caption{Transverse momentum distribution for the $D^0$ production in UPC collisions  at $\rm \sqrt{s_{\scriptscriptstyle  NN}}=5.36$ TeV in four bins of rapidity: $(-2,-1), (-1,0), (0,1), (1,2)$. FONLL and FO calculation with EPPS21 nuclear PDF, charm mass $m_c=1.3 \, \rm GeV$ with BCFY fragmentation function. Wider bands: factorization and renormalization scale variation, smaller (darker) bands: FONLL with PDF uncertainty.}
    \label{fig:ptepps21}
\end{figure}

\begin{figure}
    \centering
    \includegraphics[width=0.6\linewidth]{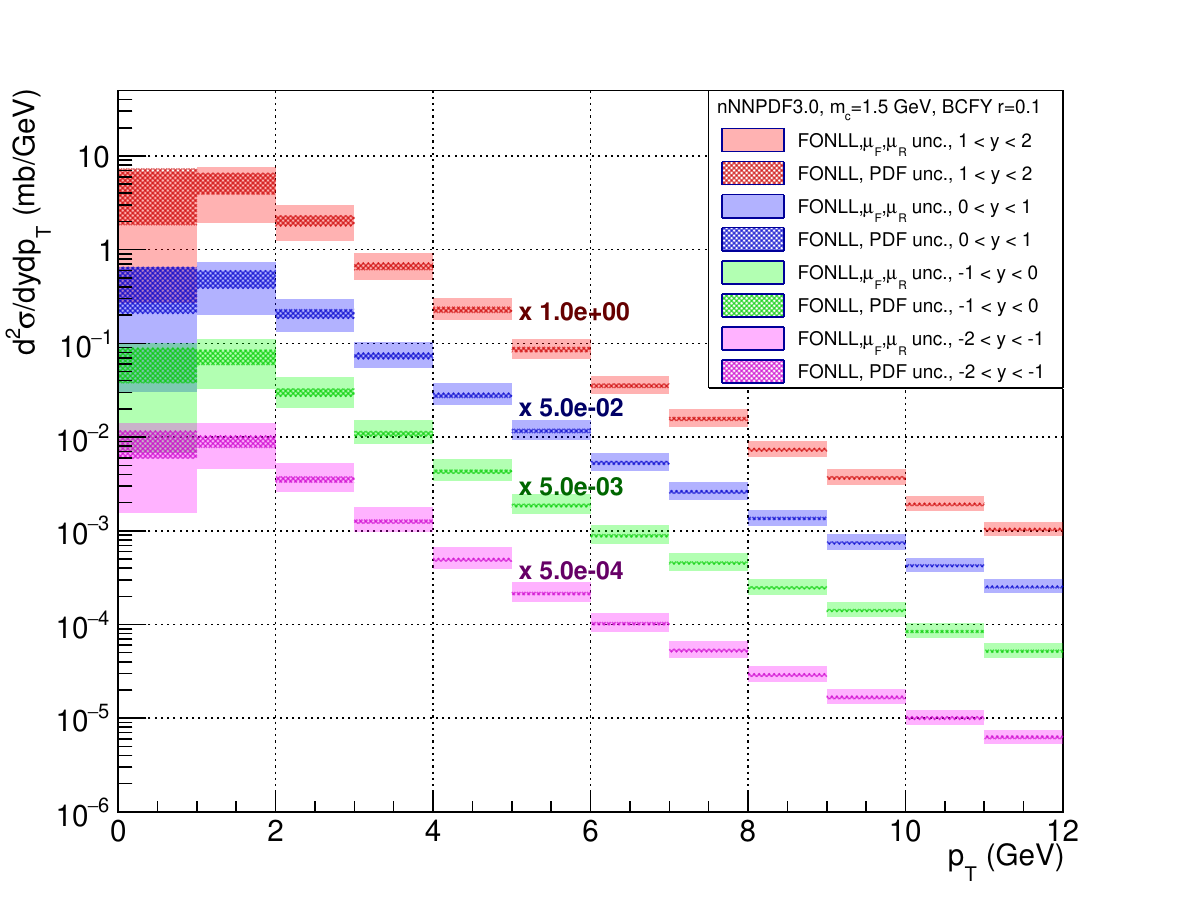}
    \caption{Transverse momentum distribution for the $D^0$ production in UPC collisions  at $\rm \sqrt{s_{\scriptscriptstyle  NN}}=5.36$ TeV in four bins of rapidity: $(-2,-1), (-1,0), (0,1), (1,2)$. FONLL  calculation with nNNPDF3.0 nuclear PDF, charm mass $m_c=1.5\,\rm GeV$ with BCFY fragmentation function. Wider bands: factorization and renormalization scale variation, smaller (darker) bands:   PDF uncertainty.}
    \label{fig:ptnnpdf30}
\end{figure}

\begin{figure}
    \centering
    \includegraphics[width=0.6\linewidth]{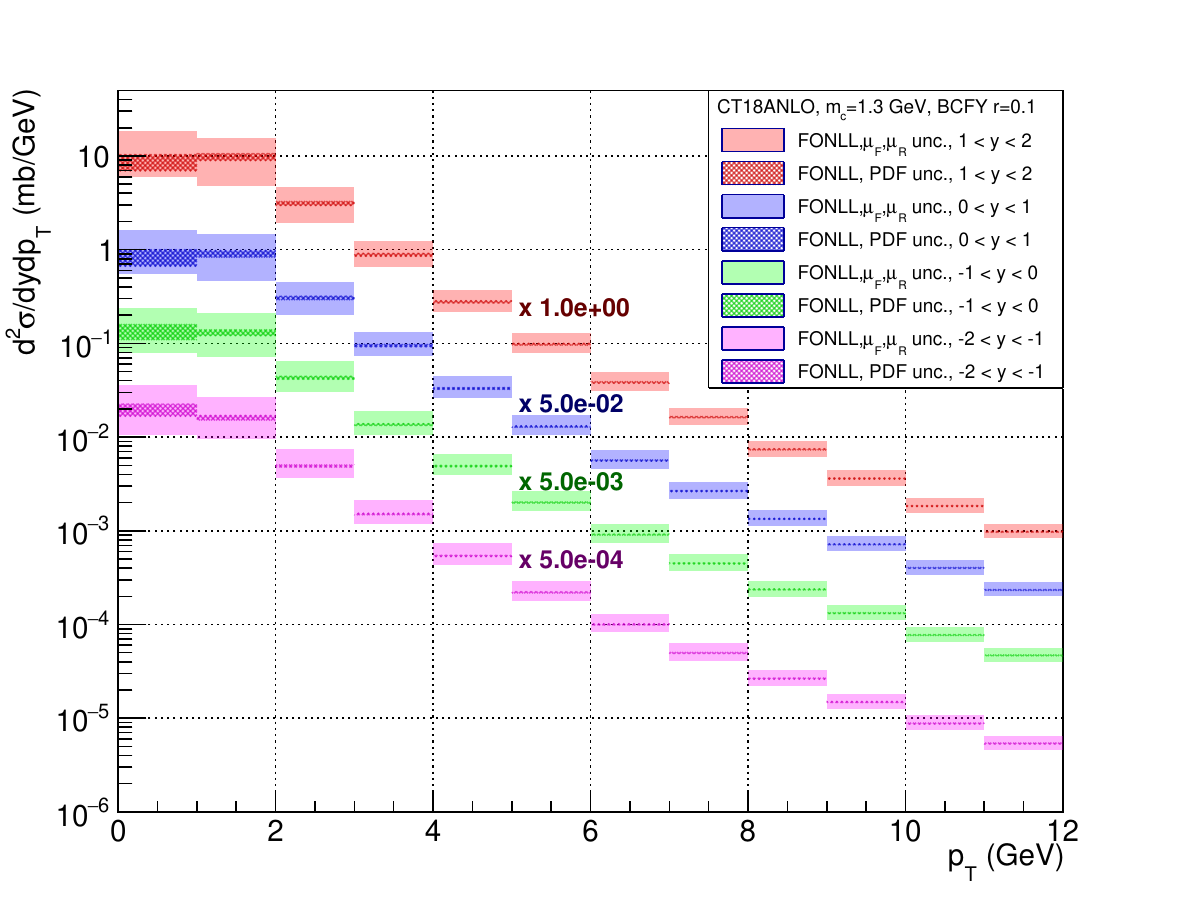}
    \caption{Transverse momentum distribution for the $D^0$ production in UPC collisions  at $\rm \sqrt{s_{\scriptscriptstyle  NN}}=5.36$ TeV in four bins of rapidity: $(-2,-1), (-1,0), (0,1), (1,2)$. FONLL  calculation with CT18ANLO proton PDF (scaled by $A=208$), charm mass $m_c=1.3 \, \rm GeV$ with BCFY fragmentation function. Wider bands: factorization and renormalization scale variation, smaller (darker) bands:   PDF uncertainty.}
    \label{fig:ptct18anlo}
\end{figure}

\begin{figure}
\centering
\begin{subfigure}{0.49\textwidth}
    \centering
    \includegraphics[width=\textwidth]{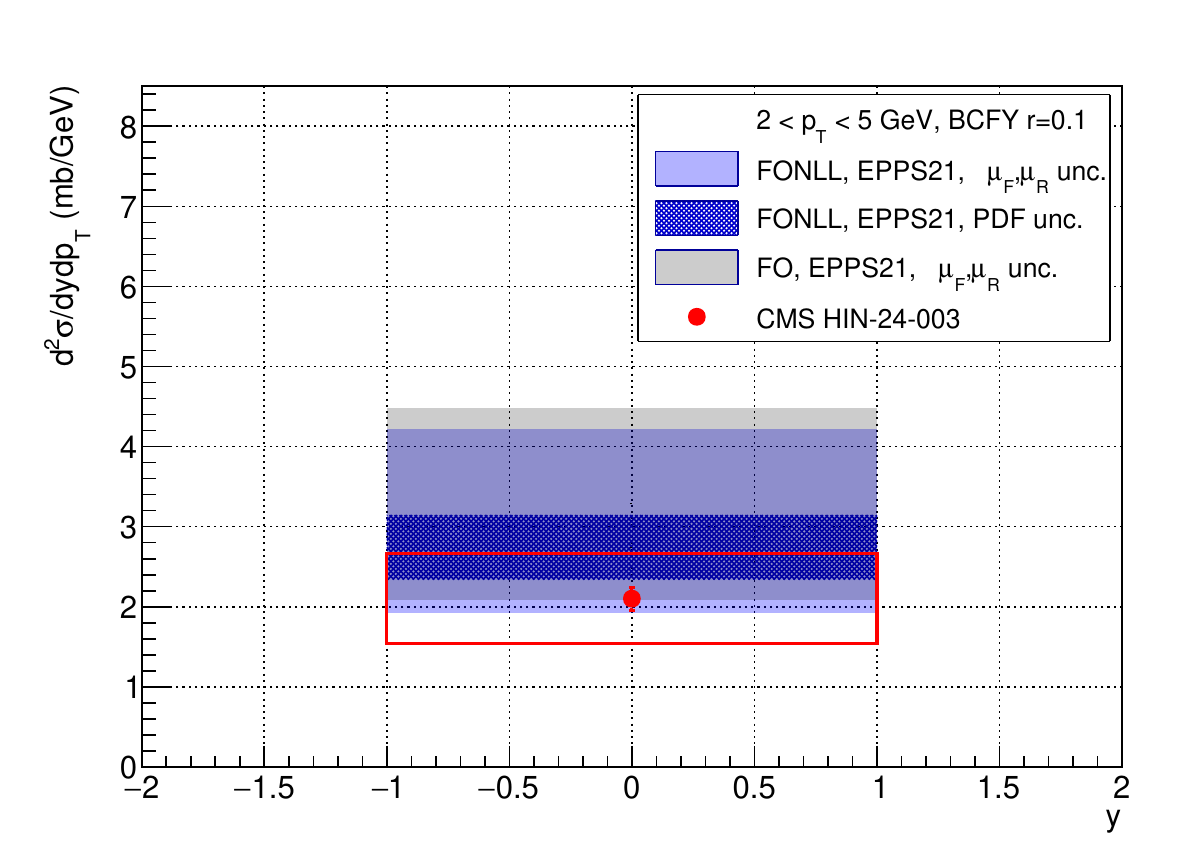}
    \label{fig:ydpt0}
\end{subfigure}
\hfill
\begin{subfigure}{0.49\textwidth}
    \centering
    \includegraphics[width=\textwidth]{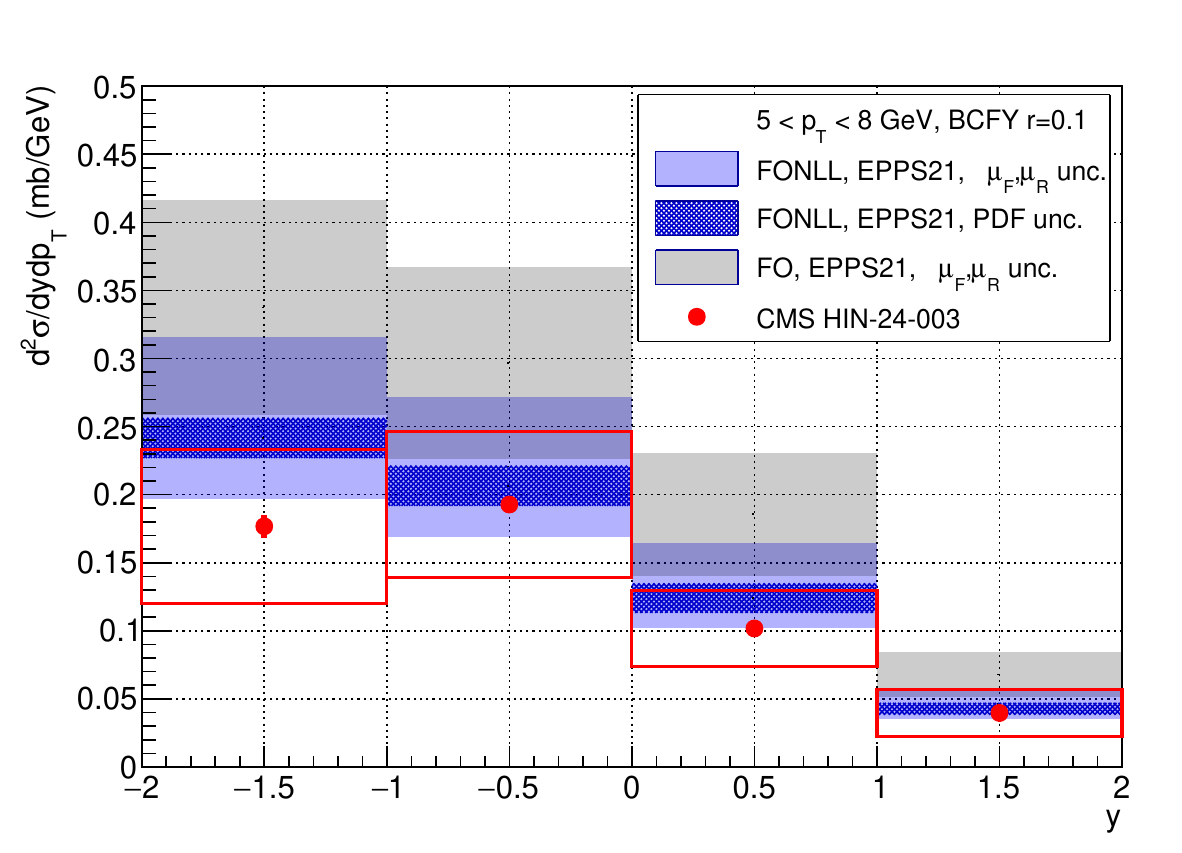}
    \label{fig:ydpt2}
\end{subfigure}
\hfill
\begin{subfigure}{0.49\textwidth}
    \centering
    \includegraphics[width=\textwidth]{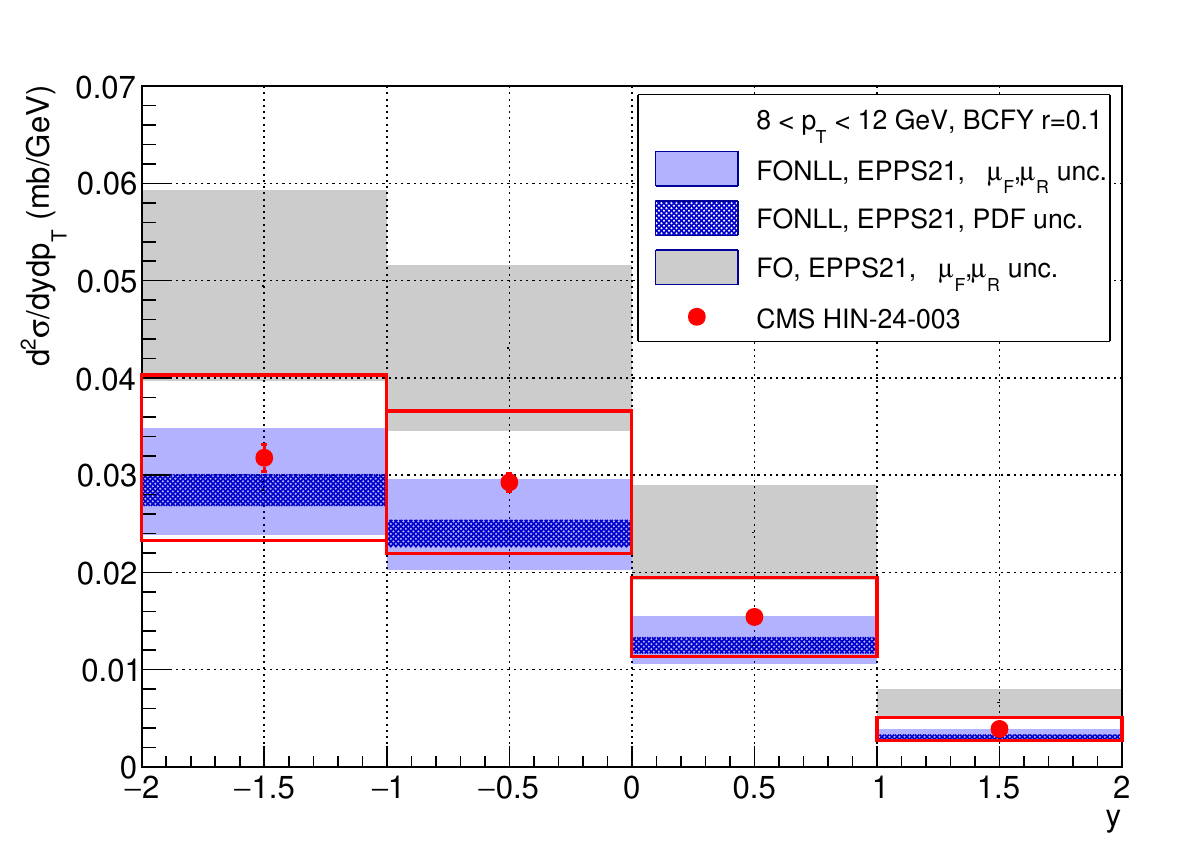}
    \label{fig:ydpt3}
\end{subfigure}
 \caption{Rapidity distribution  for the $D^0$  production in UPC PbPb collisions at $\sqrt{\rm s_{\scriptscriptstyle{NN}}}=5.36$ TeV in  $p_{T}$ bins $(2-5), (5-8), (8-12)$ GeV.  Light blue band: FONLL  calculation with factorization and renormalization scale variation, grey band: FONLL  calculation with factorization and renormalization scale variation, dark blue band: FONLL EPPS21 \cite{Eskola:2021nhw} PDF uncertainty. Both FONLL and FO calculation done with BCFY fragmentation function \cite{Braaten:1994bz,Cacciari:2003zu} with parameter $r=0.1$. Data are from CMS \cite{CMS:2025jjx}.}
 \label{fig:ydpt_epps21bcfy_cms}
\end{figure}

\noindent
In Figs.~\ref{fig:ptepps21} and \ref{fig:ptnnpdf30} we present G$\gamma$A–FONLL predictions for UPCs obtained with the EPPS21 (Fig.~\ref{fig:ptepps21}) and nNNPDF3.0 (Fig.~\ref{fig:ptnnpdf30}) nuclear PDFs, shown as functions of the $\Dzero$ transverse momentum in several rapidity intervals. Figure~\ref{fig:ptepps21} also includes the corresponding fixed-order (FO) calculation. Shaded bands represent systematic uncertainties from independent variations of the renormalization and factorization scales; for the FONLL curves the PDF uncertainty is displayed as well. The distributions are extended down to very low $p_T$, where both the scale‐variation and PDF uncertainties become large. This sizable scale dependence is correlated with the PDF uncertainty, because at low $p_T$ the calculation probes parton densities at low factorization scales, where the PDFs carry their largest intrinsic uncertainties. In Fig.~\ref{fig:ptnnpdf30}, which shows the predictions obtained with the nNNPDF3.0 set, the uncertainty band at low $p_T$ is slightly narrower than in the EPPS21 case, likely because the calculation employs a larger charm-quark mass and thus probes a correspondingly higher renormalization and factorization scale. \\[4pt] \noindent
In Fig.~\ref{fig:ptct18anlo} we show the same distribution, this time evaluated with the proton CT18ANLO PDF and scaled by the mass number $A$, therefore without nuclear-shadowing effects. Scale-variation uncertainties at low $p_T$ remain sizeble but are smaller than in the EPPS21 case, reflecting the tighter constraints on the proton PDFs.
\noindent
Figure~\ref{fig:ydpt_epps21bcfy_cms} displays the rapidity distributions obtained with G$\gamma$A–FONLL and fixed-order (FO) calculations based on EPPS21 and using the BCFY fragmentation function. The predictions are evaluated in the same $p_T$ and $y$ intervals as the CMS measurement to permit direct comparison with the data. Consistent with the pattern observed at HERA, the FO prediction lies slightly above the FONLL result at large $p_T$.

\clearpage

\bibliography{mybib}

\end{document}